\documentclass[epj]{svjour}
\usepackage{graphicx}
\usepackage{color}

\def\mb#1{\setbox0=\hbox{$#1$}\kern-.025em\copy0\kern-\wd0
\kern-0.05em\copy0\kern-\wd0\kern-.025em\raise.0233em\box0}

\newcommand {\be}{\begin{equation}}
\newcommand {\ee}{\end{equation}}

\begin{document}
   \title{The HMF model for fermions and bosons}

\author{P.H. Chavanis}

\institute{Laboratoire de Physique Th\'eorique (IRSAMC), CNRS and UPS, Universit\'e de Toulouse, F-31062 Toulouse, France}

\titlerunning{The fermionic and bosonic HMF models}

   \date{To be included later }

   \abstract{We study the thermodynamics of quantum particles with long-range interactions at $T=0$. Specifically, we generalize the Hamiltonian Mean Field (HMF) model to the case of fermions and bosons. In the case of fermions, we consider the Thomas-Fermi approximation that becomes exact in a proper thermodynamic limit. The equilibrium configurations, described by the Fermi (or waterbag) distribution, are equivalent to polytropes with index $n=1/2$.  In the case of bosons, we consider the Hartree approximation that becomes exact in a proper  thermodynamic limit. The equilibrium configurations are solutions of the mean field Schr\"odinger equation with a cosine interaction. We show that the homogeneous phase, that is unstable in the classical regime, becomes stable in the quantum regime. This takes place through a first order phase transition for fermions and through a second order phase transition for bosons where the control parameter is the normalized Planck constant. In the case of fermions, the homogeneous phase is stabilized by the Pauli exclusion principle while for bosons the stabilization is due to the Heisenberg uncertainty principle. As a result, the thermodynamic limit is different for fermions and bosons. We point out analogies between the quantum HMF model and the concepts of fermion and boson stars in astrophysics. Finally, as a by-product of our analysis, we obtain new results concerning the Vlasov dynamical stability of the waterbag distribution. We show that spatially homogeneous waterbag distributions are Vlasov stable iff $\epsilon\ge \epsilon_c=1/3$ and spatially inhomogeneous waterbag distributions are Vlasov stable iff $\epsilon\le \epsilon_*=0.379$ and $b\ge b_*=0.37$ where $\epsilon$ and $b$ are the normalized energy and magnetization. The magnetization curve displays a first order phase transition at $\epsilon_t=0.352$ and the domain of metastability ranges from $\epsilon_c$ to $\epsilon_*$. 
   \PACS{05.30.-d Quantum statistical mechanics -
   05.45.-a Nonlinear dynamics and chaos - 05.20.Dd Kinetic theory -
   64.60.De Statistical mechanics of model systems} }

   \maketitle
%

\section{Introduction}
\label{sec_model}

The dynamics and thermodynamics of systems with long-range
interactions (self-gravitating systems, geophysical flows, non neutral
plasmas,...) has recently received a particular attention from the
community of statistical mechanics \cite{houches,assisebook,oxford,cdr}.
Surprisingly, long-range interacting systems had
not been studied at a general level until now, although the main
concepts (such as ensemble inequivalence, negative specific heats,
quasi stationary states, violent collisionless relaxation and slow
collisional relaxation) had been understood early in astrophysics
and two-dimensional turbulence (see, e.g., \cite{houcheschav}
and references therein). Recently, these fundamental
concepts have been  illustrated and emphasized in the
framework of a simple toy model of systems with long-range
interactions called the Hamiltonian Mean Field (HMF) model
\cite{ar}. This model displays many analogies with self-gravitating
systems \cite{inagaki,pichon,cvb,jstat} which have proven to
be very fruitful. Until now, only the classical HMF model has been
considered \cite{cdr}. In this paper, developing further the analogy
with self-gravitating systems, we shall give a first discussion of the
quantum HMF model for fermions and bosons.  Before that, we briefly
review the main steps in the history of classical and quantum
self-gravitating systems. This will place our study in a more general
perspective.

The statistical mechanics of classical self-gravitating systems, such
as elliptical galaxies and globular clusters, started with the works
of Ogorodnikov \cite{ogorodnikov}, Antonov \cite{antonov} and
Lynden-Bell \& Wood \cite{lbw} (see reviews \cite{paddy,katzrev,ijmpb}).
These authors studied the equilibrium configurations of a self-gravitating
gas of classical particles enclosed within a box so as to prevent its
complete evaporation. They considered the microcanonical
ensemble and discovered the important phenomenon of ``gravothermal
catastrophe'' which takes place below a critical energy $E_c=-0.335GM^2/R$. This can be
viewed as a sort of phase transition between a gaseous state and a
``singular'' state in which a tightly bound binary star is surrounded
by a hot halo of stars. We call it ``singular'' because the maximum
entropy state is reached when two stars approach each other with no
limit (leading to infinite density) so that the potential energy tends to $-\infty$
and the temperature of the halo increases to $+\infty$ so as to
allow for the overall conservation of energy. This core-halo state
has infinite entropy. On the other hand, Kiessling \cite{kiessling}
and Chavanis \cite{aaiso} considered the canonical ensemble and
investigated the phenomenon of ``isothermal collapse'' which takes
place below a critical temperature $k_BT_c=GMm/2.52R$.  In that case, the final outcome
of the collapse is a Dirac peak containing all the particles.  This
compact object has infinite free energy. These singular states,
binaries stars in the microcanonical ensemble and Dirac peaks in the
canonical ensemble, are obtained by assuming that the particles are
point-like and that quantum mechanics effects can be neglected.

The case of quantum particles in gravitational interaction has
also been considered in astrophysics and led to the concepts
of fermion and boson stars.

Fermion stars have found astrophysical applications in the context of
white dwarf stars, neutron stars and dark matter models made of
massive neutrinos \cite{chandra,st}. Their equilibrium results
from a balance between the gravitational attraction and the quantum
pressure due to the electrons (in white dwarfs), the neutrons (in
neutron stars) or the neutrinos (in dark matter).  Gravitational
collapse is therefore prevented by the Pauli exclusion principle
as first realized by Fowler \cite{fowler}. At $T=0$, the gas is
completely degenerate and, in the non relativistic limit,
the system is equivalent to a polytrope of
index $n=3/2$. The density profile has a finite support leading to a
configuration with a well-defined radius $R$. The mass-radius relation
of classical fermion stars is given by $MR^3\simeq 91.9 \hbar^6/m^8 G^3$.
When special \cite{chandra1} or general
\cite{ov} relativity is taken into account, it is found
that no equilibrium state exists above a maximum mass, called the
Chandrasekhar mass, scaling like $M_{Ch}\sim M_P^3/m^2$ where
$M_P=(\hbar c/G)^{1/2}$ is the Planck mass. 
Massive stars cannot pass into
the white-dwarf stage and become neutron stars. Even more massive
stars undergo gravitational collapse and become black holes
\cite{st}. The case of self-gravitating fermions
at non-zero temperature has been considered by several authors
\cite{ht,messer,stella,bilic,prefermi}.
The equilibrium states of the non relativistic Fermi gas are obtained
by coupling the Fermi-Dirac statistics to the Poisson equation (this
corresponds to the Thomas-Fermi approximation). In that case, the
system extends to infinity and one is forced to enclose the particles
within a finite box to prevent their evaporation. The shape of the
caloric curve $T(E)$ depends on a dimensionless parameter $\mu$ which
can be viewed as a normalized system size or as an inverse normalized
Planck constant \cite{ijmpb}.  For $\mu\rightarrow +\infty$, one
recovers the classical limit in which the system undergoes
gravitational collapse and forms a singularity.  However, when quantum
mechanics is taken into account, the gravitational collapse stops when
the system feels the Pauli exclusion principle. In that case, it ends
up on a nonsingular equilibrium state which typically has the form of
a completely degenerate and very compact nucleus (fermion ball)
surrounded by a hot and almost homogeneous atmosphere (halo). One can
therefore describe interesting zeroth and first order phase
transitions between ``gaseous'' and ``condensed'' states, and evidence
microcanonical and canonical critical points that differ in the two
ensembles. The complete phase diagram of the self-gravitating Fermi
gas is given in \cite{ijmpb}.

Boson stars were introduced by Kaup \cite{kaup} and Ruffini \&
Bonazzola \cite{rb} in the sixties although no astrophysical
application of these objects was known at that time. Later on, it was
suggested that dark matter could be made of bosons and that boson
stars could have formed by Jeans gravitational instability
\cite{bianchi}. At $T=0$, bosons form a Bose-Einstein condensate (BEC)
and they are described by a single wavefunction $\psi({\bf r},t)$.
In the nonrelativistic Newtonian limit, the structure of a self-gravitating
BEC is obtained by solving the Schr\"odinger-Poisson system and in the relativistic limit one must
couple the Klein-Gordon equation to the Einstein field equations. The Newtonian
approximation is valid for sufficiently small masses and yields the
mass-radius relation $MR=9.9\hbar^2/Gm^2$ \cite{rb}. The radius decreases as mass
increases, like for classical fermion stars, but the scaling is different. This relation is valid as
long as the radius is much larger than the Schwarzschild radius $R_S=2GM/c^2$.
When relativistic effects are taken into account, there exists a maximum mass,
the Kaup mass $M_{Kaup}=0.633M_P^2/m$, above which no equilibrium configuration
exists \cite{kaup,rb}. In that case, the system collapses to a black hole.
Below the critical mass, the gravitational collapse of boson stars is prevented
by the Heisenberg uncertainty principle while the gravitational collapse of
fermion stars is prevented by the Pauli exclusion principle. This is
why the Kaup mass $M_{Kaup}\sim M_P^2/m$ for boson stars is much smaller than the Chandrasekhar mass $M_{Ch}\sim M_P^3/m^2$ for fermion stars by a factor $m/M_P\ll 1$. Such small masses (e.g. $M_{Kaup}\sim 10^{11}kg\sim 10^{-19}M_{\odot}$  for $m\sim 1 GeV/c^2$!) led to the belief that boson stars
are not very relevant astrophysical objects. However, the maximum mass
of boson stars can be considerably increased by taking into account
the self-interaction of the particles via a $\lambda \phi^4$ potential
\cite{colpi}. In that case, the maximum mass scales like
$M\sim \lambda^{1/2}M_{Ch}$ and becomes comparable with the
Chandrasekhar mass of self-gravitating fermions when $\lambda\sim
1$. Therefore, self-coupling can significantly
change the physical dimensions of boson stars, making them much more
astrophysically interesting. Recently, some authors
\cite{bohmer} have proposed that dark matter galactic halos themselves could be
gigantic self-gravitating Bose-Einstein condensates (BEC). At the galactic scale,
it is possible to neglect relativistic effects
and use the Newtonian approximation. When self-coupling
is taken into account, the structure of a self-gravitating BEC is obtained by solving the Gross-Pitaevskii-Poisson system (this corresponds to the Hartree approximation). The short-range interaction produces an effective pressure described by a barotropic equation of
state equivalent to a polytrope of index $n=1$.  The equilibrium  of the
system then results from the balance between the gravitational
attraction, the pressure due to the self-interaction and the quantum pressure due to the Heisenberg uncertainty principle. A detailed study of the equilibrium configurations is made in \cite{prep}.

In this paper, we shall apply the same concepts to the HMF model in
which the gravitational interaction in $d=3$ dimensions is replaced by
a cosine interaction in $d=1$ dimension. This generalization has two
main interests. First, it will allow us to study the thermodynamics of
quantum particles with long-range interactions in a simpler model and
to derive several analytical results. Secondly, it
will be possible in future works to perform very efficient numerical
simulations of this system since the HMF model is one dimensional and
the potential of interaction is smooth. This will allow us to explore the
dynamical properties of quantum particles with long-range interactions
in a simple setting. Generalizing the HMF model to the
quantum regime is the natural next step in the systematic exploration
of the properties of this model since its introduction  in 1995 \cite{ar}.

The statistical mechanics of the classical HMF model has been treated
by different methods (see review \cite{cdr}). Among
them, the approach of \cite{inagaki,cvb,etapes} based on the
maximization of the Boltzmann entropy at fixed mass and energy closely
follows the standard methodology developed in astrophysics
\cite{antonov,lbw,paddy,katzrev,ijmpb}.  For the classical HMF model, there exists a
second order phase transition between a homogeneous phase (for $E>E_c$
and $T>T_c$) and a clustered phase (for $E<E_c$ and $T<T_c$). Contrary
to 3D self-gravitating systems, the ensembles are equivalent for all
accessible energies and temperatures.  At $T=0$ or $E=E_{min}$, a limit that will be
considered in the following, the density profile
$\rho(\theta)=M\delta(\theta)$ is a Dirac peak centered at $\theta=0$ (say)
containing all the mass. This ``singular'' state corresponds to a
complete condensation of the system in which the maximum magnetization
is $b=1$. We shall here study how this result is modified when quantum
mechanics effects are taken into account.

In the case of fermions, the statistical equilibrium state is obtained
by maximizing the Fermi-Dirac entropy at fixed mass and energy. This
leads to the mean field Fermi-Dirac distribution with a cosine
interaction. In fact, this problem has already been treated in a
different context. Indeed, in the collisionless regime of the
dynamics, the classical HMF model is governed by the Vlasov equation.  Now, the
Vlasov equation can undergo a process of violent collisionless
relaxation leading to a quasi stationary state (QSS). In the case
where the fine-grained distribution function takes only two values
$f=\eta_0$ and $f=0$, the statistical mechanics of the Vlasov equation
developed by Lynden-Bell \cite{lb} predicts that the QSS is obtained
by maximizing a ``fermionic'' mixing entropy at fixed mass and
energy. Here, the ``degeneracy'' is due to dynamical effects
(Liouville theorem), not to quantum mechanics (Pauli's exclusion
principle). Still, the mathematical problem is the same provided that
the maximum value of the distribution function $f_0$ fixed by the
initial condition is interpreted as the maximum value of the
distribution function $2/h$ fixed by the Pauli exclusion principle (see \cite{csmnras}
in the astrophysical context). The phase transitions associated with the Lynden-Bell or Fermi-Dirac
distributions are very rich and subtle and they have been described in detail in
\cite{epjb,antoniazzi1,antoniazzi2,antoniazzi3,marseille,staniscia1,staniscia2}. Due to the analogy
with the quantum problem, these studies directly apply to a gas of fermions
with cosine interaction.
Therefore, we shall not repeat this analysis here. We shall,
however, study the case $T=0$ (corresponding to the
ground state energy $E_{ground}$) that was not treated in detail in the previous
works. In that limit, the system is completely degenerate and the
distribution function reduces to the Fermi distribution or,
equivalently, to the (spatially inhomogeneous) waterbag distribution. This
distribution is equivalent to a polytrope of index $n=1/2$. The
general theory of polytropes in the context of the HMF model has been
recently developed  by Chavanis \& Campa \cite{cc} and we shall make use
of their results in the specific case $n=1/2$ which presents
interesting features. In terms of the normalized Planck constant $\chi$ for
fermions  defined in equation (\ref{cph1}), we show that the homogeneous phase is unstable for $\chi<\chi_c=\sqrt{2}$, metastable for $\chi_c<\chi<\chi_t\simeq 1.45$ and fully stable for $\chi>\chi_t$. In parallel, the inhomogeneous phase is fully stable for $\chi<\chi_t$, metastable for $\chi_t<\chi<\chi_*\simeq 1.48$ and disappears for $\chi>\chi_*$ (see Figure \ref{bVShN0.5}). This shows that quantum effects can stabilize the homogeneous phase
because of the Pauli exclusion principle, exactly like for fermion stars in astrophysics. For the HMF model, this stabilization happens through a first order phase transition where the control parameter is the normalized Planck constant. As a by product of our analysis, we obtain new results concerning the Vlasov dynamical stability of the waterbag distribution. We show that spatially homogeneous waterbag distributions are Vlasov stable iff $\epsilon\ge \epsilon_c=1/3$ and spatially inhomogeneous waterbag distributions are Vlasov stable iff $\epsilon\le \epsilon_*=0.379$ and $b\ge b_*=0.37$ where $\epsilon$ and $b$ are the normalized energy and magnetization. The magnetization curve displays a first order phase transition at $\epsilon_t=0.352$ and the domain of metastability ranges from $\epsilon_c$ to $\epsilon_*$ (see Figure \ref{bVSepsilonN0.5}).

At $T=0$, bosons form a Bose-Einstein condensate (BEC) and the
equilibrium state is obtained by solving the mean field Schr\"odinger
equation with a cosine interaction (the pressure derived from the
Bose-Einstein statistics is zero at $T=0$ so that a hydrodynamical
description is not appropriate). The normalized Planck constant $\chi$
for bosons is defined in equation (\ref{planckb}). For $\chi=0$, we
recover the classical result: the density profile forms a Dirac peak
at $\theta=0$. For $\chi\rightarrow 0$, we can make a harmonic
approximation and obtain quantum corrections at the order
$O(\chi)$. More generally, by solving the problem numerically for any
value of $\chi$, we evidence a second order phase transition between
magnetized states for $\chi<\chi_c=\sqrt{2}$ and non magnetized states
for $\chi>\chi_c$. This shows that quantum effects can stabilize the
homogeneous phase because of the Heisenberg uncertainty principle,
exactly like for boson stars in astrophysics.

The paper is organized as follows. In Section \ref{sec_hmfmodel}, we introduce the HMF model. In Section \ref{sec_fermions} we consider the case of fermions at $T=0$ and study the cosine Fermi distribution. We show that our study also determines the Vlasov dynamical stability of the  (possibly inhomogeneous) Lynden-Bell and  waterbag distributions, independently on the quantum mechanics context. In Section \ref{sec_bosons} we consider the case of bosons at $T=0$ and study the cosine Schr\"odinger equation. In Section \ref{sec_mfgp}, we consider the mean field Gross-Pitaevskii equation describing BECs with short-range interactions or fermions beyond the Thomas-Fermi approximation. In this context, we study the stability of a homogeneous distribution with respect to the quantum Euler and Vlasov (or Wigner) equations. The appropriate thermodynamic limits for fermions and bosons are discussed in Appendix \ref{sec_tlfb}. Technical results are given in the other Appendices.

\section{The HMF model}
\label{sec_hmfmodel}

We consider a system of $N$ particles of unit mass $m=1$ moving on a ring of unit radius $R=1$ and interacting via a cosine potential of the form $u=-\frac{k}{2\pi}\cos(\theta-\theta')$, where $k>0$ is the coupling constant. The Hamiltonian reads
\begin{equation}
\label{fa1}
H=\frac{1}{2}\sum_{i=1}^N v_i^2-\frac{k}{2\pi}\sum_{i<j}\cos(\theta_i-\theta_j),
\end{equation}
where $\theta_i$ and $v_i=\dot\theta_i$ are the position (angle) and velocity of particle $i$. We introduce the magnetization vector ${\bf b}=(b_x,b_y)$ where $b_x=\frac{1}{N}\sum_i \cos\theta_i$ and $b_y=\frac{1}{N}\sum_i \sin\theta_i$.

We assume that the system can be described by a distribution function $f(\theta,v,t)$ such that $f(\theta,v,t)d\theta dv$ gives the density of particles with position $\theta$ and velocity $v$ at time $t$. It is normalized such that $M=\int f\, d\theta dv$. As we shall see, this description applies to classical particles and fermions (in the Thomas-Fermi approximation) but not to bosons at $T=0$ forming Bose-Einstein condensates. In the mean field approximation, the energy (kinetic $+$ potential) is given by
\begin{equation}
\label{fa2}
E=K+W=\int f\frac{v^2}{2}\, d\theta dv+\frac{1}{2}\int\rho\Phi\, d\theta,
\end{equation}
where
\begin{equation}
\label{fa3}
\Phi(\theta,t)=-\frac{k}{2\pi}\int_0^{2\pi} \cos(\theta-\theta')\rho(\theta',t)\, d\theta',
\end{equation}
is the self-consistent potential generated by the density of particles $\rho(\theta,t)=\int f(\theta,v,t)\, dv$. Expanding the cosine function, the potential can be rewritten
\begin{equation}
\label{fa4}
\Phi(\theta,t)=-B_x\cos\theta-B_y\sin\theta,
\end{equation}
where
\begin{equation}
\label{fa5}
B_x=\frac{k}{2\pi}\int_0^{2\pi} \rho(\theta,t)\cos\theta\, d\theta,
\end{equation}
\begin{equation}
\label{fa6}
B_y=\frac{k}{2\pi}\int_0^{2\pi} \rho(\theta,t)\sin\theta\, d\theta,
\end{equation}
are proportional to the two components of the average magnetization $b_x=\frac{1}{N}\int \rho\cos\theta\, d\theta$ and $b_y=\frac{1}{N}\int \rho\sin\theta\, d\theta$. The potential energy can be expressed in terms of the magnetization as
\begin{equation}
\label{fa7}
W=-\frac{\pi B^2}{k}.
\end{equation}
On the other hand, the kinetic energy can be expressed in term of the pressure $p(\theta,t)=\int f v^2\, dv$ as
\begin{equation}
\label{fa8}
E=\frac{1}{2}\int p\, d\theta.
\end{equation}

\section{Fermions at zero temperature}
\label{sec_fermions}

In this section, we consider the HMF model at $T=0$ in the case where the particles are fermions of spin $s=1/2$. We use a mean field approximation that becomes exact in a proper thermodynamic limit $N\rightarrow +\infty$ defined in Appendix \ref{sec_tlfb}. This mean field approximation could be rigorously justified like in the astrophysical problem \cite{ht,messer}.

\subsection{The Fermi-Dirac distribution}
\label{sec_fd}

According to the Pauli exclusion principle, there are at most $2s+1=2$ fermions in a phase space cell of size $h$, where $h$ is the Planck constant\footnote{Since the units of length and  mass have been fixed to unity, the Planck constant $h$ so defined is dimensionless (see Appendix \ref{sec_tlfb} for the restoration of dimensional variables). Therefore, it can be viewed as an external parameter that can take values between $0$ (classical regime) and $+\infty$ (quantum regime).}. Therefore, the maximum value of the distribution function fixed by the Pauli exclusion principle is
\begin{equation}
\label{fd1}
\eta_0=\frac{2}{h}.
\end{equation}
The statistical equilibrium state in the microcanonical ensemble is obtained by maximizing the Fermi-Dirac entropy
\begin{equation}
\label{fd2}
S=-\int \left\lbrace \frac{f}{\eta_0}\ln\frac{f}{\eta_0}+\left (1-\frac{f}{\eta_0}\right )\ln \left (1-\frac{f}{\eta_0}\right )\right\rbrace\, d\theta dv,
\end{equation}
at fixed mass and energy (see, e.g., \cite{ijmpb}). Writing the variational principle in the form $\delta S-\beta\delta E-\alpha\delta M=0$, where $\beta$ and $\alpha$ are Lagrange multipliers, we obtain the mean field  Fermi-Dirac distribution
\begin{equation}
\label{fd3}
f=\frac{\eta_0}{1+\lambda e^{\beta \eta_0 (\frac{v^2}{2}+\Phi(\theta))}}.
\end{equation}
This distribution function satisfies the constraint $f\le \eta_0$. The statistical equilibrium state in the canonical ensemble is obtained by minimizing the Fermi-Dirac free energy $F=E-TS$ at fixed mass, where  $T\ge 0$ is the temperature. The critical points of constrained entropy and constrained free energy are the same, given by equation (\ref{fd3}), but their stability may differ in microcanonical and canonical ensembles in case of ensemble inequivalence (see, e.g., \cite{ijmpb}).

\subsection{The Fermi distribution}
\label{sec_f}

At  $T=0$, the Fermi gas is completely degenerate since the states with individual energy $\epsilon=v^2/2+\Phi(\theta)$  smaller than the Fermi energy $\epsilon_F$ are completely filled.  This leads to the Fermi distribution
\begin{eqnarray}
\label{f1}
f=\eta_0, \quad {\rm if}\quad \epsilon\le \epsilon_F,\nonumber\\
f=0,\quad {\rm if}\quad \epsilon> \epsilon_F.
\end{eqnarray}
It can be rewritten
\begin{eqnarray}
\label{f2}
f=\eta_0, \quad {\rm if}\quad |v|\le v_F(\theta),\nonumber\\
f=0,\quad {\rm if}\quad |v|> v_F(\theta),
\end{eqnarray}
where
\begin{eqnarray}
\label{f3}
v_F(\theta)\equiv \sqrt{2(\epsilon_F-\Phi(\theta))},
\end{eqnarray}
is the space dependent Fermi velocity. This zero temperature limit corresponds to the ground state of the Fermi gas (minimum energy state $E_{min}$).

The density and the pressure corresponding to the distribution function (\ref{f2}) are given by
\begin{eqnarray}
\label{f4}
\rho(\theta)=\int_{-v_F(\theta)}^{v_F(\theta)}f(v)\, dv=2\eta_0 v_F(\theta),
\end{eqnarray}
\begin{eqnarray}
\label{f5}
p(\theta)=\int_{-v_F(\theta)}^{v_F(\theta)}f(v)v^2\, dv=\frac{2}{3}\eta_0 v_F^3(\theta).
\end{eqnarray}
Eliminating the Fermi velocity between these two expressions, we obtain the equation of state
\begin{eqnarray}
\label{f6}
p=\frac{1}{12\eta_0^2}\rho^3.
\end{eqnarray}
This is the equation of state of a polytrope
\begin{eqnarray}
\label{f7}
p=K\rho^{\gamma},\qquad \gamma=1+\frac{1}{n},
\end{eqnarray}
with a polytropic constant
\begin{eqnarray}
\label{f8}
K=\frac{1}{12\eta_0^2}=\frac{\pi^2\hbar^2}{12},
\end{eqnarray}
and a polytropic index
\begin{eqnarray}
\label{f9}
\gamma=3, \qquad {\rm i.e.}\qquad n=\frac{1}{2}.
\end{eqnarray}
As is well-known in astrophysics, the Fermi gas at $T=0$ (e.g. a white dwarf star) is equivalent to a polytrope with index $n=d/2$ (in our case, the dimension of space is $d=1$) \cite{wdD}. Accordingly, we can study the Fermi gas at $T=0$ by using the theory of polytropes developed by Chavanis \& Campa \cite{cc} in the context of the HMF model. In this analogy, the polytropic constant $K$ can be interpreted as a polytropic temperature.

{\it Remark:} the Fermi distribution (\ref{f1}) is a particular steady state of the Vlasov equation called the {\it waterbag} distribution. Therefore, as a by-product, our study will also determine the structure and the stability of the (spatially inhomogeneous) waterbag distribution, independently of the quantum mechanics interpretation.

\subsection{The homogeneous polytrope $n=1/2$}
\label{sec_hp}

Let us first consider the case of a spatially homogeneous distribution. In that case, $\Phi=0$, and the Fermi distribution can be rewritten
\begin{eqnarray}
\label{hp1}
f=\eta_0, \quad {\rm if}\quad |v|\le v_F,\nonumber\\
f=0,\quad {\rm if}\quad |v|> v_F,
\end{eqnarray}
with $v_F=\sqrt{2\epsilon_F}$. The density is given by $\rho=2\eta_0 v_F$. The Fermi velocity is related to the mass and to the maximum value of the distribution function by
\begin{eqnarray}
\label{hp3}
v_F=\frac{M}{4\pi \eta_0}=\frac{M\hbar}{4}.
\end{eqnarray}
The energy is
\begin{eqnarray}
\label{hp4}
E=\frac{1}{2}\int p\, d\theta=\pi K\rho^3=\pi K\left (\frac{M}{2\pi}\right )^3.
\end{eqnarray}
Using equation (\ref{f8}), we obtain
\begin{eqnarray}
\label{hp4b}
E=\frac{M^3}{96\pi^2\eta_0^2}=\frac{M^3\hbar^2}{96}.
\end{eqnarray}
If we define the normalized energy and the  normalized inverse polytropic temperature by \cite{cc}:
\begin{eqnarray}
\label{hp5}
\epsilon\equiv\frac{8\pi E}{kM^2},\qquad \eta\equiv \frac{k\pi}{KM},
\end{eqnarray}
the relation (\ref{hp4}) can be rewritten
\begin{eqnarray}
\label{hp7}
\epsilon=\frac{1}{\eta}.
\end{eqnarray}
Of course, the magnetization vanishes in the homogeneous phase: $b=0$.

\subsection{The inhomogeneous polytrope $n=1/2$}
\label{sec_ip}

Let us now consider the case of spatially inhomogeneous distributions. Combining equations (\ref{f4}) and (\ref{f3}), the density profile is given by
\begin{eqnarray}
\label{ip1}
\rho(\theta)=2\eta_0 \sqrt{2(\epsilon_F-\Phi(\theta))}.
\end{eqnarray}
We need to consider two cases, depending on whether $\epsilon_F$ is positive or negative.

\subsubsection{The case $\epsilon_F>0$}
\label{sec_ipA}

\begin{figure}[!h]
\begin{center}
\includegraphics[clip,scale=0.3]{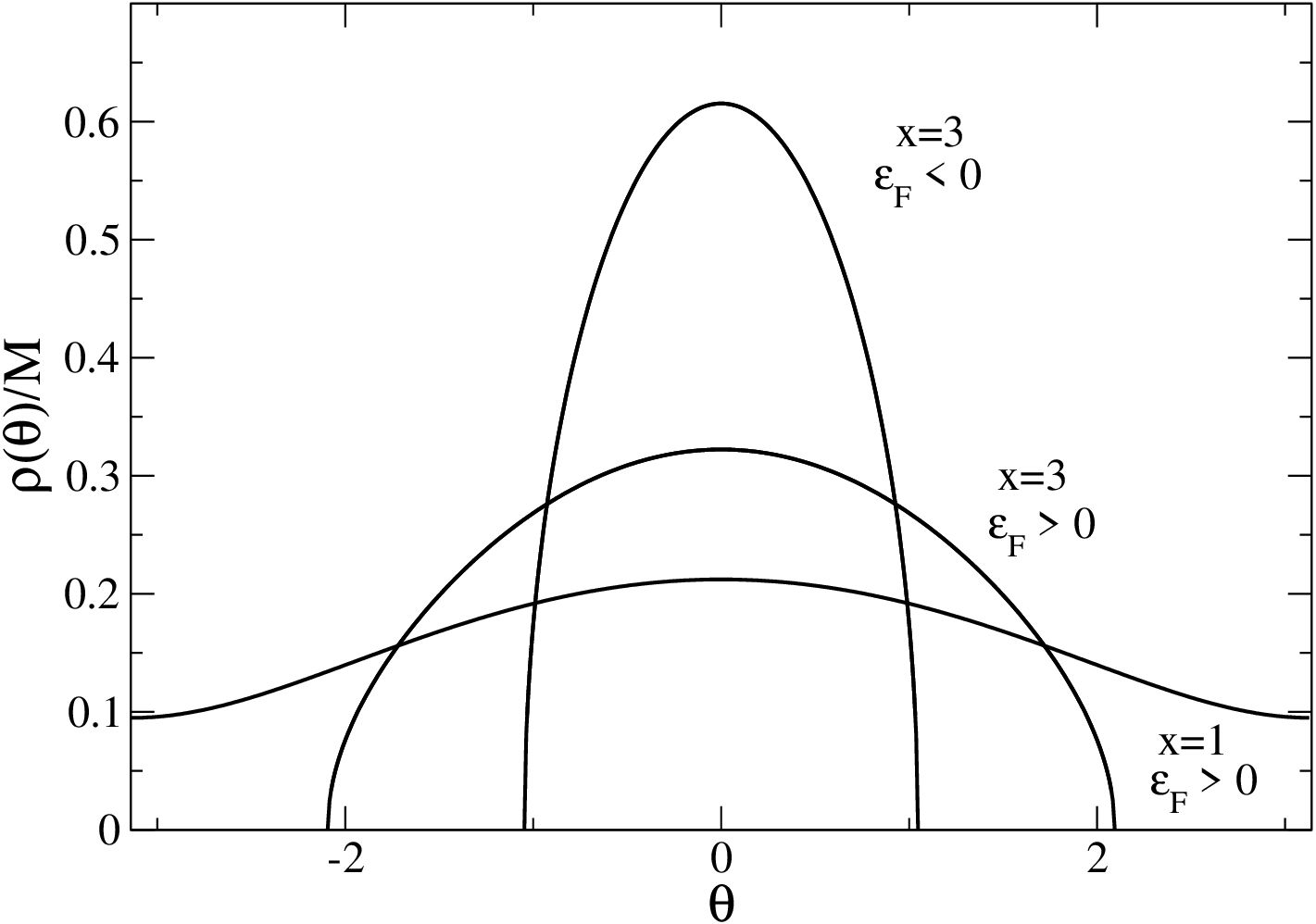}
\caption{Density profile of $n=1/2$ polytropes for different values of $x$. For $\epsilon_F>0$ we have taken $x=1$ and $x=3$. For $\epsilon_F<0$ we have taken $x=3$. }
\label{profiles}
\end{center}
\end{figure}

Let us first assume that $\epsilon_F>0$. In that case, defining $A=2\eta_0\sqrt{2\epsilon_F}$ and using equation (\ref{f8}), the density profile (\ref{ip1}) can be rewritten
\begin{eqnarray}
\label{ip2}
\rho(\theta)=A\left \lbrack 1-\frac{2}{3A^2K}\Phi(\theta)\right \rbrack_+^{1/2}.
\end{eqnarray}
We can assume, without loss of generality, that the distribution is symmetric with respect to the axis $\theta=0$. This implies that $B_y=0$ and $B_x=B$. Then, using equation (\ref{fa4}), we obtain
\begin{eqnarray}
\label{ip5}
\rho(\theta)=A\left ( 1+\frac{2}{3}x\cos\theta\right )_+^{1/2},
\end{eqnarray}
with
\begin{eqnarray}
\label{ip6}
x=\frac{B}{KA^2}.
\end{eqnarray}
The amplitude $A$ is determined by the mass $M=\int\rho\, d\theta$ according to
\begin{eqnarray}
\label{ip7}
A=\frac{M}{2\pi I_{3,0}(x)},
\end{eqnarray}
where we have introduced the integrals
\begin{eqnarray}
\label{ip8}
I_{\gamma,m}(x)=\frac{1}{2\pi}\int_0^{2\pi}\left (1+\frac{\gamma-1}{\gamma}x\cos\theta\right )_+^{\frac{1}{\gamma-1}}\cos(m\theta)\, d\theta,\nonumber\\
\end{eqnarray}
that can be interpreted as deformed Bessel functions \cite{cc}. By symmetry, we can restrict ourselves to the interval $0\le\theta\le \pi$. We need to distinguish two cases \cite{cc}. If $x<x_c=3/2$, the polytrope is {\it incomplete} in the sense that the density is strictly positive at $\theta=\pi$. If $x>x_c=3/2$, the polytrope is {\it complete} in the sense that the density vanishes at $\theta=\theta_c<\pi$ where
\begin{eqnarray}
\label{ip9}
\theta_c=\arccos \left (-\frac{3}{2x}\right ).
\end{eqnarray}
In that case, the profile has a compact support: $\rho=0$ for $\theta_c\le\theta\le\pi$. Some typical density profiles are represented in Figure \ref{profiles}. Since $v_F(\theta)=\frac{1}{2\eta_0}\rho(\theta)$, the density profile also gives the shape of the Fermi distribution in the phase space $(\theta,v)$. The velocity distribution $\phi(v)=\int f\, d\theta$ is $\phi(v)=2\eta_0\arccos\lbrack \frac{3}{2x}(\frac{4\eta_0^2v^2}{A^2}-1)\rbrack$ if $v\le \frac{A}{2\eta_0}(1+\frac{2}{3}x)^{1/2}$ and $\phi(v)=0$ otherwise.

Substituting equation (\ref{ip5}) in equation (\ref{fa5}), we find that the magnetization is related to $x$ by
\begin{eqnarray}
\label{ip10}
b\equiv \frac{2\pi B}{kM}=\frac{I_{3,1}(x)}{I_{3,0}(x)}.
\end{eqnarray}
Combining equations (\ref{ip6}), (\ref{ip7}) and (\ref{ip10}), we find that the normalized inverse polytropic temperature is related to $x$ by
\begin{eqnarray}
\label{ip11}
\eta\equiv \frac{k\pi}{KM}=\frac{x}{2I_{3,0}(x)I_{3,1}(x)}.
\end{eqnarray}
Finally, it is shown in \cite{cc} that the normalized energy can be written
\begin{eqnarray}
\epsilon\equiv\frac{8\pi E}{kM^2}=-\frac{2}{3}\frac{I_{3,1}(x)^2}{I_{3,0}(x)^2}-\frac{4}{3}\frac{I_{3,1}(x)}{I_{3,0}(x)}\cos\theta_c\nonumber\\
+\frac{2}{x}\frac{I_{3,1}(x)}{I_{3,0}(x)}\left (1-\frac{2}{3}x\right )_{+}.
\label{ip12}
\end{eqnarray}
For incomplete polytropes ($x<x_c$), it takes the form
\begin{eqnarray}
\epsilon=-\frac{2}{3}\frac{I_{3,1}(x)^2}{I_{3,0}(x)^2}+\frac{4}{3}\frac{I_{3,1}(x)}{I_{3,0}(x)}
+\frac{2}{x}\frac{I_{3,1}(x)}{I_{3,0}(x)}\left (1-\frac{2}{3}x\right ).\nonumber\\
\label{ip13}
\end{eqnarray}
For complete polytropes ($x>x_c$), it reduces to
\begin{eqnarray}
\epsilon=-\frac{2}{3}\frac{I_{3,1}(x)^2}{I_{3,0}(x)^2}+\frac{2I_{3,1}(x)}{xI_{3,0}(x)}.
\label{ip14}
\end{eqnarray}

\subsubsection{The case $\epsilon_F<0$}
\label{sec_ipB}

We now assume that $\epsilon_F<0$. In that case, defining
$A=2\eta_0\sqrt{2|\epsilon_F|}$ and using equation (\ref{f8}), the
density profile (\ref{ip1}) can be rewritten
\begin{eqnarray}
\label{ip15}
\rho(\theta)=A\left \lbrack -1-\frac{2}{3A^2K}\Phi(\theta)\right \rbrack_+^{1/2}.
\end{eqnarray}
Using equations (\ref{fa4}) and (\ref{ip6}), we obtain
\begin{eqnarray}
\label{ip16}
\rho(\theta)=A\left (-1+\frac{2}{3}x\cos\theta\right )_+^{1/2}.
\end{eqnarray}
The amplitude $A$ is determined by the mass $M=\int\rho\, d\theta$ according to
\begin{eqnarray}
\label{ip18}
A=\frac{M}{2\pi {\cal I}_{3,0}(x)},
\end{eqnarray}
where we have introduced the integrals
\begin{eqnarray}
\label{ip19}
{\cal I}_{\gamma,m}(x)=\frac{1}{2\pi}\int_0^{2\pi}\left (-1+\frac{\gamma-1}{\gamma}x\cos\theta\right )_+^{\frac{1}{\gamma-1}}\cos(m\theta)\, d\theta.\nonumber\\
\end{eqnarray}
The central density is defined iff $x>x_c=3/2$, so that we shall restrict ourselves to this range of parameters. In that case, the polytrope is always {\it complete} in the sense that the density vanishes at $\theta=\theta_c<\pi$ where
\begin{eqnarray}
\label{ip17}
\theta_c=\arccos \left (\frac{3}{2x}\right ).
\end{eqnarray}
The density has a compact support: $\rho=0$ for $\theta_c\le\theta\le\pi$ (see Figure \ref{profiles}). Furthermore, $\phi(v)=2\eta_0\arccos\lbrack \frac{3}{2x}(\frac{4\eta_0^2v^2}{A^2}+1)\rbrack$ if $v\le \frac{A}{2\eta_0}(-1+\frac{2}{3}x)^{1/2}$ and $\phi(v)=0$ otherwise.

Substituting equation (\ref{ip16}) in equation  (\ref{fa5}), we find that the normalized magnetization is related to $x$ by
\begin{eqnarray}
\label{ip20}
b\equiv \frac{2\pi B}{kM}=\frac{{\cal I}_{3,1}(x)}{{\cal I}_{3,0}(x)}.
\end{eqnarray}
Combining equations (\ref{ip6}), (\ref{ip18}) and (\ref{ip20}), the normalized inverse polytropic temperature is related to $x$ by
\begin{eqnarray}
\label{ip21}
\eta\equiv \frac{k\pi}{KM}=\frac{x}{2{\cal I}_{3,0}(x){\cal I}_{3,1}(x)}.
\end{eqnarray}
Finally, extending the results of \cite{cc}, we find that the normalized energy can be written
\begin{eqnarray}
\epsilon\equiv\frac{8\pi E}{kM^2}=-\frac{2}{3}\frac{{\cal I}_{3,1}(x)^2}{{\cal I}_{3,0}(x)^2}-\frac{4}{3}\frac{{\cal I}_{3,1}(x)}{{\cal I}_{3,0}(x)}\cos\theta_c\nonumber\\
+\frac{2}{x}\frac{{\cal I}_{3,1}(x)}{{\cal I}_{3,0}(x)}\left (1-\frac{2}{3}x\right )_{+}.
\label{ip22}
\end{eqnarray}
Since the polytropes are always complete, it reduces to
\begin{eqnarray}
\epsilon=-\frac{2}{3}\frac{{\cal I}_{3,1}(x)^2}{{\cal I}_{3,0}(x)^2}-\frac{2{\cal I}_{3,1}(x)}{x{\cal I}_{3,0}(x)}.
\label{ip23}
\end{eqnarray}

{\it Remark:} the case $\epsilon_F<0$ was forgotten in
\cite{cc} where it was implicitly assumed that $\lambda>0$ in equation (33) of that paper. Therefore, it was found that the series of equilibria
suddenly stops at some finite magnetization, inverse temperature and
energy corresponding to $x\rightarrow +\infty$ in equations
(\ref{ip10})-(\ref{ip12}). In fact, if we take the case $\lambda<0$
into account, the series of equilibria continues until the point
$b\rightarrow 1$, $\eta\rightarrow +\infty$ and $\epsilon\rightarrow
\epsilon_{min}=-2$ corresponding to $x\rightarrow x_c=3/2$ in
equations (\ref{ip20})-(\ref{ip22}). In that limit, the density
profile tends to a Dirac peak $\rho(\theta)=M\delta(\theta)$ (see
Appendix \ref{sec_class}). This is more satisfactory on a physical
point of view. Fortunately, this part of the branch does not change
the nature of the phase transitions described in \cite{cc} so that the
conclusions of this study are unaltered.

\subsection{Vlasov dynamical stability of the waterbag distribution}
\label{sec_vlasov}

Before investigating the thermodynamical stability of the Fermi distribution (see Section \ref{sec_ts}), we shall make a digression and investigate its dynamical stability with respect to the Vlasov equation. In that context, the Fermi distribution will be referred to as the waterbag distribution. On general grounds, a thermodynamically stable distribution must be dynamically stable. Therefore, it makes sense to first study its dynamical stability. Furthermore, the Vlasov dynamical stability of the waterbag distribution  is interesting in its own right, independently of the quantum mechanics interpretation.

\subsubsection{Formal nonlinear stability}
\label{sec_formal}

Let us first recall general results of formal nonlinear dynamical stability.
A distribution function of the form $f=f(\epsilon)$, depending only on
the individual energy $\epsilon=v^2/2+\Phi(\theta)$ of the particles,
is a steady state of the Vlasov equation. If $f=f(\epsilon)$ with
$f'(\epsilon)\le 0$, then it extremizes a functional $S=-\int C(f)\,
d\theta dv$, where $C$ is convex ($C''>0$), at fixed mass $M$ and
energy $E$. Furthermore, it can be shown that if $f$ is a {\it
maximum} of $S$ at fixed mass and energy, then it is formally
nonlinearly dynamically stable with respect to the Vlasov equation
\cite{campachav}. It can also be shown that formal stability implies
linear stability although the converse is wrong in general. We are
thus led to investigating the maximization problem
\begin{eqnarray}
\label{dyn1}
\max_f\left\lbrace S\lbrack f\rbrack\, |\, E\lbrack f\rbrack=E, \, M\lbrack f\rbrack=M\right\rbrace,
\end{eqnarray}
which is similar to a criterion of microcanonical stability in
thermodynamics. A {\it less refined} condition of formal nonlinear
stability  is provided by the minimization problem
\begin{eqnarray}
\label{dyn2}
\min_f\left\lbrace F\lbrack f\rbrack=E[f]-T S[f]\, |\, M\lbrack f\rbrack=M\right\rbrace,
\end{eqnarray}
which is similar to a criterion of canonical stability in
thermodynamics. This corresponds to the classical Casimir-energy stability
criterion \cite{holm}. The variational principles (\ref{dyn1}) and (\ref{dyn2}) have
the same critical points. Furthermore, a solution of (\ref{dyn2}) is
always a solution of the more constrained problem (\ref{dyn1}) but
the converse is wrong in case of ``ensemble inequivalence''
\cite{ellis} (a notion applied here in a {\it dynamical} context).

A distribution function of the form $f=f(\epsilon)$ with
$f'(\epsilon)\le 0$ determines a ``gas'' characterized by a density
profile $\rho(\theta)$ and a barotropic equation of state $p=p(\rho)$
\cite{cvb}. Furthermore, it can be shown \cite{cvb,aaantonov,assiseChav} that the
minimization problem (\ref{dyn2}) is equivalent to the minimization
problem
\begin{eqnarray}
\label{dyn3}
\min_\rho\left\lbrace F\lbrack \rho\rbrack\, |\, M\lbrack \rho\rbrack=M\right\rbrace,
\end{eqnarray}
where
\begin{eqnarray}
\label{dyn4}
F[\rho]=\frac{1}{2}\int \rho\Phi\, d\theta+\int\rho\int^\rho \frac{p(\rho')}{\rho^{'2}}\, d\rho'd\theta.
\end{eqnarray}
It can also be shown that this minimization problem provides a
necessary and sufficient condition of formal nonlinear dynamical
stability of the ``gas'' with respect to the Euler
equation \cite{bt}. Therefore, a distribution function $f(\epsilon)$ with
$f'(\epsilon)\le 0$ is stable with respect to the Vlasov equation if the
corresponding barotropic gas $\rho(\theta)$ is stable with respect to
the Euler equation, but the reciprocal is wrong in case of ensemble
inequivalence (since (\ref{dyn2}) is not equivalent to
(\ref{dyn1})). This provides a new interpretation of the (nonlinear)
Antonov first law in terms of ensemble inequivalence \cite{aaantonov}.

If we consider a functional of the form
\begin{equation}
S=-\frac{1}{q-1}\int (f^q-f)\, d\theta dv,  \label{dyn5}
\end{equation}
then the critical points of ``entropy''\footnote{As explained in \cite{cc}, we use a {\it thermodynamical analogy} to investigate a dynamical stability problem. By an abuse of language we adopt a common vocabulary.} $S$ at fixed mass and energy, satisfying $\delta S-\beta\delta E-\alpha\delta M=0$, where $\beta$ and $\alpha$ are Lagrange multipliers, are given by the polytropic distribution function
\begin{equation}
f(\theta,v)=\left\lbrace \mu-\frac{(q-1)\beta}{q}\left\lbrack \frac{v^2}{2}+\Phi(\theta)\right\rbrack\right\rbrace_+^{1/(q-1)}.  \label{dyn7}
\end{equation}
It is convenient to introduce the index $n$ by the definition
\begin{equation}
n=\frac{1}{2}+\frac{1}{q-1}. \label{dyn8}
\end{equation}
Then, it can be shown \cite{cc} that the equation of state associated with the distribution function (\ref{dyn7}) is the polytropic one
\begin{equation}
p=K\rho^{\gamma}, \qquad \gamma=1+\frac{1}{n},\label{dyn9}
\end{equation}
where the polytropic temperature $K$ is an increasing function of
$\beta^{-1}$. On the other hand, the density profile is of the form
\begin{equation}
\rho(\theta)=\left\lbrack \lambda-\frac{\gamma-1}{K\gamma}\Phi(\theta)\right
\rbrack_+^{\frac{1}{\gamma-1}},
\label{dyn10}
\end{equation}
where $\lambda$ can be related to $\mu$. Finally, the free energy (\ref{dyn4}) becomes
\begin{eqnarray}
\label{dyn11}
F[\rho]=\frac{1}{2}\int \rho\Phi\, d\theta+\frac{K}{\gamma-1}\int(\rho^{\gamma}-\rho)\, d\theta.
\end{eqnarray}

In the context of the HMF model, the Vlasov dynamical stability of
polytropic distribution functions with arbitrary index $n$ has been
studied in \cite{cc}. The waterbag distribution corresponds to a
polytropic index $n=1/2$, i.e. $\gamma=3$ and $q=\infty$. Indeed,
equation (\ref{dyn7}) reduces to the waterbag distribution (\ref{f1}),
equation (\ref{dyn10}) reduces to the density profile (\ref{ip2}) and
equation (\ref{dyn9}) reduces to the equation of state (\ref{f6}). However,
the formal stability criterion based on the maximization problems
(\ref{dyn1}) or (\ref{dyn2}) is not directly applicable
since the waterbag distribution (\ref{f1}) is a singular case for
which no associated functional $S$ exists (for $q=\infty$ the
functional (\ref{dyn5}) is undetermined). In order to investigate its
Vlasov dynamical stability, we shall use a ``ruse'' and view the waterbag
distribution (\ref{f1}) as a limit of a polytropic distribution with
$n\rightarrow 1/2$. Thus, we shall use stability results established for
polytropes of index $n>1/2$ and pass to the limit $n\rightarrow 1/2$.

{\it Remark 1:} although $S[f]$ and $F[f]$ are ill-defined for $n=1/2$, the functional $F[\rho]$ is well-defined and takes the form
\begin{eqnarray}
\label{dyn12}
F[\rho]=\frac{1}{2}\int \rho\Phi\, d{\bf r}+\frac{K}{2}\int \rho^{3}\, d\theta.
\end{eqnarray}
We could thus study the Vlasov dynamical stability of the Fermi distribution  through the minimization problem (\ref{dyn3}). This will be considered in Section \ref{sec_ts} where another interpretation of this functional will be given in relation to the thermodynamical stability of the Fermi distribution.

{\it Remark 2:} when we study the Vlasov stability of a polytrope of index $n$ and mass $M$, the proper control parameter is the energy $E$ (for the maximization problem (\ref{dyn1})) or the polytropic temperature $K$ (for the minimization problem (\ref{dyn2})). We shall therefore use these control parameters in a first step and express the conditions of stability in terms of these parameters. Then, when we focus on the waterbag distribution $n=1/2$, it is more convenient to take $\eta_0$ (which is a function of $E$ or $K$) as a control parameter and express the conditions of stability in terms of $\eta_0$. This is particularly important in relation to the Lynden-Bell theory where $\eta_0$ is the natural control parameter \cite{epjb,staniscia1,staniscia2}. Finally, when we come back to the initial quantum problem, it is more relevant to take $h$ as a control parameter (which is related to $\eta_0$) and express the conditions of stability in terms of $h$. Although these problems are closely related, it is important, for clarity, to consider them successively. This is what we shall do in the sequel.

\subsubsection{Homogeneous distributions}
\label{sec_hd}

Let us first consider the Vlasov dynamical stability of a spatially homogeneous distribution. We assume that $f=f(v)$ is an even distribution function with a single maximum at $v=0$ (this corresponds to $f=f(\epsilon)$ with $f'(\epsilon)\le 0$ and $\Phi=0$). Such a distribution is stable iff
\begin{eqnarray}
\label{hd1}
1+\frac{k}{2}\int_{-\infty}^{+\infty}\frac{f'(v)}{v}\ge 0,
\end{eqnarray}
or, equivalently, iff
\begin{eqnarray}
\label{hd2}
c_s^2\ge \frac{kM}{4\pi},
\end{eqnarray}
where $c_s^2=p'(\rho)$ is the velocity of sound in the corresponding
barotropic gas \cite{cvb}. Equations (\ref{hd1}) and (\ref{hd2}) are
equivalent and they provide criteria of linear and nonlinear
dynamical stability with respect to the Vlasov and Euler
equations \cite{cvb,campachav,yama,cd}.

For the waterbag (or Fermi) distribution characterized by the equation
of state (\ref{f6}), we find that
$c_s(\theta)=\rho(\theta)/(2\eta_0)$. We note that the
velocity of sound coincides with the Fermi velocity:
$c_s(\theta)=v_F(\theta)$. For a spatially homogeneous distribution
$\rho=M/(2\pi)$, we obtain
\begin{eqnarray}
\label{hd3}
c_s^2=v_F^2=\left (\frac{M}{4\pi\eta_0}\right )^2=\left (\frac{M\hbar}{4}\right )^2=\frac{3KM^2}{4\pi^2}.
\end{eqnarray}
Using equations (\ref{hd3}) and (\ref{hp7}), the stability criterion (\ref{hd2}) can be rewritten in terms of the normalized  inverse polytropic temperature and normalized energy as
\begin{eqnarray}
\label{hd5}
\eta\le \eta_c=3, \qquad \epsilon\ge \epsilon_c=\frac{1}{3}.
\end{eqnarray}
This is a particular case of the general result (156) of \cite{cvb} corresponding to a polytropic index $\gamma=3$. Of course, the same results can be obtained directly from the criterion (\ref{hd1}) using the distribution function (\ref{hp1}) for which $f'(v)=\eta_0\lbrack \delta(v+v_F)-\delta(v-v_F)\rbrack$.

\subsubsection{Inhomogeneous distributions}
\label{sec_in}

The Vlasov dynamical stability of spatially inhomogeneous
distributions is more difficult to investigate. Stability criteria can
be obtained by solving eigenvalue equations \cite{cvb,cc,cd},
variational principles \cite{etapes,campachav} or dispersion relations
\cite{barre}. On the other hand, as explained in \cite{cc}, we can
study the formal stability of a steady state of the Vlasov equation by
using the Poincar\'e theory of linear series of equilibria (see, e.g.,
\cite{katzrev,ijmpb}). This is a very simple and powerful graphical
method that only requires to solve the first order variational problem
and to know the stability of at least one point in the series of
equilibria by another method. Let us apply it to the present situation
(for more details see \cite{cc}).

\begin{figure}[!h]
\begin{center}
\includegraphics[clip,scale=0.3]{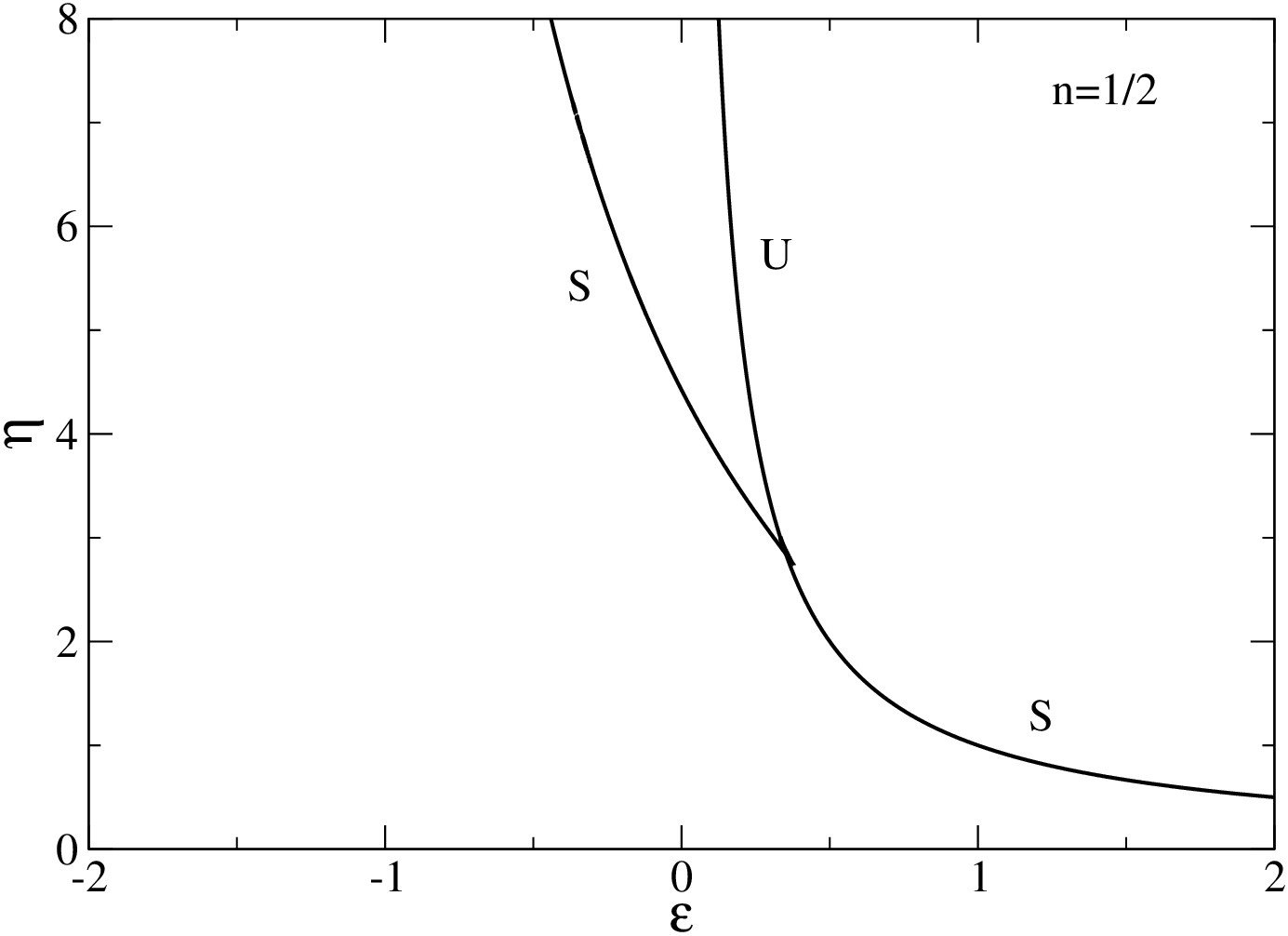}
\caption{Series of equilibria of the waterbag distribution corresponding to a polytrope $n=1/2$. It gives the inverse polytropic temperature as a function of the energy. The homogeneous branch (unstable) tends towards $\epsilon=0$ and the inhomogeneous branch (stable) tends towards $\epsilon=-2$.}
\label{caloN0.5GLOBALE}
\end{center}
\end{figure}

\begin{figure}[!h]
\begin{center}
\includegraphics[clip,scale=0.3]{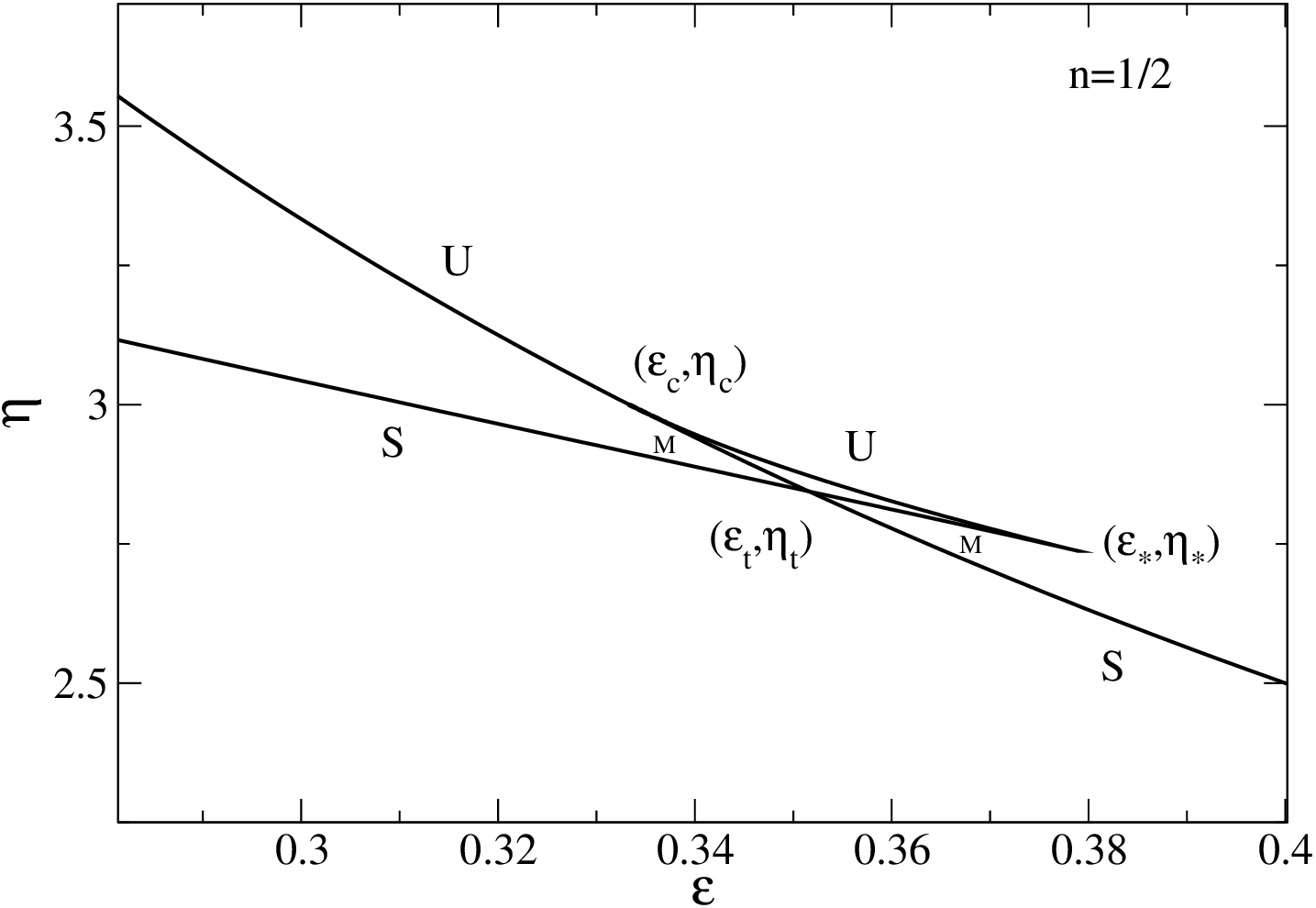}
\caption{Zoom of Figure \ref{caloN0.5GLOBALE} near the bifurcation point.}
\label{caloN0.5}
\end{center}
\end{figure}

The formal stability of a steady state of the Vlasov equation is based
on the optimization problems (\ref{dyn1}) and (\ref{dyn2}). To solve
the maximization problem (\ref{dyn1}) corresponding to the
``microcanonical'' ensemble, we just need to plot $\beta$ (conjugate
variable) as a function of $E$ (conserved quantity) since
$\beta=(\partial S/\partial E)_M$. It is shown in \cite{cc} that
$\eta$ is a monotonic function of $\beta$ so that it is equivalent to
plot $\eta$ as a function of $\epsilon$. This can be done by
eliminating $x$ between equations (\ref{ip11}) and (\ref{ip12}) and
between equations (\ref{ip21}) and (\ref{ip22}). The series of
equilibria is represented in Figures \ref{caloN0.5GLOBALE} and
\ref{caloN0.5}. The homogeneous phase exists for $\epsilon\ge 0$
and the inhomogeneous phase for $\epsilon_{min}=-2\le \epsilon\le \epsilon_*\simeq
0.379$. It bifurcates from the homogeneous phase at
$\epsilon_c=1/3$. According to the Poincar\'e theorem, a change of
stability can occur only at a bifurcation point or at a turning
point. We have already shown, by a direct calculation, that the
homogeneous phase is unstable for $\epsilon<\epsilon_c$ and stable for
$\epsilon>\epsilon_c$ (see Section \ref{sec_hd}). Therefore, the inhomogeneous branch that
appears at $\epsilon=\epsilon_c$ is necessarily unstable. At
$\epsilon=\epsilon_*$, there is a turning point of energy
so that the inhomogeneous branch becomes
stable. Since there is no other turning point or bifurcation point in
the series of equilibria, it remains stable until the end. We can use
the same method in the ``canonical'' ensemble. To solve the
minimization problem (\ref{dyn2}), we just need to plot $E$ as a
function of $\beta$ since $E=(\partial (\beta F)/\partial
\beta)_M$. As explained previously, this is equivalent to plotting
$\epsilon$ as a function of $\eta$. Therefore, we just need to rotate
the curves of Figures \ref{caloN0.5GLOBALE} and \ref{caloN0.5} by $90$
degrees. The homogeneous phase exists for $\eta\ge 0$ and the
inhomogeneous phase for $\eta\ge \eta_*\simeq 2.737$. It bifurcates
from the homogeneous phase at $\eta_c=3$. We have already shown that the
homogeneous phase is unstable for $\eta>\eta_c$ and stable for
$\eta<\eta_c$ (see Section \ref{sec_hd}). Therefore, the inhomogeneous branch that appears at
$\eta=\eta_c$ is necessarily unstable. At $\eta=\eta_*$, there is a turning point
of polytropic temperature so that
the inhomogeneous branch becomes stable. Since there is no other
turning point or bifurcation point in the series of equilibria, it
remains stable until the end. We note that the ensembles are
equivalent\footnote{Since the ``ensembles'' are equivalent, the stability criteria that we obtain determine both the Vlasov stability of the waterbag distribution (\ref{f1}) and the Euler stability of the corresponding ``barotropic gas'' described by the density profile (\ref{ip1}).} for the polytropic index $n=1/2$ although there is a small
region of ensemble inequivalence for $n>1/2$ \cite{cc}. This is
because, for $n=1/2$, the series of equilibria makes a {\it spike} at
$(\epsilon,\eta)=(\epsilon_*,\eta_*)$ while for $n>1/2$ the turning
points of energy and polytropic temperature are distinct. The case
$n=1/2$ is therefore very singular in this respect. We shall see below
another consequence of this ``singular'' behavior.

We can also study the  stability of a polytrope by plotting the ``entropy'' $S$ versus the energy $E$, and by comparing the entropies of the solutions having the same energy. It is shown in \cite{cc} that, for $n>1/2$, the normalized entropy is given by
\begin{eqnarray}
\label{in1}
s=-\left (n-\frac{1}{2}\right )\epsilon_{kin} \eta^{\frac{2n}{2n-1}},
\end{eqnarray}
where $\epsilon_{kin}$ is the normalized kinetic energy. For $n>1/2$, the study of the curve $s(\epsilon)$ is equivalent to the study of the curve $\sigma(\epsilon)$ where
\begin{eqnarray}
\label{in2}
\sigma=-\epsilon_{kin}^{2n-1} \eta^{2n}.
\end{eqnarray}
If we pass to the limit $n\rightarrow 1/2$, we find that
\begin{eqnarray}
\label{in3}
\sigma=-\eta.
\end{eqnarray}
Therefore, for the polytrope $n=1/2$, the entropy $\sigma$ is equal to
the opposite of the inverse polytropic temperature $\eta$. This is a
very singular situation. Indeed, the caloric curve $\eta(\epsilon)$
coincides with the entropic curve $\sigma(\epsilon)$! This is the
reason why the caloric curve of Figure \ref{caloN0.5}, which has a
``triangular'' shape and presents ``cusps'', looks similar to the familiar entropy vs energy
curve characteristic of first order phase transitions (see, e.g.,
Figure 9 of \cite{cc}). Therefore, the intersection between the
homogeneous and inhomogeneous branches in Figure \ref{caloN0.5}
determines the transition energy $\epsilon_t$ at which these phases
have the same ``entropy''. Similar results are obtained in the canonical ensemble. In that case,
we can study the stability of a polytrope by plotting the free energy $F$ as a
function of the polytropic temperature $K$, and by comparing the free energies of the
solutions having the same polytropic temperature. Now, using the results of \cite{cc}, we see that
for $n=1/2$ the free energy coincides with the energy, i.e.
\begin{eqnarray}
\label{in4}
f=\epsilon.
\end{eqnarray}
Therefore, the free energy curve $f(\eta)$ coincides with the caloric curve $\epsilon(\eta)$.
Therefore, the intersection between the
homogeneous and inhomogeneous branches in Figure \ref{caloN0.5}
determines the transition temperature $\eta_t$ at which these phases
have the same ``free energy''\footnote{We recall that we are studying here the Vlasov dynamical stability problem. Therefore, when we use the word ``entropy'', ``free energy'', ``microcanonical ensemble'', ``canonical ensemble'', ``temperature'' etc. we are invoking the {\it thermodynamical analogy} explained in \cite{cc}. To be more correct, we should add the prefix ``pseudo'' in front of these terms.}.

\begin{figure}[!h]
\begin{center}
\includegraphics[clip,scale=0.3]{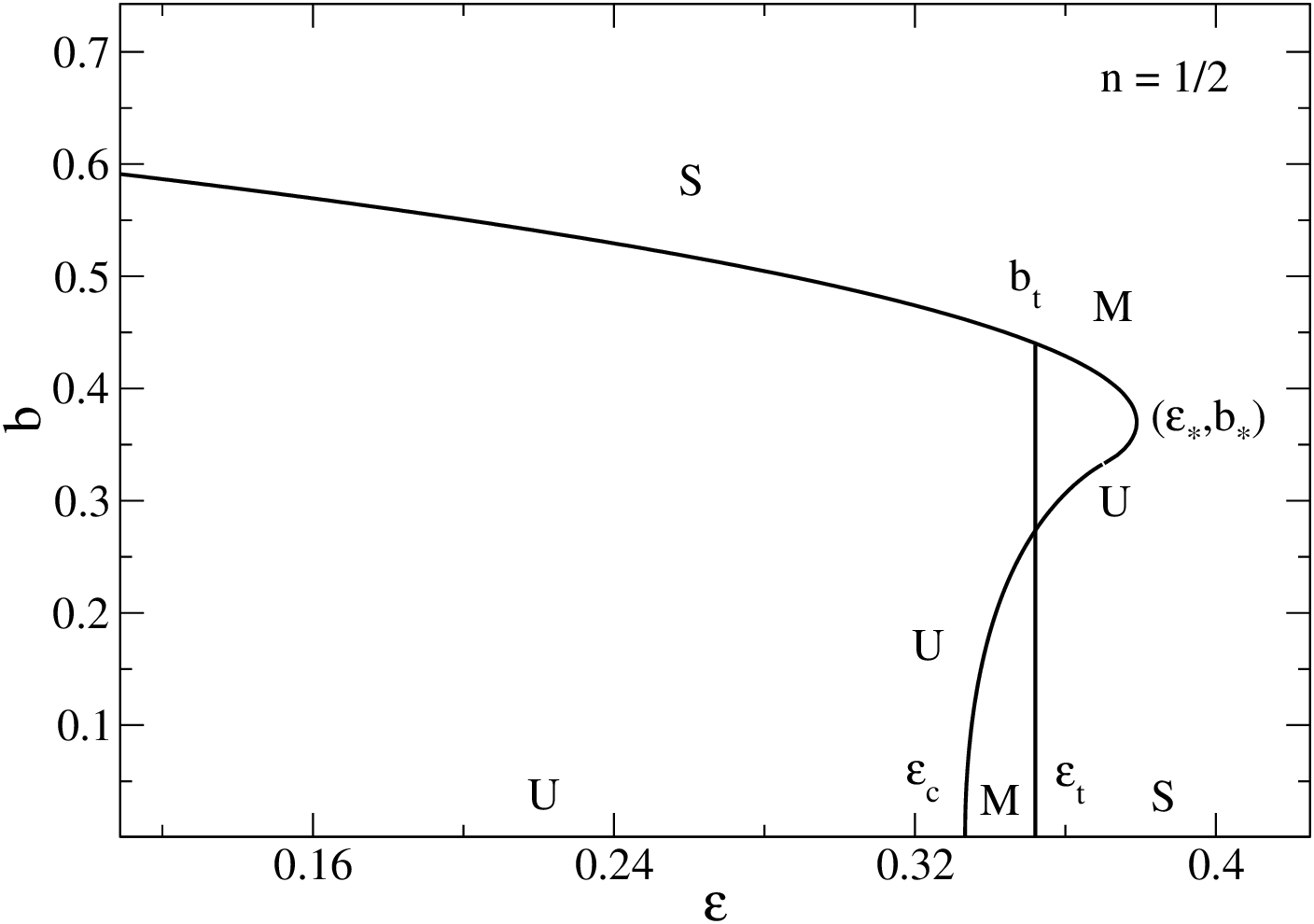}
\caption{Magnetization as a function of the energy.}
\label{bVSepsilonN0.5}
\end{center}
\end{figure}

\begin{figure}[!h]
\begin{center}
\includegraphics[clip,scale=0.3]{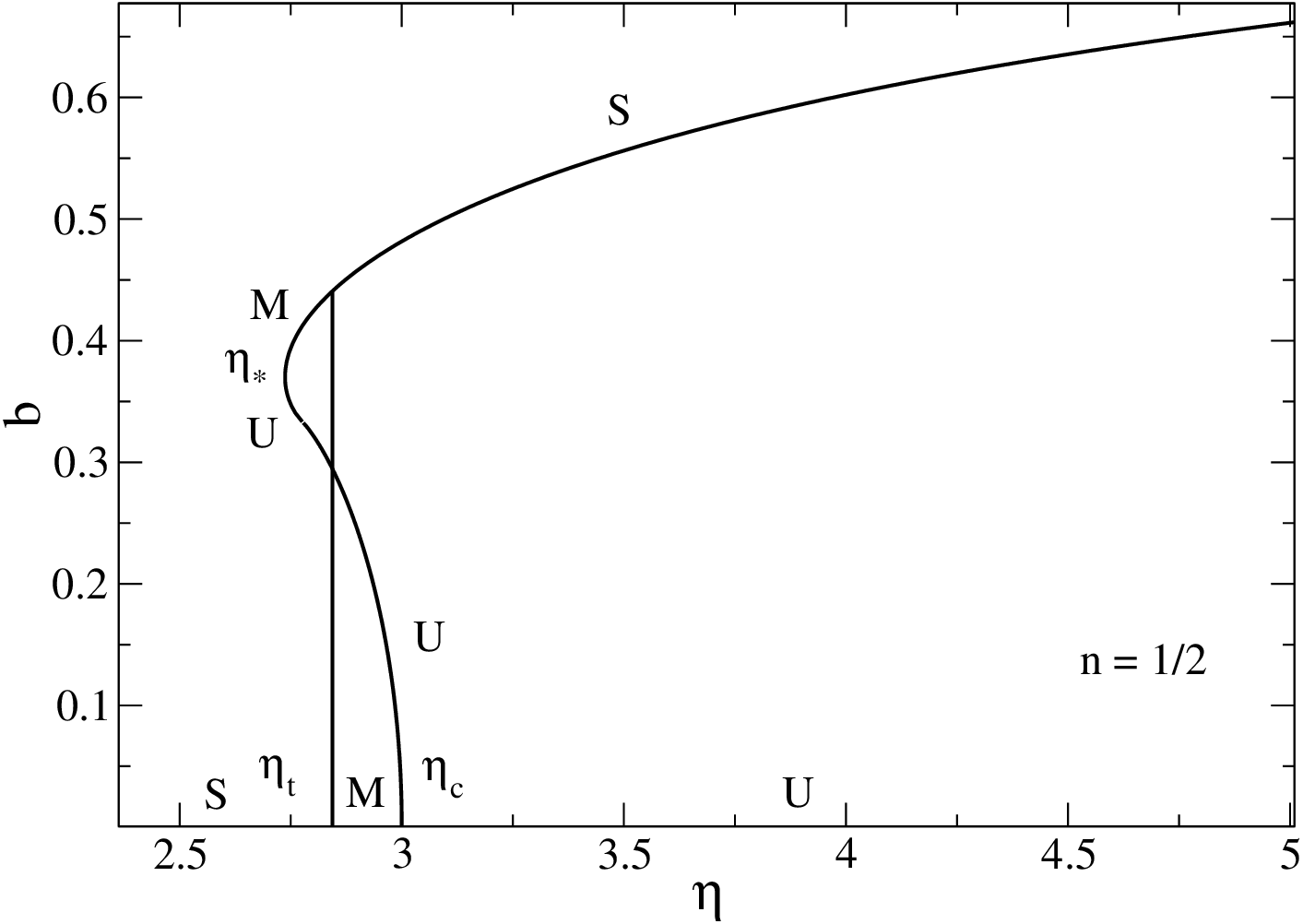}
\caption{Magnetization as a function of the inverse polytropic temperature.}
\label{bVSetaN0.5}
\end{center}
\end{figure}

We can now describe more precisely the
dynamical stability of the waterbag distribution with respect to the
variational problems (\ref{dyn1}) and (\ref{dyn2}) in the limit
$n\rightarrow 1/2^+$. In the microcanonical ensemble, a global entropy
maximum will be called fully stable (S), a local entropy maximum will
be called metastable (M) and a saddle point will be called unstable
(U). Considering Figure \ref{caloN0.5}, we conclude that the
homogeneous phase is fully stable for
$\epsilon>\epsilon_t\simeq 0.352$, metastable for
$\epsilon_c<\epsilon<\epsilon_t$ and unstable for
$0<\epsilon<\epsilon_c$. The upper part of the inhomogeneous branch is
unstable. The lower part of the inhomogeneous branch is metastable for
$\epsilon_t<\epsilon<\epsilon_*$ and fully stable for
$\epsilon_{min}<\epsilon<\epsilon_t$. The energy $\epsilon_*$ can be
interpreted as a spinodal point at which the metastable inhomogeneous branch
disappears. In the canonical ensemble, a global free energy minimum
will be called fully stable (S), a local free energy minimum will be
called metastable (M) and a saddle point will be called unstable
(U). Considering Figure \ref{caloN0.5}, we conclude that the
homogeneous phase is fully stable (S) for $\eta<\eta_t\simeq
2.844$, metastable (M) for $\eta_t<\eta<\eta_c$ and unstable (U) for
$\eta>\eta_c$. The upper part of the inhomogeneous branch is
unstable. The lower part is metastable for $\eta_*<\eta<\eta_t$ and
fully stable for $\eta>\eta_t$. The inverse temperature $\eta_*$ can
be interpreted as a spinodal point at which the metastable inhomogeneous branch
disappears. As noted previously, the ensembles are equivalent. If we
regard the curve of Figure \ref{caloN0.5} as a caloric curve
$\eta(\epsilon)$, we see that $\eta$ is continuous at the transition
$\epsilon=\epsilon_t$ while  $\eta'(\epsilon)$ is discontinuous. This
looks like a second order phase transition. However,
if we regard the curve of Figure \ref{caloN0.5} as an entropic curve
$\sigma(\epsilon)$, we see that it displays a first order phase
transition marked by the discontinuity of $\sigma'(\epsilon)$ at
$\epsilon_t$ and the existence of metastable states. For $n>1/2$, this
leads to a discontinuity of $\eta(\epsilon)$ at $\epsilon_t$ (see
Figure 21 of \cite{cc}).  Therefore, it is better to say that the
caloric curve $\eta(\epsilon)$ displays a first order phase transition
(as for $n>1/2$) in which the jump of temperature tends to zero! The
polytrope $n=1/2$ (waterbag distribution) presents therefore very peculiar
features.

Finally, we can plot the magnetization $b$ as a function of the
energy $\epsilon$ and polytropic inverse temperature $\eta$ by
eliminating $x$ between equations (\ref{ip10}), (\ref{ip11}) and
(\ref{ip12}) and between equations (\ref{ip20}), (\ref{ip21}) and
(\ref{ip22}). The corresponding curves are represented in Figures
\ref{bVSepsilonN0.5} and \ref{bVSetaN0.5}. According to the previous
results, the inhomogeneous phase is unstable for $b\le b_*\simeq
0.37$, metastable for $b_*\le b\le b_t\simeq 0.44$ and fully stable
for $b>b_t$. Note that the magnetization curves present a
discontinuity at $\epsilon_t$ and $\eta_t$ confirming that the
transition is first order.

\subsubsection{The control parameter $\epsilon$}
\label{sec_epsilon}

Let us summarize the previous discussion by expressing the dynamical stability of the waterbag distribution with respect to the Vlasov equation in terms of its energy $\epsilon$ and magnetization $b$:

(i) The spatially homogeneous waterbag distributions ($b=0$) are fully stable for $\epsilon>\epsilon_t=0.352$, metastable  for $\epsilon_c=1/3<\epsilon<\epsilon_t=0.352$ and unstable for $\epsilon<\epsilon_c=1/3$.

(ii) The spatially inhomogeneous waterbag distributions are fully stable for $\epsilon_{min}=-2\le \epsilon\le \epsilon_t=0.352$ and $b>b_t=0.44$, metastable for $\epsilon_{t}=0.352\le \epsilon\le \epsilon_*=0.379$ and $b_*=0.37\le b\le b_t=0.44$.  They do not exist for $\epsilon\ge \epsilon_*=0.379$.

In principle, we cannot conclude that the spatially inhomogeneous waterbag distributions  with $\epsilon_{c}=1/3\le \epsilon\le \epsilon_*=0.379$ and $b\le b_*=0.37$ are Vlasov unstable since the criteria (\ref{dyn1}) and (\ref{dyn2}) provide just {\it sufficient} conditions of dynamical stability \cite{campachav}. However, it seems natural that these solutions are unstable.

\subsubsection{The control parameter $\eta_0$}
\label{sec_mr}

A polytrope of index $n$ is characterized by its energy $E$ (in dimensionless form $\epsilon$) and its inverse polytropic temperature $1/K$ (in dimensionless form $\eta$). These are the control parameters that we have used in the previous sections. For the waterbag distribution, equivalent to a polytrope $n=1/2$, it is better to use the maximum value of the distribution function $\eta_0$ as a control parameter instead of the polytropic temperature $K$. Let us introduce the normalized maximum value of the distribution function
\begin{eqnarray}
\label{mr1}
\mu=\eta_0 \left (\frac{2\pi k}{M}\right )^{1/2}.
\end{eqnarray}
According to equation (\ref{f8}), it is related to the normalized inverse
polytropic temperature (\ref{hp5}) by
\begin{eqnarray}
\label{mr2}
\mu^2=\frac{\eta}{6}.
\end{eqnarray}

\begin{figure}[!h]
\begin{center}
\includegraphics[clip,scale=0.3]{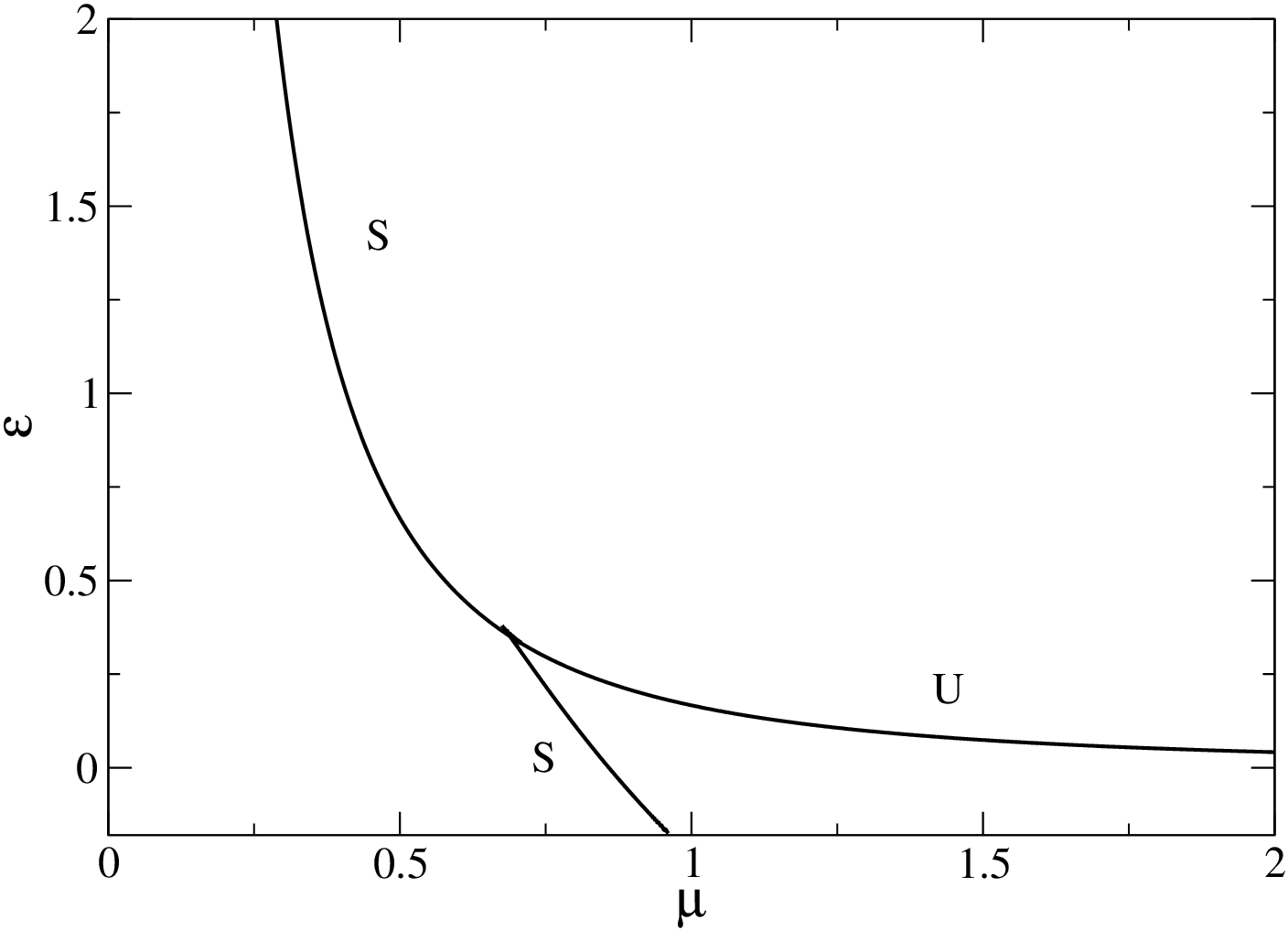}
\caption{Energy $\epsilon$ of the polytrope $n=1/2$ as a function of  the  maximum value of the distribution function $\mu$.  This corresponds to the ground state energy $\epsilon_{ground}$ in the Lynden-Bell theory when the initial condition has only two levels $0$ and $\eta_0$. The inhomogeneous branch tends towards $\epsilon=-2$ for $\mu\rightarrow +\infty$.}
\label{epsilonVSmuN0.5GLOBALE}
\end{center}
\end{figure}

\begin{figure}[!h]
\begin{center}
\includegraphics[clip,scale=0.3]{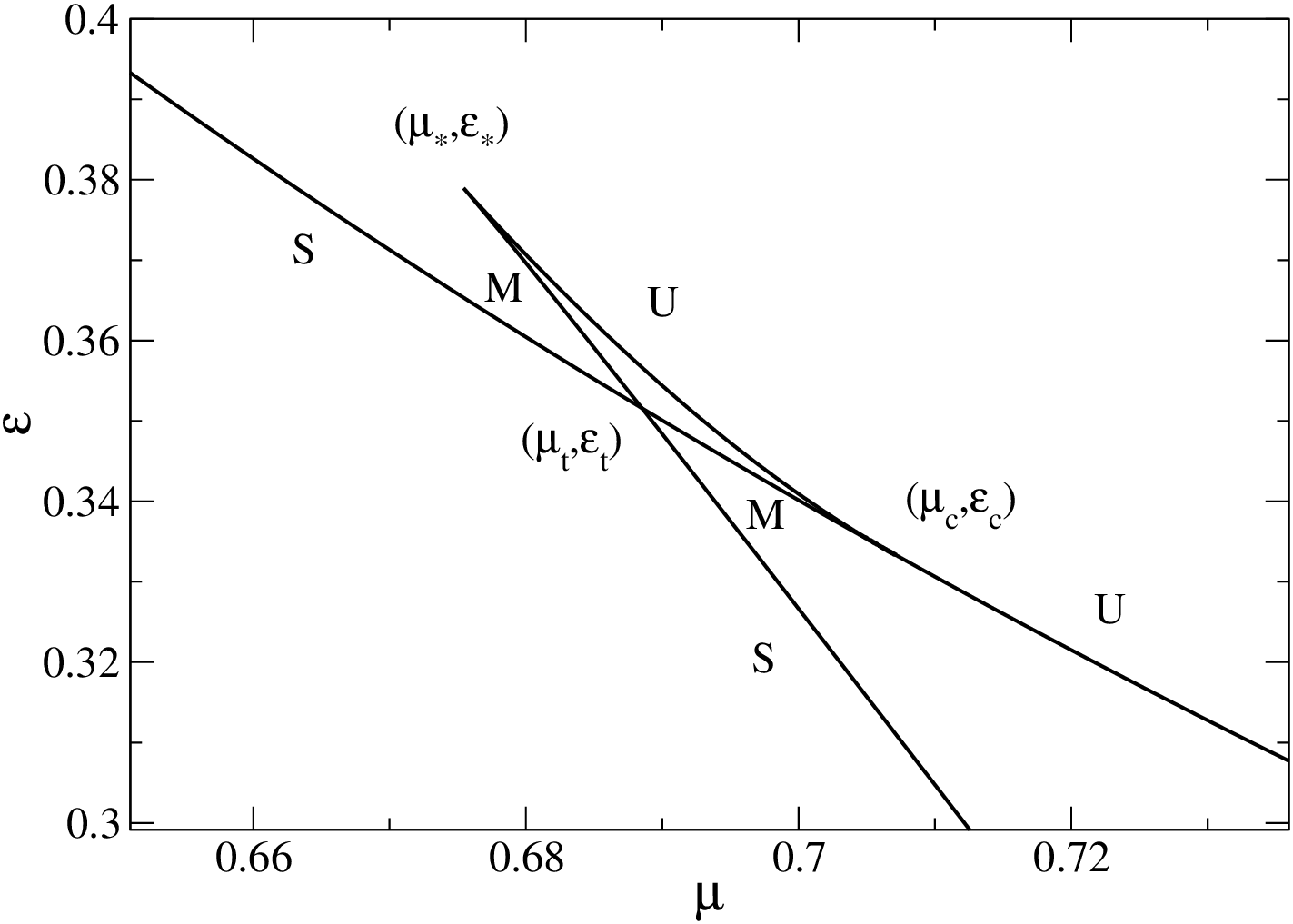}
\caption{Zoom of Figure \ref{epsilonVSmuN0.5GLOBALE} near the bifurcation point.}
\label{epsilonVSmuN0.5}
\end{center}
\end{figure}

\begin{figure}[!h]
\begin{center}
\includegraphics[clip,scale=0.3]{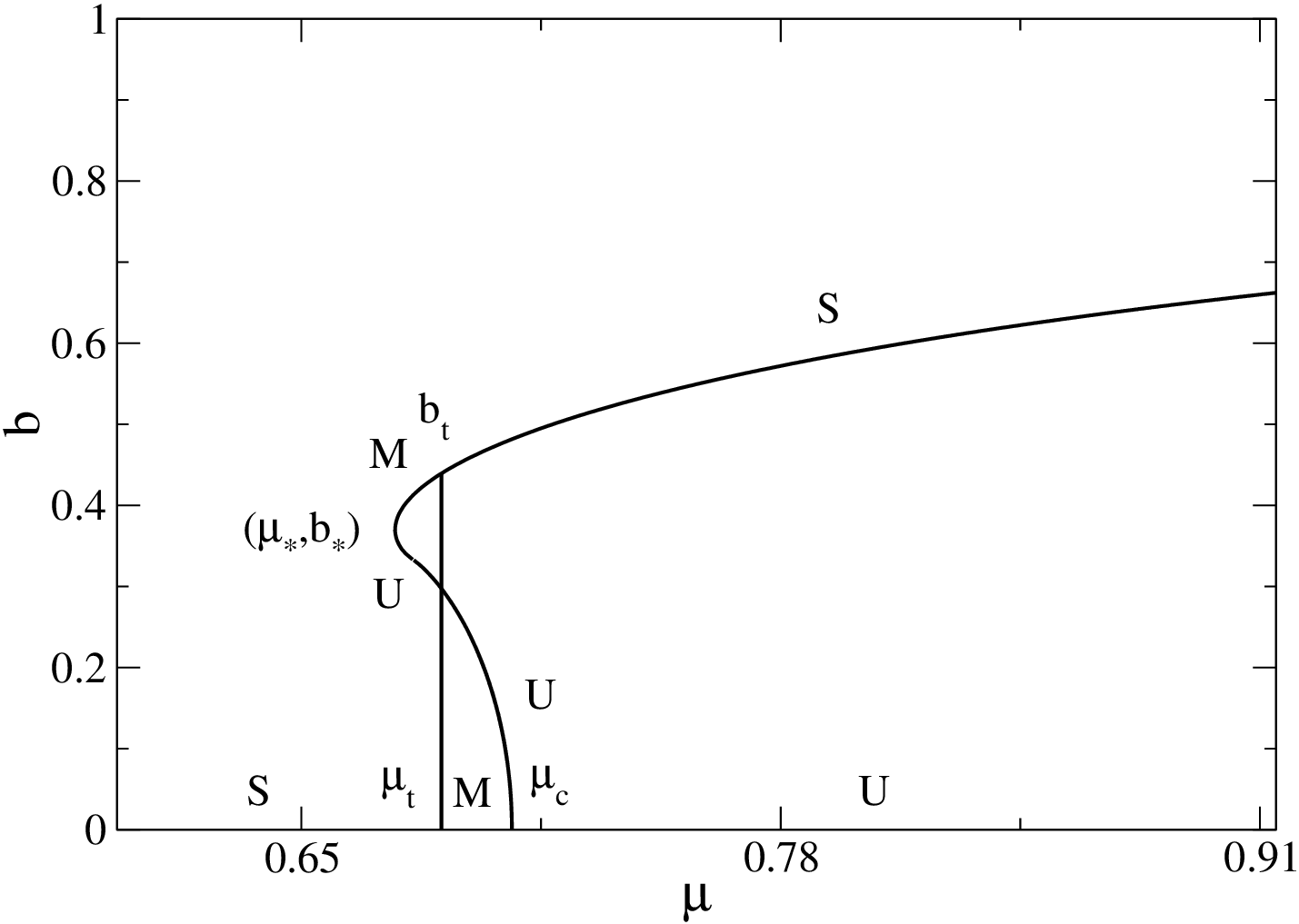}
\caption{Magnetization $b$ as a function of  the  maximum value of the distribution function $\mu$.}
\label{bVSmuN0.5}
\end{center}
\end{figure}

We can now reformulate the preceding results in terms of the variables
$\epsilon$ and $\mu$. The curve giving the energy $\epsilon$ as a
function of the maximum value of the distribution function $\mu$ is
plotted in Figures \ref{epsilonVSmuN0.5GLOBALE} and
\ref{epsilonVSmuN0.5}. According to equations  (\ref{hp7}) and (\ref{mr2}), the energy of
the homogeneous phase is related to the maximum value of the
distribution function by
\begin{eqnarray}
\label{mr3}
\epsilon=\frac{1}{6\mu^2}.
\end{eqnarray}
The stability criterion (\ref{hd2}) expressed in terms of the  maximum value of the distribution function becomes
\begin{eqnarray}
\label{mr4}
\mu\le \mu_c=\frac{1}{\sqrt{2}}.
\end{eqnarray}
The homogeneous phase exists for $\mu\ge 0$ and $\epsilon\ge 0$. It is
fully stable for $\mu<\mu_t\simeq 0.688$ (i.e. $\epsilon>\epsilon_t$),
metastable for $\mu_t<\mu<\mu_c$
(i.e. $\epsilon_c<\epsilon<\epsilon_t$) and unstable for $\mu>\mu_c$
(i.e. $\epsilon<\epsilon_c$).  The inhomogeneous phase exists for
$\mu\ge \mu_*\simeq 0.675$ and $-2\le
\epsilon\le \epsilon_*$. It bifurcates from the homogeneous phase at
$\mu_c=1/\sqrt{2}$ and $\epsilon_c=1/3$. The upper part of the
inhomogeneous branch is unstable. The lower part of the inhomogeneous
branch is metastable for $\mu_*\le \mu \le \mu_t$
(i.e. $\epsilon_t\le\epsilon\le\epsilon_*$) and fully stable for
$\mu>\mu_t$ (i.e. $\epsilon_{min}\le \epsilon\le\epsilon_t$).  The curve giving the
magnetization $b$ as a function of the maximum value of the distribution
function $\mu$ is plotted in Figure \ref{bVSmuN0.5}. The inhomogeneous
phase is unstable for $b\le b_*\simeq 0.37$, metastable for $b_*\le
b\le b_t\simeq 0.44$ and fully stable for $b>b_t$.

{\it Remark:} recalling that the waterbag distribution (\ref{f1}) corresponds to the minimum energy state (i.e. $T=0$) in the two levels approximation of the Lynden-Bell theory \cite{epjb}, the curve of Figure \ref{epsilonVSmuN0.5GLOBALE} gives the ground state energy $\epsilon_{ground}$ as a function of the initial value $\mu$ of the distribution function. As will be shown in the next section, the stable states (S) correspond to global minima of energy while the metastable states (M) correspond to local energy minima and the unstable states (U) to saddle points. In previous works \cite{epjb,marseille,staniscia1}, only the minimum energy $\epsilon_{min}(\mu)=1/(6\mu^2)$ of the homogeneous phase, and the minimum energy $\epsilon_{MIN}(\mu)$ of the inhomogeneous phase in the particular case where the initial condition is a rectangular waterbag distribution, had been determined. Here, we have determined the {\it absolute} minimum energy $\epsilon_{ground}(\mu)$ of the inhomogeneous phase. For $\mu\rightarrow +\infty$ (non degenerate limit), we recover the classical result $\epsilon_{ground}(\mu)\rightarrow -2$. Therefore, the curve of Figure \ref{epsilonVSmuN0.5GLOBALE} completes the phase diagram of \cite{staniscia1} by giving the minimum accessible energy (ground state). This point is specifically  discussed in Appendix A of \cite{staniscia2}.

\subsection{Thermodynamical stability of the Fermi distribution}
\label{sec_ts}

In the previous sections, we have studied the dynamical stability of the Fermi (or waterbag) distribution function (\ref{f1}) and density profile (\ref{ip1})  with respect to the Vlasov and Euler equations. We shall now consider the thermodynamical stability of the Fermi distribution.

\subsubsection{The minimum energy state}
\label{sec_me}

In the mean field approximation, the statistical equilibrium state of a gas of fermions in interaction is obtained by maximizing the Fermi-Dirac entropy at fixed mass and energy in the microcanonical ensemble or by minimizing the Fermi-Dirac free energy at fixed mass in the canonical ensemble (see Section \ref{sec_fd}). Here, we are interested by the ground state $E=E_{ground}$  corresponding to $T=0$. We thus have to determine the minimum energy state for a given value of mass while respecting the Pauli exclusion principle $f\le \eta_0$.

To solve this minimization problem, we can proceed in two steps. We first minimize the energy at fixed mass {\it and} density profile $\rho(\theta)=\int f(\theta,v)\, dv$. Since the specification of the density profile determines the mass and the potential energy, this is equivalent to minimizing the kinetic energy at fixed density profile. This is achieved by populating the lowest kinetic energy states with the maximum of fermions allowed by the Pauli exclusion principle. This leads to the optimal distribution function
\begin{eqnarray}
\label{me1}
f_*=\eta_0, \quad {\rm if}\quad |v|\le v_F(\theta),\nonumber\\
f_*=0,\quad {\rm if}\quad |v|> v_F(\theta),
\end{eqnarray}
where the Fermi velocity $v_F(\theta)$ is determined by the density profile according to $v_F(\theta)=\rho(\theta)/({2\eta_0})$. We can now express  the energy (\ref{fa2}) as a functional of $\rho$ by writing $E[\rho]=E[f_*]$. This yields
\begin{equation}
\label{me3}
E=\frac{K}{2}\int \rho^3\, d\theta+\frac{1}{2}\int\rho\Phi\, d\theta,
\end{equation}
where $K$ is given by equation (\ref{f8}). Finally, the ground state corresponds to the optimal distribution function $f_*$ with the optimal density profile $\rho_*$ that minimizes $E[\rho]$ at fixed mass $M$.

{\it Remark 1:} we can proceed differently by considering the limit $T\rightarrow 0^+$ of the Fermi-Dirac free energy. For $T>0$, the Fermi-Dirac free energy is given by $F=E-TS$ where $E$ is given by equation (\ref{fa2}) and $S$ by equation (\ref{fd2}). Now, for a general entropic functional of the form $S=-\int C(f)\, d\theta dv$ where $C$ is convex, it is shown in \cite{cvb,aaantonov,assiseChav} that the minimization of $F[f]$ at fixed mass is equivalent to the minimization of $F[\rho]$ at fixed mass, where $F[\rho]$ is the free energy (\ref{dyn4}). In the case of fermions, $p=p(\rho)$ is given by the Fermi-Dirac equation of state (see, e.g., \cite{fermiD}). For $T=0$, it reduces to the polytropic equation of state  (\ref{f6}) and the free energy $F[\rho]$ reduces to the energy functional (\ref{me3}).

{\it Remark 2:} for a polytrope of index $n=1/2$, the free energy (\ref{dyn4}) is equivalent to the energy functional (\ref{me3}). We can therefore directly use the results of Section \ref{sec_vlasov} to investigate the thermodynamical stability problem. In particular, the stability of the homogeneous phase is determined  by the criteria given in Section  \ref{sec_hd} since they are precisely obtained by minimizing $F[\rho]$ at fixed mass \cite{cvb,campachav,yama}. The stability of the inhomogeneous states can be determined by using the results of Section \ref{sec_in} or by proceeding as in the next section.

{\it Remark 3:} a maximum of entropy at fixed mass and energy, a minimum of free energy at fixed mass and a  minimum of energy at fixed mass with the constraint $f\le \eta_0$ are  guaranteed to be dynamically stable with respect to the Euler and Vlasov equations. Therefore, thermodynamical stability implies dynamical stability. However, the converse is not always true since the criteria (\ref{dyn1}) and (\ref{dyn2}) provide just {\it sufficient} conditions of dynamical stability \cite{campachav}.

\subsubsection{The series of equilibria}
\label{sec_se}

\begin{figure}[!h]
\begin{center}
\includegraphics[clip,scale=0.3]{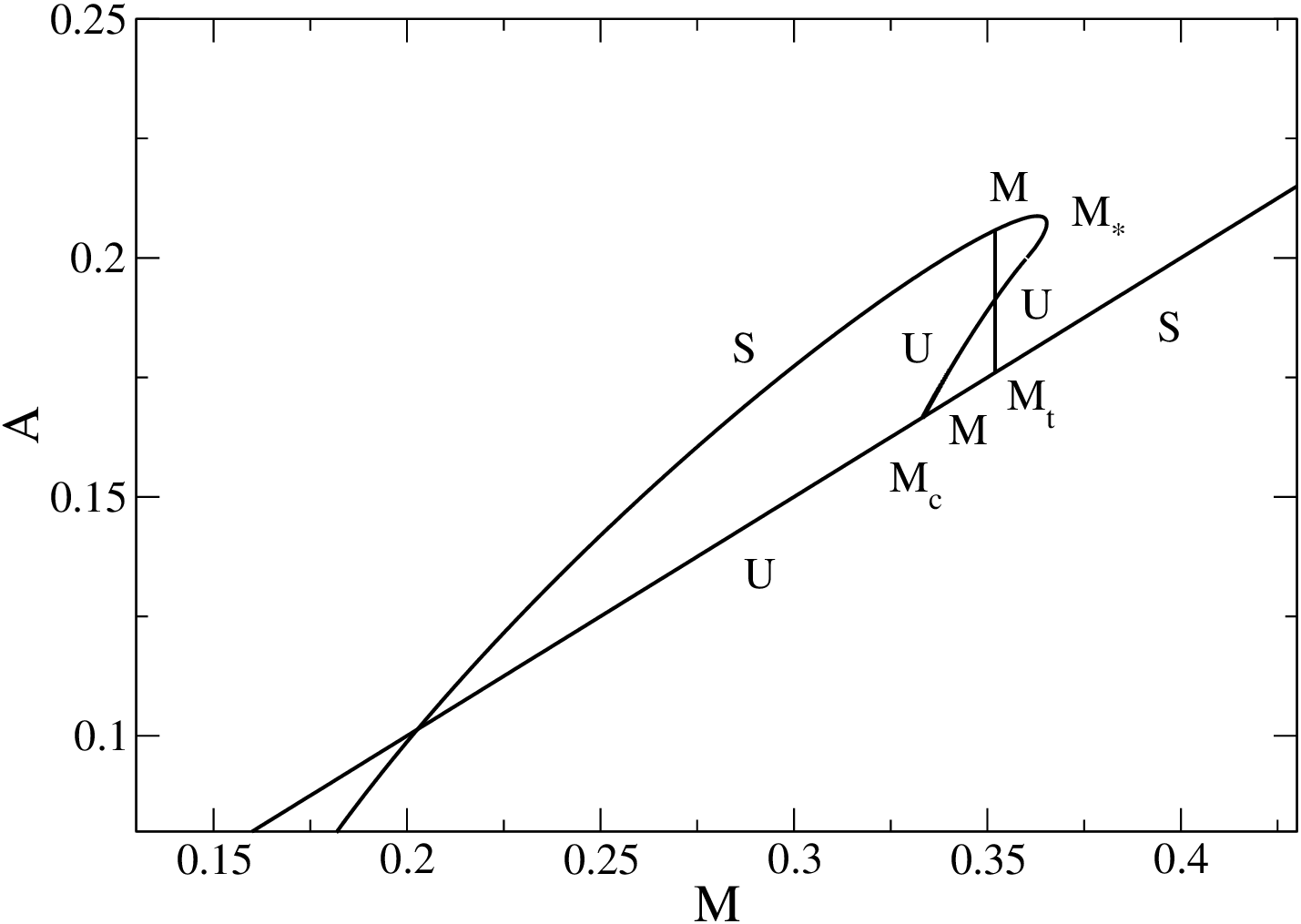}
\caption{The chemical potential ${\cal A}$ as a function of the mass ${\cal M}$.}
\label{m-a}
\end{center}
\end{figure}

The critical points of energy $E$, given by equation (\ref{me3}), at fixed mass $M$ are determined by the variational principle $\delta E-\alpha\delta M=0$ where $\alpha$ is a Lagrange multiplier associated with the
conservation of mass.  This returns the density profile (\ref{ip1})
with $\alpha=\epsilon_F$. In the following, we shall consider the case $\epsilon_F>0$
since it corresponds to the region where the phase transition
occurs. To determine whether a critical point of energy is a
minimum or a saddle point, we can use the Poincar\'e
theorem. To that purpose, we have to plot $\alpha$ as a function of
$M$ (for fixed $\hbar$) since $\alpha=\partial E/\partial
M$.   Since $\alpha$ is an increasing function of $A$, this is
equivalent to plotting $A$ as a function of $M$ (for fixed
$\hbar$). For the inhomogeneous phase, using the results of Section \ref{sec_ip},
we obtain the relations
\begin{equation}
\label{se2}
{\cal M}\equiv \frac{M\pi\hbar^2}{12k}=\frac{1}{\eta(x)},
\end{equation}
\begin{equation}
\label{se3}
{\cal A}\equiv \frac{A\pi^2\hbar^2}{12k}=\frac{I_{3,1}(x)}{x},
\end{equation}
which determine ${\cal A}({\cal M})$ in parametric form (with parameter $x$).
For the homogeneous phase,
we simply have ${\cal A}={\cal
M}/2$. The curve ${\cal A}({\cal M})$ is represented in Figure
\ref{m-a}. We know from the study of Section \ref{sec_hd} that
the homogeneous phase is stable for ${\cal M}>{\cal M}_{c}=1/3$ and
unstable for ${\cal M}<{\cal M}_{c}$. Then, using the Poincar\'e
theorem, we deduce that the inhomogeneous phase is unstable close to
the bifurcation point but that it becomes, and remains, stable after
the turning point of mass at $M_*\simeq 0.365$.

\begin{figure}[!h]
\begin{center}
\includegraphics[clip,scale=0.3]{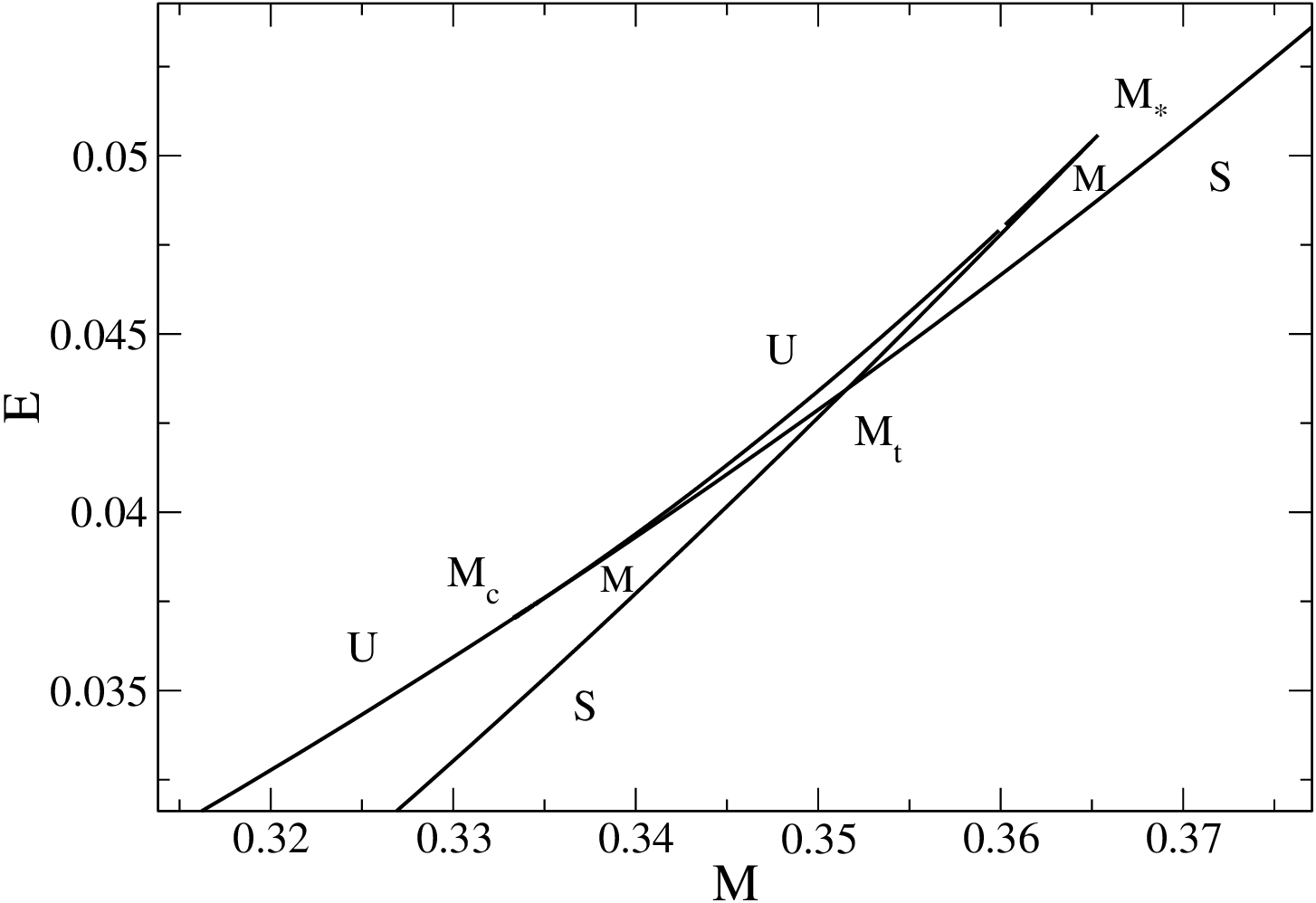}
\caption{The energy ${\cal E}$ as a function of the mass ${\cal M}$.}
\label{m-e}
\end{center}
\end{figure}

To study the stability of the Fermi distribution, we can also plot the
energy $E$ as a function of the mass $M$ and compare the energy of the
different solutions that have the same mass (for fixed
$\hbar$). For the inhomogeneous phase, using the results of Section \ref{sec_ip},
we obtain the relations
\begin{equation}
\label{se4}
{\cal E}\equiv \frac{E\pi^3\hbar^4}{18 k^3}=\frac{\epsilon(x)}{\eta(x)^2},
\end{equation}
\begin{equation}
\label{se5}
{\cal M}\equiv \frac{M\pi\hbar^2}{12k}=\frac{1}{\eta(x)},
\end{equation}
which determines ${\cal E}({\cal M})$ in parametric form (with parameter $x$). For the homogeneous phase, using equation (\ref{hp4b}), we simply have
${\cal E}={\cal M}^3$. The curve ${\cal E}({\cal M})$ is represented
in Figure \ref{m-e}. This curve determines the transition
mass ${\cal M}_t\simeq 0.352$ separating stable and metastable states (here, fully stable states 
are global minima of energy and metastable states are local minima of energy). The curves of
Figures \ref{m-a} and \ref{m-e} display a first order phase transition
marked by the discontinuity of $\alpha(M)=E'(M)$ at $M=M_t$ and the
occurrence of metastable states. On the other hand, $M_*$ can be
interpreted as a spinodal point marking the end of the inhomogeneous  metastable
phase.

In conclusion: the homogeneous phase exists for ${\cal M}\ge 0$. It is fully stable for ${\cal M}>{\cal M}_t$, metastable for  ${\cal M}_c<{\cal M}<{\cal M}_t$ and unstable for ${\cal M}<{\cal M}_c$. The inhomogeneous phase exists for $0\le {\cal M}\le {\cal M}_*$. It bifurcates from the homogeneous phase at ${\cal M}={\cal M}_c$. The lower part of the inhomogeneous branch is unstable. The upper part of the inhomogeneous branch is fully stable for ${\cal M}<{\cal M}_t$ and  metastable for  ${\cal M}_t<{\cal M}<{\cal M}_*$. Since ${\cal M}=1/\eta$, we recover the same stability results as in the dynamical approach of Section \ref{sec_vlasov} but from a different point of view. This equivalence was expected in view of Remark 2 of Section \ref{sec_me}.

\subsubsection{The control parameter $\hbar$}
\label{sec_cph}

\begin{figure}[!h]
\begin{center}
\includegraphics[clip,scale=0.3]{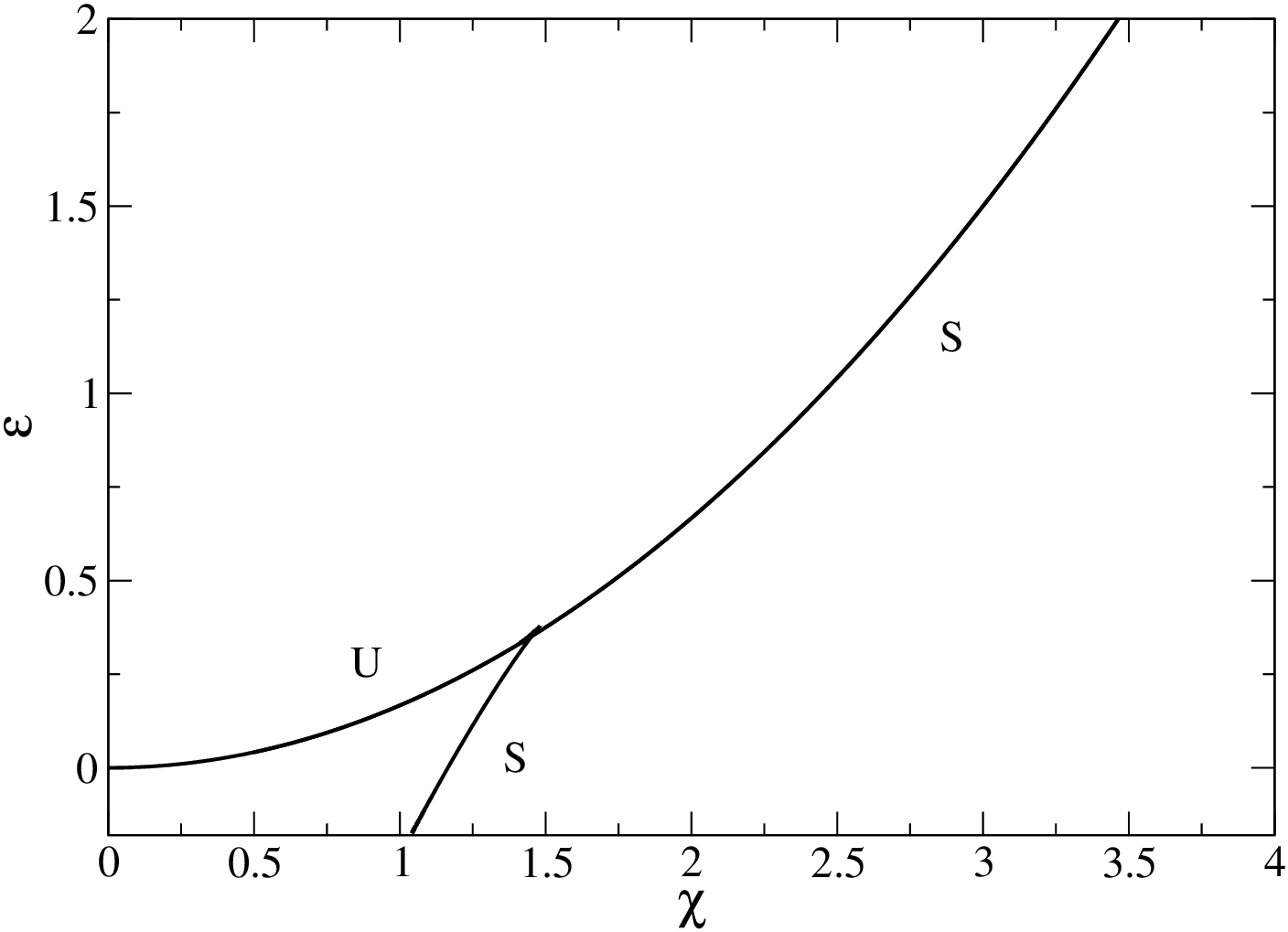}
\caption{Ground state energy $\epsilon$ as a function of the Planck constant $\chi$. The inhomogeneous branch (stable) tends towards $\epsilon=-2$ for $\chi\rightarrow 0$.}
\label{epsilonVShN0.5GLOBALE}
\end{center}
\end{figure}

\begin{figure}[!h]
\begin{center}
\includegraphics[clip,scale=0.3]{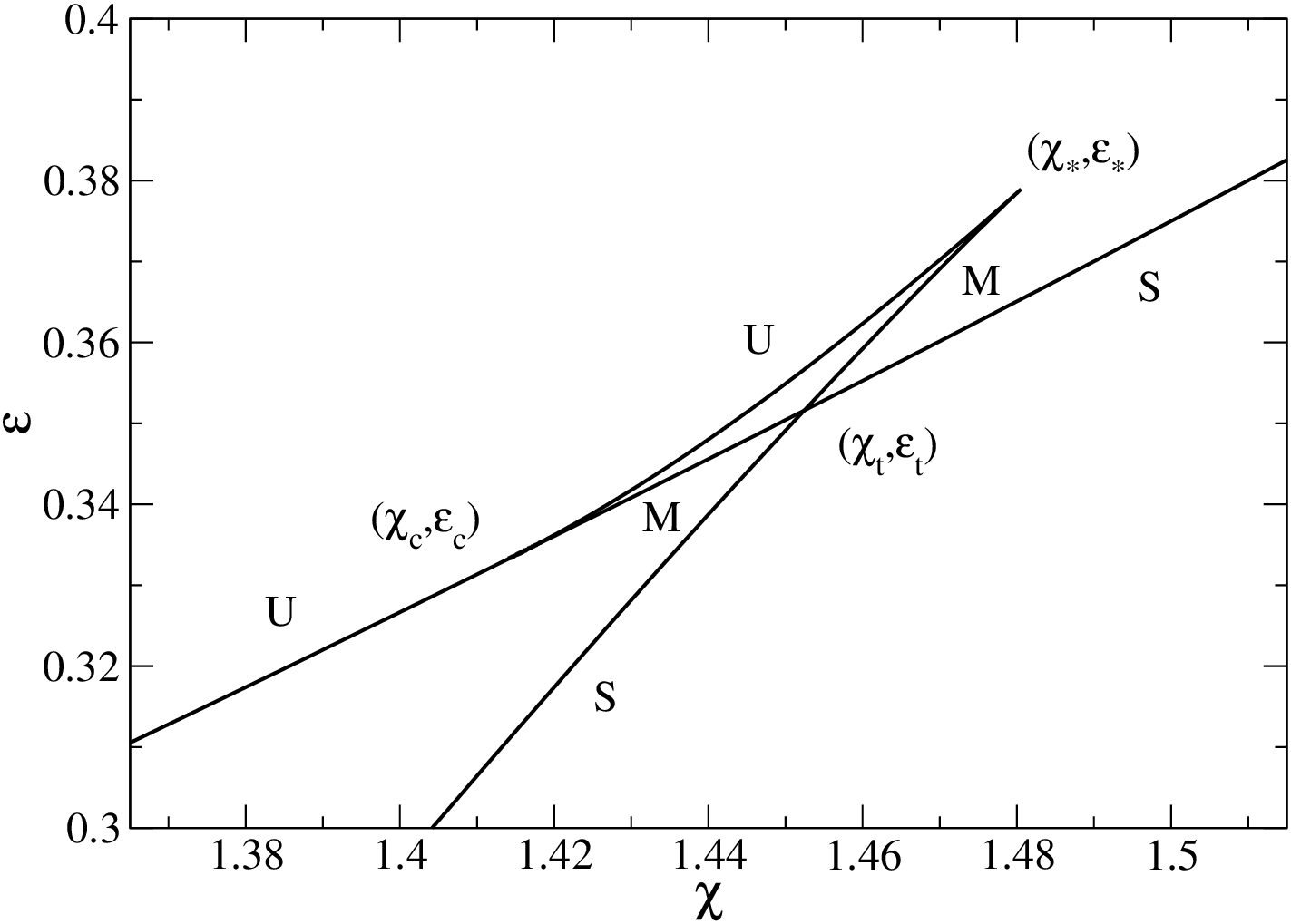}
\caption{Zoom of Figure \ref{epsilonVShN0.5GLOBALE} near the bifurcation point.}
\label{epsilonVShN0.5}
\end{center}
\end{figure}

\begin{figure}[!h]
\begin{center}
\includegraphics[clip,scale=0.3]{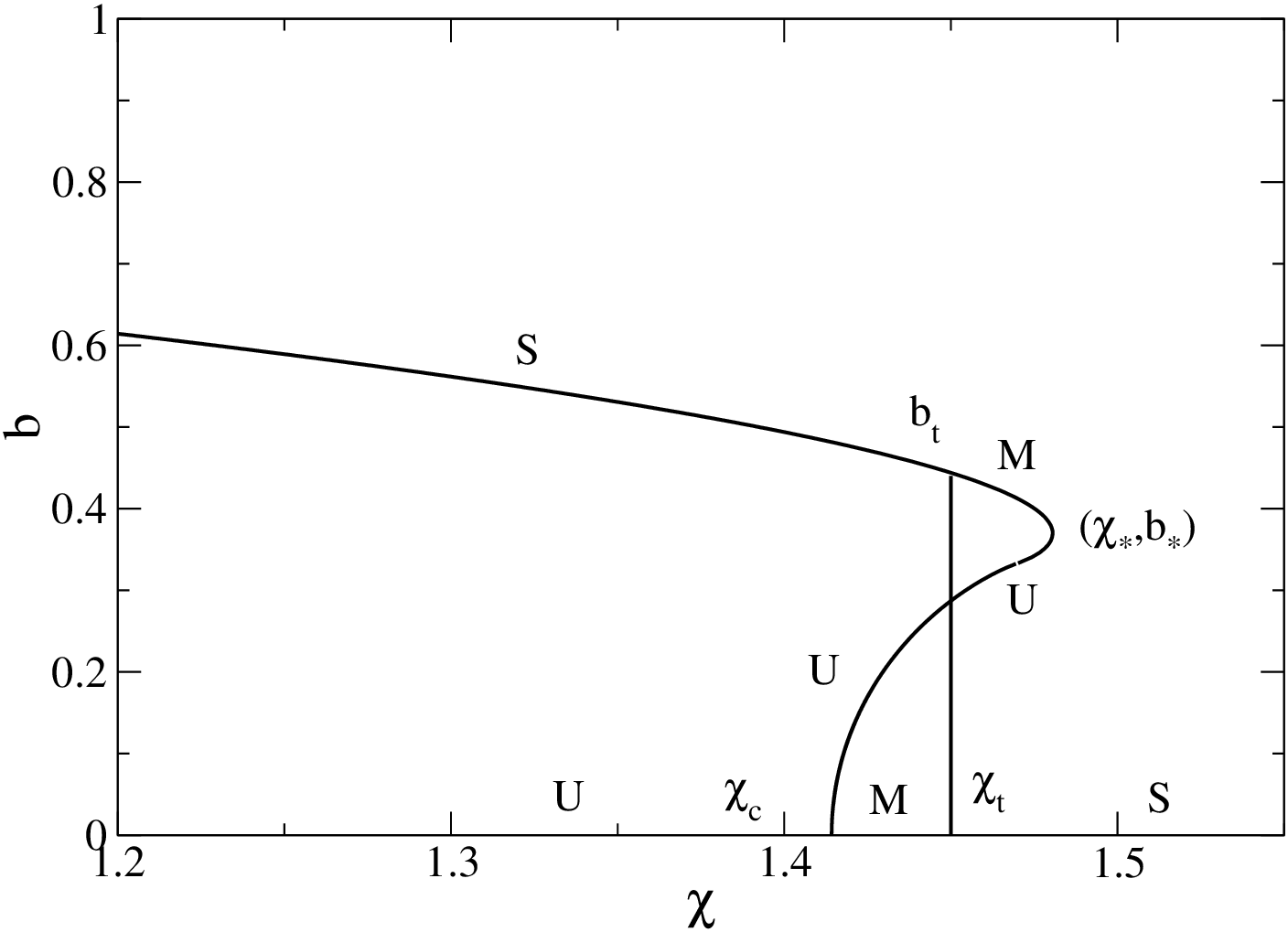}
\caption{Magnetization $b$ as a function of the Planck constant $\chi$. It displays a first order phase transition between the classical regime (inhomogeneous state $b\neq 0$) and the quantum regime (homogeneous state $b=0$).}
\label{bVShN0.5}
\end{center}
\end{figure}

In the previous section, we have fixed $\hbar$ and taken the mass $M$ as a control parameter. This is the right way to solve the thermodynamical  problem corresponding to the minimization of energy at fixed mass. Now that we have established the conditions of stability, in order to present the final results, it is  more relevant to fix $M$ and take $\hbar$ as a control parameter. We therefore introduce the normalized Planck constant for fermions
\begin{eqnarray}
\label{cph1}
\chi=\hbar \left (\frac{\pi M}{2k}\right )^{1/2}.
\end{eqnarray}
According to equations (\ref{f8}) and (\ref{hp5}), it is related to the parameter $\eta$ by
\begin{eqnarray}
\label{cph2}
\chi=\left (\frac{6}{\eta}\right )^{1/2}.
\end{eqnarray}
Since $M$ is fixed, the curves giving the energy and the magnetization as a function of the normalized Planck constant correspond to $\epsilon(\chi)$ and $b(\chi)$.  They are plotted in Figures \ref{epsilonVShN0.5GLOBALE}, \ref{epsilonVShN0.5} and \ref{bVShN0.5}. We can now reformulate the preceding results in terms of the normalized Planck constant $\chi$. Before that, we note that the normalized Planck constant is related to the normalized maximum value of the distribution function by
\begin{eqnarray}
\label{cph3}
\chi=\frac{1}{\mu}.
\end{eqnarray}
Therefore, the results will coincide with those obtained in Section \ref{sec_mr} up to a slight reinterpretation of the parameters. However, we describe the results in detail in order to facilitate the comparison with the case of bosons in Section \ref{sec_cos}.

The curve giving the ground state energy $\epsilon$ as a function of
the Planck constant $\chi$ (in dimensionless units) is plotted in
Figures \ref{epsilonVShN0.5GLOBALE} and
\ref{epsilonVShN0.5}. According to equations (\ref{mr3}) and
(\ref{cph3}), the ground state energy of the homogeneous phase is
related to the Planck constant by
\begin{eqnarray}
\label{cph4}
\epsilon=\frac{\chi^2}{6}.
\end{eqnarray}
On the other hand, the stability criterion (\ref{hd2}) can be expressed in terms of the Planck constant as
\begin{eqnarray}
\label{cpg5}
\chi\ge \chi_c=\sqrt{2}.
\end{eqnarray}
The homogeneous phase exists for $\chi\ge 0$  and $\epsilon\ge 0$. It is unstable for $\chi<\chi_c$ (i.e. $\epsilon<\epsilon_c$), metastable for $\chi_c\le \chi\le \chi_t\simeq 1.45$  (i.e. $\epsilon_c<\epsilon<\epsilon_t\simeq 0.352$) and fully stable for $\chi>\chi_t$ (i.e. $\epsilon>\epsilon_t$). The inhomogeneous phase exists for $\chi\le \chi_*\simeq 1.48$ and  $-2\le \epsilon\le \epsilon_*\simeq 0.379$. It bifurcates from the homogeneous phase at $\chi_c=\sqrt{2}$ and  $\epsilon_c=1/3$. The upper part of the inhomogeneous branch is unstable. The lower part of the inhomogeneous branch is metastable for $\chi_t\le \chi\le \chi_*$ (i.e. $\epsilon_t\le\epsilon\le\epsilon_*$) and fully stable for $0\le \chi<\chi_t$ (i.e. $-2\le \epsilon\le\epsilon_t$). The curve giving the magnetization $b$ versus the Planck constant $\chi$ is plotted in Figure \ref{bVShN0.5}. The inhomogeneous phase is unstable for $b\le b_*\simeq 0.37$, metastable for $b_*\le b\le b_t\simeq 0.44$ and fully stable for $b>b_t$. It displays a first order phase transition at $\chi=\chi_t$ marked by the discontinuity of the magnetization.

Summarizing, we have determined the ground state ($T=0$) of the Fermi gas with cosine interaction and investigated its thermodynamical stability. In the classical limit $\chi=0$, the ground state energy is $\epsilon=-2$ and the magnetization is $b=1$. This corresponds to a classical inhomogeneous gas at $T=0$ whose density profile is a Dirac peak $\rho=M\delta(\theta)$. The homogeneous phase with energy $\epsilon=0$ and magnetization $b=0$ is unstable since $T=0<T_c$. This returns the classical results \cite{cvb}. However, we note that when quantum mechanics is taken into account, there exists a critical value of the normalized Planck constant above which the homogeneous phase becomes stable (more precisely, it is metastable for $\chi_c<\chi<\chi_t$ and fully stable for $\chi>\chi_t$). In parallel, the inhomogeneous phase becomes metastable for $\chi_t<\chi<\chi_*$ and disappears for $\chi>\chi_*$. Therefore, in the quantum regime, the homogeneous phase is stabilized against clustering by the Pauli exclusion principle. This is similar to the stabilization of fermion stars against gravitational collapse in astrophysics due to quantum mechanics.
In the HMF model, this stabilization takes place through a {\it first order} phase transition.

{\it Remark:} the various approaches developed in Sections \ref{sec_vlasov} and \ref{sec_ts} show the importance of the parameter $\eta$ in the stability analysis. This parameter can have several interpretations: inverse polytropic temperature (\ref{hp5}), maximum value of the initial condition (\ref{mr2}), inverse mass (\ref{se2}) and inverse Planck constant (\ref{cph3}). The change of stability corresponds to the turning point of $\eta(x)$ that takes place at $\eta_*\simeq 2.737$. This is relatively similar to  results obtained in astrophysics \cite{fermiD}.

\section{Bosons at zero temperature}
\label{sec_bosons}

In this section, we consider the HMF model at $T=0$ in the case where the particles are bosons. We use a mean field approximation that becomes exact in a proper thermodynamic limit $N\rightarrow +\infty$ defined in Appendix  \ref{sec_tlfb}.

\subsection{The mean field Schr\"odinger equation}
\label{sec_mfs}

Let us consider a system of bosons  interacting via a long-range binary potential $u(|{\bf r}-{\bf r}'|)$. At zero temperature $T=0$, these particles form a Bose-Einstein condensate (BEC) described by a wave function $\psi({\bf r},t)$. For $N\rightarrow +\infty$, the wave function is solution of the mean field Schr\"odinger equation
\begin{equation}
\label{mfs1}
i\hbar \frac{\partial\psi}{\partial t}({\bf r},t)=-\frac{\hbar^2}{2m}\Delta\psi({\bf r},t)+m\Phi({\bf r},t)\psi({\bf r},t),
\end{equation}
\begin{eqnarray}
\label{mfs2}
\Phi({\bf r},t)=\int \rho({\bf r}',t) u(|{\bf r}-{\bf r}'|)\, d{\bf r}',
\end{eqnarray}
\begin{eqnarray}
\label{mfs3}
\rho({\bf r},t)=Nm|\psi({\bf r},t)|^2,
\end{eqnarray}
\begin{eqnarray}
\label{mfs4}
\int |\psi({\bf r},t)|^2\, d{\bf r}=1.
\end{eqnarray}
Equation (\ref{mfs4}) is the normalization condition, equation (\ref{mfs3}) gives the density profile of the BEC, equation (\ref{mfs2}) determines the associated potential and equation (\ref{mfs1}) determines the wavefunction. For the gravitational potential, these equations reduce to the Schr\"odinger-Poisson system which describes boson stars and self-gravitating BECs \cite{rb,bohmer,prep}. The validity of the Schr\"odinger-Poisson system has been justified rigorously by Lieb \& Yau \cite{yau}. In this paper, we shall study the bosonic HMF model at $T=0$ that is described by the mean field  Schr\"odinger equation (\ref{mfs1}) with a cosine potential of interaction (\ref{fa3}).

\subsection{The Madelung transformation}
\label{sec_mad}

Let use the Madelung \cite{madelung} transformation to rewrite the Schr\"odinger equation in the form of hydrodynamic equations. We first set
\begin{equation}
\label{mad1}
\psi({\bf r},t)=A({\bf r},t) e^{iS({\bf r},t)/\hbar}
\end{equation}
where $A({\bf r},t)$ and $S({\bf r},t)$ are real functions. We clearly have
\begin{equation}
\label{mad2}
A=\sqrt{|\psi|^2},\qquad S=\frac{\hbar}{2i}\ln\left (\frac{\psi}{\psi^*}\right ),
\end{equation}
where $\psi^*$ denotes the complex conjugate. Substituting equation (\ref{mad1}) in equation (\ref{mfs1}) and separating real and imaginary parts, we obtain
\begin{equation}
\label{mad3}
\frac{\partial S}{\partial t}+\frac{1}{2m}(\nabla S)^2+m\Phi-\frac{\hbar^2}{2m}\frac{\Delta A}{A}=0,
\end{equation}
\begin{equation}
\label{mad4}
\frac{\partial A^2}{\partial t}+\nabla\cdot \left (\frac{A^2 \nabla S}{m}\right )=0.
\end{equation}
The first equation has a form similar to the classical Hamilton-Jacobi equation with an additional potential term  $Q=-\frac{\hbar^2}{2m}\frac{\Delta A}{A}$ called the quantum potential. Following Madelung, we introduce the density and velocity fields
\begin{equation}
\label{mad5}
\rho=NmA^2=Nm|\psi|^2, \qquad  {\bf u}=\frac{1}{m}\nabla S.
\end{equation}
We note that the flow defined in this way is irrotational since $\nabla\times {\bf u}={\bf 0}$.  With these notations, equation (\ref{mad4}) becomes similar to the equation of continuity in hydrodynamics. On the other hand, equation (\ref{mad3}) can be interpreted as a generalized Bernouilli equation for a potential flow. Taking the gradient of equation (\ref{mad3}) and using the well-known identity $({\bf u}\cdot \nabla){\bf u}=\nabla (\frac{{\bf u}^2}{2})-{\bf u}\times (\nabla\times {\bf u})$ which reduces to $({\bf u}\cdot \nabla){\bf u}=\nabla (\frac{{\bf u}^2}{2})$ for an irrotational flow, we obtain an equation similar to the pressureless  Euler equation with an additional quantum potential. In conclusion, the mean field Schr\"odinger equation is equivalent to the ``hydrodynamic'' equations
\begin{equation}
\label{mad6}
\frac{\partial\rho}{\partial t}+\nabla\cdot (\rho {\bf u})=0,
\end{equation}
\begin{equation}
\label{mad7}
\frac{\partial {\bf u}}{\partial t}+({\bf u}\cdot \nabla){\bf u}=-\nabla\Phi-\frac{1}{m}\nabla Q,
\end{equation}
where
\begin{equation}
\label{mad8}
Q=-\frac{\hbar^2}{2m}\frac{\Delta \sqrt{\rho}}{\sqrt{\rho}}=-\frac{\hbar^2}{4m}\left\lbrack \frac{\Delta\rho}{\rho}-\frac{1}{2}\frac{(\nabla\rho)^2}{\rho^2}\right\rbrack,
\end{equation}
is the quantum potential. We shall refer to equations (\ref{mad6}), (\ref{mad7}) and (\ref{mfs2}) as the  quantum Euler equations.  In the classical limit $\hbar\rightarrow 0$, the quantum potential disappears and we recover the ordinary Euler equations.

{\it Remark:} the quantum potential first appeared in the work of Madelung \cite{madelung} and was rediscovered by Bohm \cite{bohm} (it is sometimes called the Bohm potential). We note the identity
\begin{equation}
\label{bohm1}
-\frac{1}{m}\nabla Q\equiv -\frac{1}{\rho}\partial_j P_{ij},
\end{equation}
where $P_{ij}$ is the quantum stress tensor
\begin{equation}
\label{bohm2}
P_{ij}=-\frac{\hbar^2}{4m^2}\rho\partial_i\partial_j\ln\rho,
\end{equation}
or
\begin{equation}
\label{bohm3}
P_{ij}=\frac{\hbar^2}{4m^2}\left (\frac{1}{\rho}\partial_i\rho\partial_j\rho-\delta_{ij}\Delta\rho\right ).
\end{equation}
This shows that the quantum potential is equivalent to an anisotropic pressure.

\subsection{The time independent Schr\"odinger equation}
\label{sec_tis}

If we consider a wavefunction of the form
\begin{equation}
\label{tis1}
\psi({\bf r},t)=A({\bf r})e^{-i\frac{Et}{\hbar}},
\end{equation}
we obtain the  time independent Schr\"odinger equation
\begin{eqnarray}
-\frac{\hbar^2}{2m}\Delta\psi({\bf r})+m\Phi({\bf r})\psi({\bf r})=E\psi({\bf r}),
\label{tis2}
\end{eqnarray}
where $\psi({\bf r})\equiv A({\bf r})$ is real and $\rho({\bf r})=Nm\psi^2({\bf r})$. Dividing equation (\ref{tis2}) by $\psi({\bf r})$, we get
\begin{equation}
\label{tis3}
m\Phi-\frac{\hbar^2}{2m}\frac{\Delta\sqrt{\rho}}{\sqrt{\rho}}=E,
\end{equation}
which can be written in terms of the quantum potential as
\begin{equation}
\label{tis4}
m\Phi+Q=E.
\end{equation}
This relation can also be derived from the Hamilton-Jacobi equation (\ref{mad3}) by setting  $S=-Et$.  

Using the Madelung transformation, equation (\ref{tis4}) represents the steady state of the quantum Euler equations (\ref{mad6})-(\ref{mad8}). Indeed, taking $\partial_t=0$ and ${\bf u}={\bf 0}$ in equation (\ref{mad7}), we get
\begin{equation}
\label{tis5}
\nabla\Phi+\frac{1}{m}\nabla Q={\bf 0},
\end{equation}
which can be interpreted as a condition of quantum hydrostatic equilibrium. It describes the balance between the long-range interaction and the quantum pressure due to the Heisenberg uncertainty principle. Integrating this relation, we recover equation (\ref{tis4}) where the energy $E$ appears as a constant of integration.

\subsection{The total energy}
\label{sec_te}

The energy functional associated with the mean field Schr\"odinger equation, or equivalently with the  quantum Euler equations, is
\begin{equation}
\label{te1}
E_{tot}=\Theta_c+\Theta_Q+W.
\end{equation}
The first two terms correspond to the total kinetic energy
\begin{eqnarray}
\label{te1b}
\Theta=\frac{N\hbar^2}{2m}\int |\nabla\psi|^2 \, d{\bf r}.
\end{eqnarray}
Using the Madelung transformation, it can be decomposed into the ``classical'' kinetic energy
\begin{equation}
\label{te2}
\Theta_c=\int\rho\frac{{\bf u}^2}{2}\, d{\bf r}.
\end{equation}
and the ``quantum'' kinetic energy
\begin{equation}
\label{te4}
\Theta_Q=\frac{1}{m}\int \rho Q\, d{\bf r}.
\end{equation}
Substituting the expression of the quantum potential (\ref{mad8}) in equation (\ref{te4}), we equivalently have
\begin{eqnarray}
\label{te5}
\Theta_Q&=&-\frac{\hbar^2}{2m^2}\int \sqrt{\rho}\Delta\sqrt{\rho}\, d{\bf r}\nonumber\\
&=&\frac{\hbar^2}{2m^2}\int (\nabla\sqrt{\rho})^2\, d{\bf r}=\frac{\hbar^2}{8m^2}\int \frac{(\nabla\rho)^2}{\rho}\, d{\bf r}.
\end{eqnarray}
The third term is the potential energy
\begin{equation}
\label{te3}
W=\frac{1}{2}\int\rho\Phi\, d{\bf r}.
\end{equation}
The total energy can be expressed in terms of the wavefunction as
\begin{eqnarray}
\label{vf1}
E_{tot}=\int \left\lbrack \frac{N\hbar^2}{2m}|\nabla\psi|^2+\frac{1}{2}Nm\Phi|\psi|^{2}\right\rbrack \, d{\bf r},
\end{eqnarray}
and the  mean field Schr\"odinger  equation can be written
\begin{eqnarray}
\label{vf2}
i\hbar \frac{\partial \psi}{\partial t}=\frac{\delta E_{tot}}{\delta\psi^*}.
\end{eqnarray}

It is easy to show that the energy functional (\ref{te1}) and the total mass $M=\int\rho\, d{\bf r}$ are conserved by the quantum Euler equations. This implies that a minimum of $E_{tot}$ at fixed mass $M$ is a nonlinearly dynamically stable steady state of the quantum Euler equations \cite{holm}. Writing the first variations as $\delta E_{tot}-\alpha\delta M=0$, where $\alpha$ is a Lagrange multiplier, we get ${\bf u}={\bf 0}$ and
\begin{equation}
\label{te6}
m\Phi+Q=m\alpha.
\end{equation}
Taking the gradient of this relation, we obtain equation (\ref{tis5}) which characterizes a steady state of the quantum Euler equations. Equation (\ref{te6}) also  coincides with the time independent Schr\"odinger equation (\ref{tis4}) provided that the Lagrange multiplier $\alpha$ and the energy $E$ are related to each other by $E=m\alpha$. Therefore, the energy $E$ arising in the time-independent Schr\"odinger equation can be interpreted as a chemical potential. At equilibrium (${\bf u}={\bf 0}$, $\Theta_c=0$), the total energy reduces to
\begin{equation}
\label{te9}
E_{tot}=\Theta_Q+W.
\end{equation}
Multiplying equation (\ref{tis4}) by $\rho$ and integrating over the entire domain, we obtain
\begin{equation}
\label{te10}
2W+\Theta_Q=NE.
\end{equation}
Therefore,
\begin{equation}
\label{te11}
E_{tot}=NE-W.
\end{equation}
From this equation we note that $E_{tot}\neq NE$ when $W\neq 0$. In the classical limit $\hbar\rightarrow 0$, $\Theta_Q=0$, so that
\begin{equation}
\label{te12}
E_{tot}=W=\frac{1}{2}NE.\quad ({\rm classical})
\end{equation}

\subsection{The cosine interaction}
\label{sec_cos}

For the bosonic HMF model at $T=0$, corresponding to a cosine interaction, the time independent Schr\"odinger equation (\ref{tis2}) takes the form
\begin{eqnarray}
\label{cos1}
-\frac{\hbar^2}{2}\psi''(\theta)+\Phi(\theta)\psi(\theta)=E\psi(\theta),
\end{eqnarray}
where
\begin{eqnarray}
\label{cos2}
\Phi(\theta)=-\frac{k}{2\pi}\int_0^{2\pi} \rho(\theta') \cos(\theta-\theta')\, d\theta',
\end{eqnarray}
\begin{eqnarray}
\label{cos3}
\rho(\theta)=N\psi^2(\theta),
\end{eqnarray}
\begin{eqnarray}
\label{cos4}
\int_0^{2\pi} \psi^2(\theta)\, d\theta=1.
\end{eqnarray}
We can suppose without loss of generality that $\psi(\theta)$ and $\rho(\theta)$ are symmetric with respect to $\theta=0$. In that case, $\Phi(\theta)$ is given by
\begin{eqnarray}
\label{cos5}
\Phi(\theta)=-B\cos\theta,
\end{eqnarray}
with
\begin{eqnarray}
\label{cos6}
B=\frac{k}{2\pi}\int_0^{2\pi} \rho(\theta) \cos\theta\, d\theta.
\end{eqnarray}
Therefore, the time independent Schr\"odinger equation with a cosine potential can be rewritten
\begin{eqnarray}
\label{cos7}
-\frac{\hbar^2}{2}\psi''-\frac{kM}{2\pi}b\cos\theta\, \psi=E\psi,
\end{eqnarray}
\begin{eqnarray}
\label{cos8}
b=\int_0^{2\pi} \psi^2 \cos\theta\, d\theta,
\end{eqnarray}
\begin{eqnarray}
\label{cos9}
\int_0^{2\pi} \psi^2\, d\theta=1,
\end{eqnarray}
\begin{eqnarray}
\label{cos10}
\rho(\theta)=N\psi^2(\theta),
\end{eqnarray}
where $b=2\pi B/kM$ is the magnetization. The boundary conditions are
\begin{eqnarray}
\label{cos11}
\psi'(0)=\psi'(\pi)=0.
\end{eqnarray}
This system of equations defines an eigenvalue problem for the wave function $\psi(\theta)$ where the eigenvalue $E$ is the energy. In the following, we shall be interested by the fundamental eigenmode corresponding to the smallest value of $E$. For this mode, the wave function $\psi(\theta)$ has no node so that the density profile decreases monotonically with the distance.

\begin{figure}[!h]
\begin{center}
\includegraphics[clip,scale=0.3]{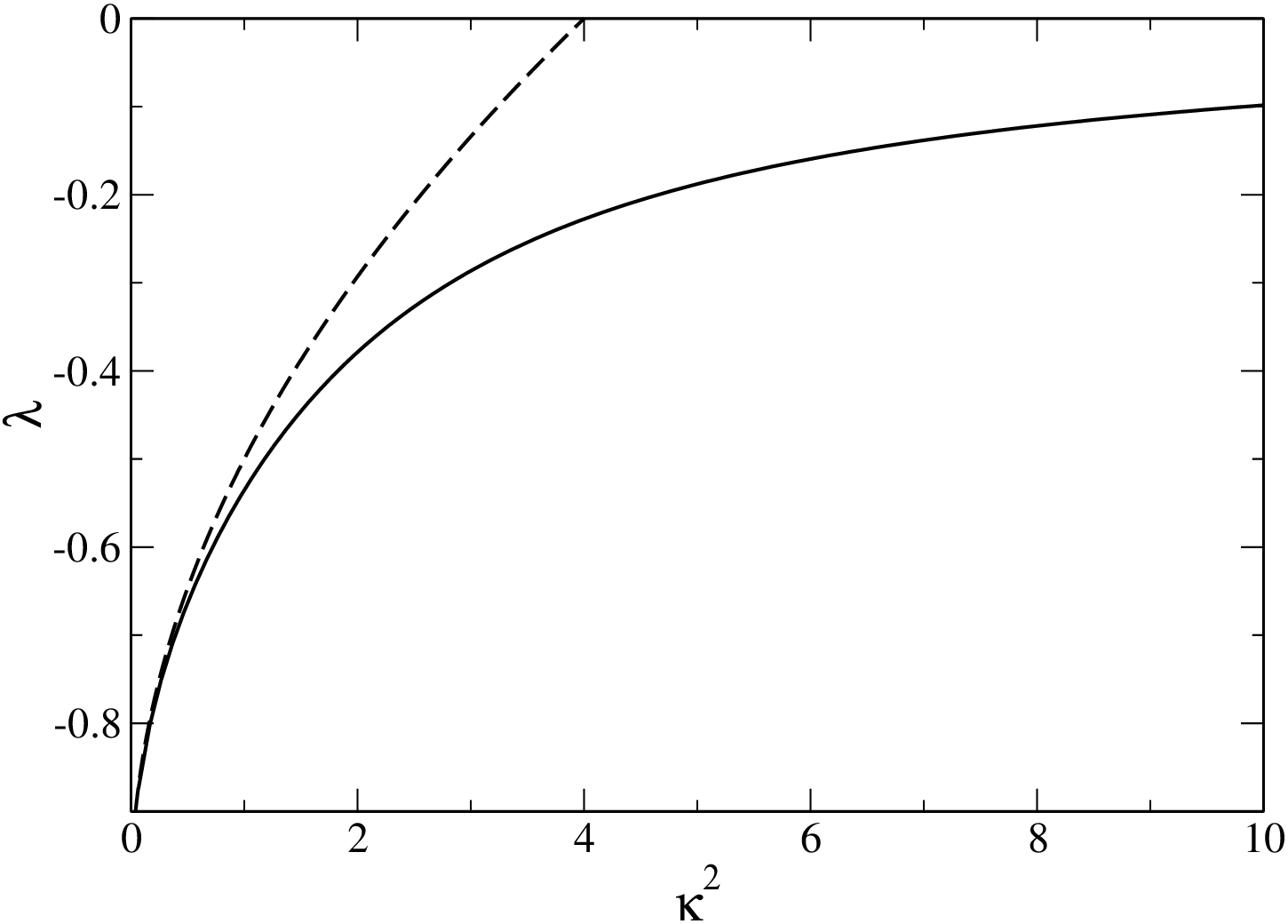}
\caption{Eigenvalue $\lambda$ as a function of $\kappa^2$. The dashed line starting at $\kappa^2=0$ corresponds to the harmonic approximation of Appendix \ref{sec_class}.}
\label{lam-eps}
\end{center}
\end{figure}

\begin{figure}[!h]
\begin{center}
\includegraphics[clip,scale=0.3]{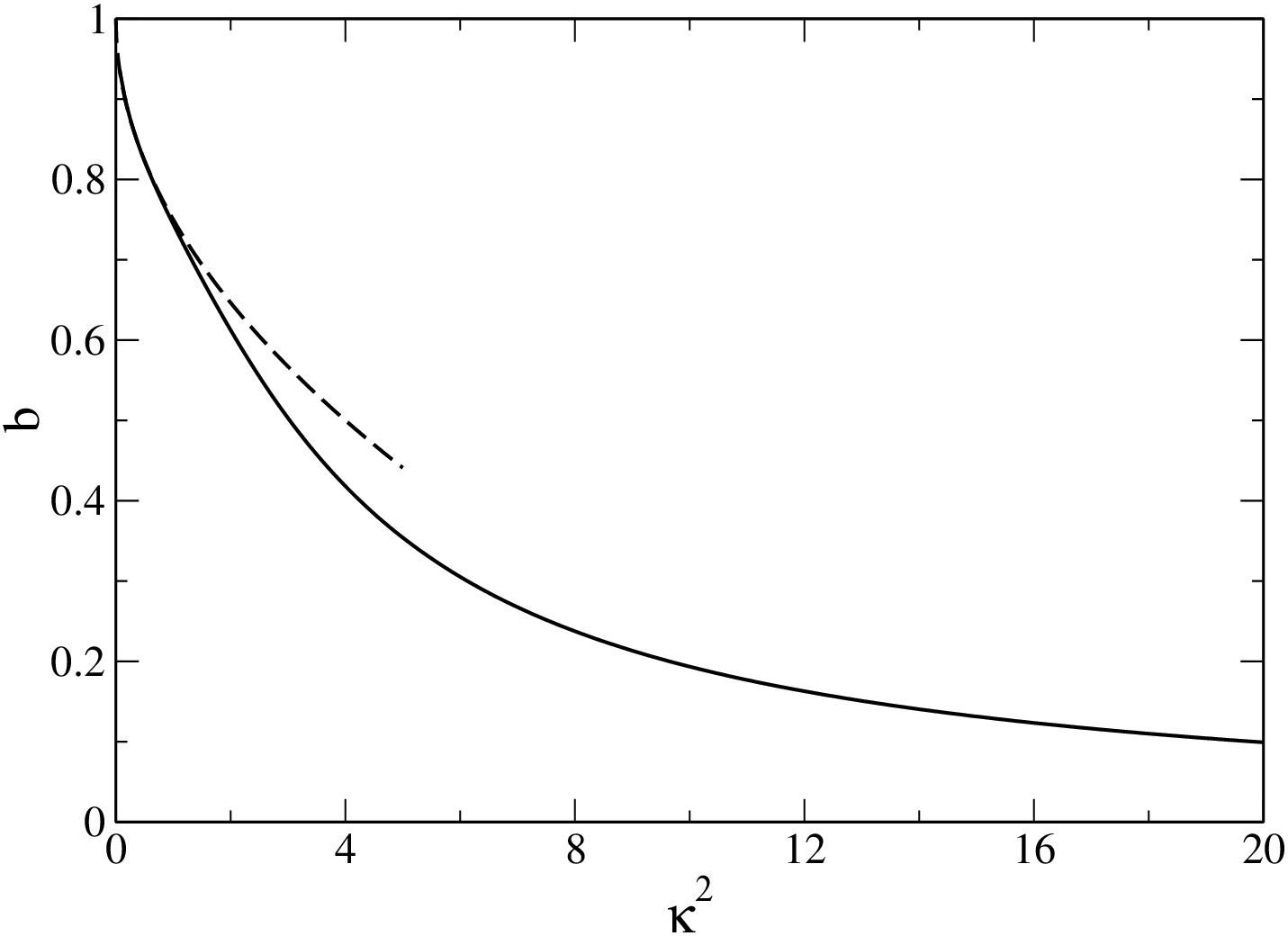}
\caption{Magnetization $b$ as a function of $\kappa^2$.}
\label{b-eps}
\end{center}
\end{figure}

\begin{figure}[!h]
\begin{center}
\includegraphics[clip,scale=0.3]{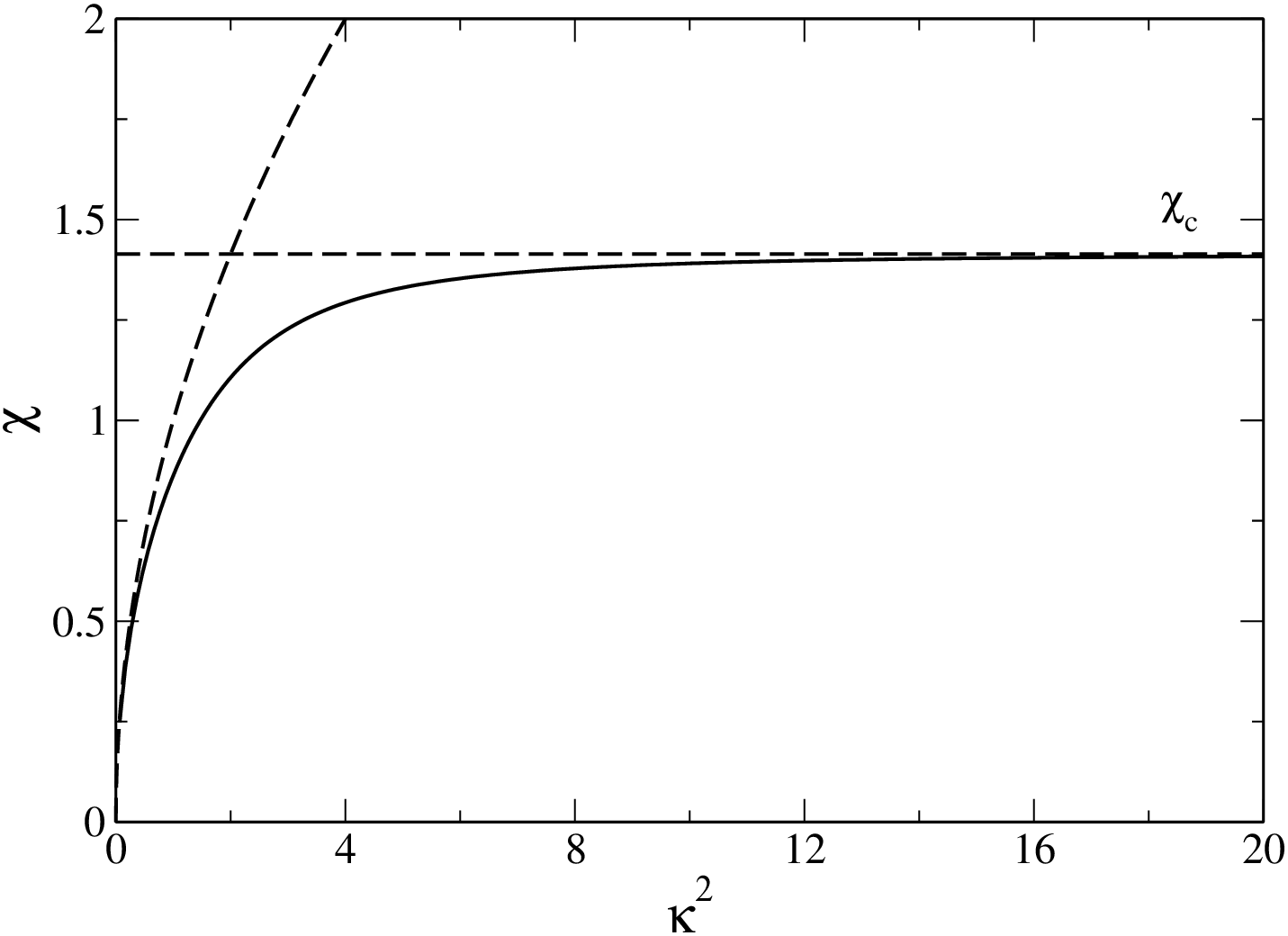}
\caption{Normalized Planck constant $\chi$ as a function of $\kappa^2$. The normalized Planck constant tends to the critical value $\chi_c=\sqrt{2}$ when $\kappa\rightarrow \infty$.}
\label{h-eps}
\end{center}
\end{figure}

\begin{figure}[!h]
\begin{center}
\includegraphics[clip,scale=0.3]{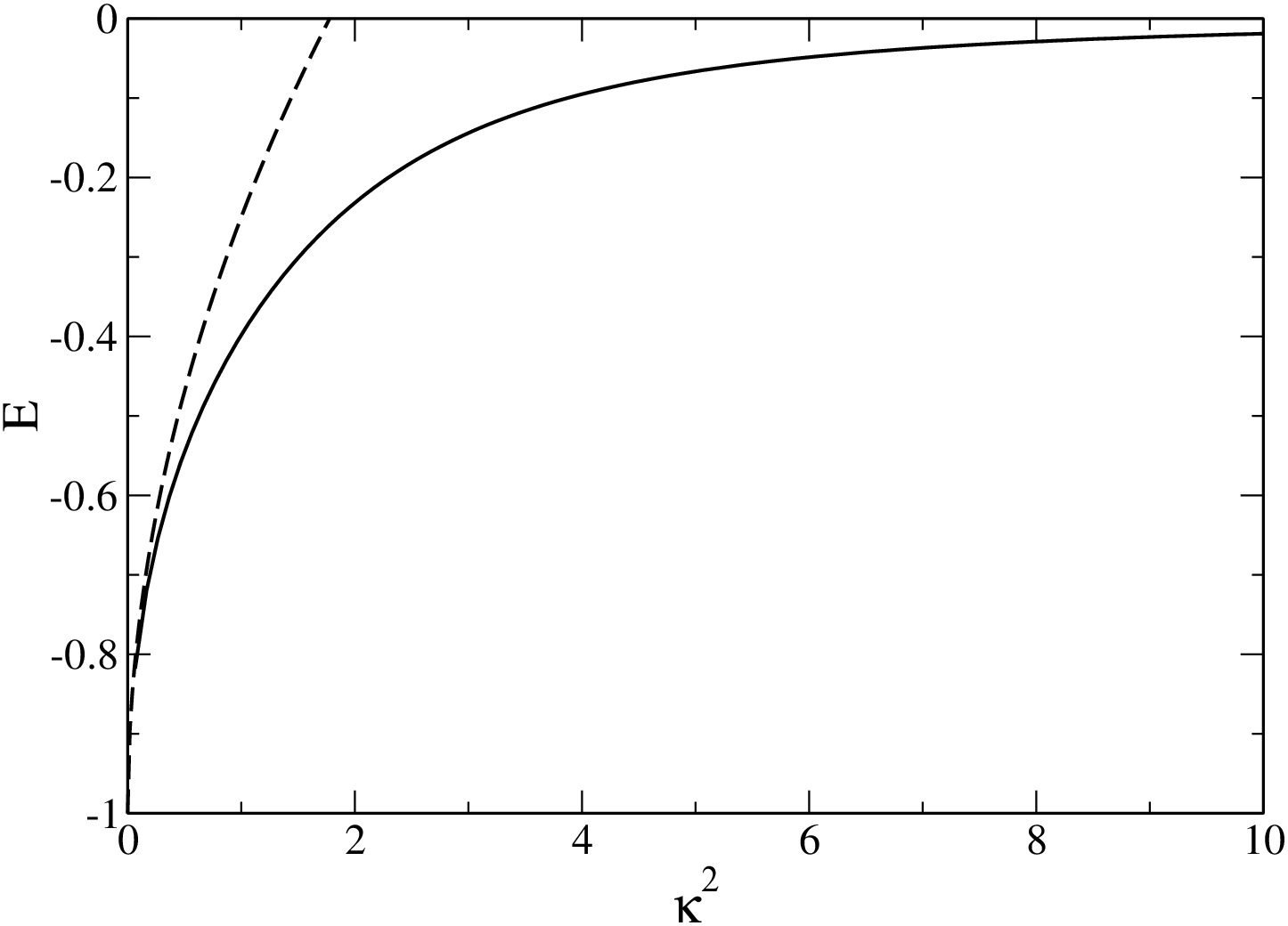}
\caption{Normalized energy ${\cal E}$ as a function of $\kappa^2$.}
\label{ene-eps}
\end{center}
\end{figure}

We note that $\psi(\theta)={1}/{\sqrt{2\pi}}$ is a particular solution of these equations. It corresponds to a spatially homogeneous distribution $\rho(\theta)=M/(2\pi)$ with $b=E=0$. Its stability is investigated in Appendix \ref{sec_sh}. Introducing the normalized Planck constant for bosons
\begin{eqnarray}
\label{planckb}
\chi\equiv \hbar \left (\frac{2\pi}{kM}\right )^{1/2},
\end{eqnarray}
we find that the homogeneous phase is stable if
\begin{eqnarray}
\label{planckb2}
\chi\ge \chi_c=\sqrt{2},
\end{eqnarray}
and unstable otherwise. We stress that the normalized Planck constant is different for fermions and bosons (see Appendix \ref{sec_tlfb}).

Inhomogeneous solutions with $b\neq 0$ must be obtained numerically.
Dividing equation (\ref{cos7}) by $kMb/(2\pi)$ and setting $\kappa^2=2\pi\hbar^2/(kMb)$ and $\lambda=2\pi E/(kMb)$, we obtain
\begin{eqnarray}
\label{cos12}
-\frac{\kappa^2}{2}\psi''-\cos\theta\psi=\lambda\psi,
\end{eqnarray}
\begin{eqnarray}
\label{cos13}
\int_0^{2\pi} \psi^2\, d\theta=1.
\end{eqnarray}
For given $\kappa$, this is just the ordinary Schr\"odinger equation for an anharmonic oscillator.  We can therefore determine the fundamental eigenvalue $\lambda=\lambda(\kappa)$ corresponding to the wavefunction that has no node. Then, we can obtain the magnetization from the equation
\begin{eqnarray}
\label{cos14}
b=\int_0^{2\pi} \psi^2 \cos\theta\, d\theta=b(\kappa).
\end{eqnarray}
Finally, we get
\begin{eqnarray}
\label{cos15}
\chi\equiv \hbar \left (\frac{2\pi}{kM}\right )^{1/2}=\kappa \sqrt{b(\kappa)},
\end{eqnarray}
\begin{eqnarray}
\label{cos16}
{\cal E}\equiv \frac{2\pi E}{kM}=\lambda(\kappa)b(\kappa).
\end{eqnarray}
These equations  determines the normalized energy ${\cal E}={\cal E}(\chi)$ and the magnetization $b=b(\chi)$ as a function of the normalized Planck constant $\chi$ in parametric form (with parameter $\kappa$). In fact, we can simplify the problem even further. Indeed, if we define $\phi(\theta)=\psi(\theta)/\psi(0)$, we obtain
\begin{eqnarray}
\label{cos17}
-\frac{\kappa^2}{2}\phi''-\cos\theta\phi=\lambda\phi,
\end{eqnarray}
\begin{eqnarray}
\label{cos18}
\phi(0)=1,\qquad \phi'(0)=\phi'(\pi)=0.
\end{eqnarray}
This is a simple shooting problem. For given $\kappa$, solving equation (\ref{cos17}) with the initial condition $\phi(0)=1$ and $\phi'(0)=0$, we have to find the eigenvalue $\lambda(\kappa)$ so as to satisfy the boundary condition $\phi'(\pi)=0$. Then, $\psi(0)$ is given by the normalization condition (\ref{cos13}) leading to
\begin{eqnarray}
\label{cos19}
\psi(0)=\frac{1}{\sqrt{\int_0^{2\pi} \phi^2\, d\theta}}.
\end{eqnarray}
This completely determines the wave function $\psi(\theta)=\psi(0)\phi(\theta)$ and the density profile $\rho(\theta)/N=\psi(\theta)^2$. Finally, the magnetization is given by
\begin{eqnarray}
\label{cos14b}
b=\frac{\int_0^{2\pi} \phi^2 \cos\theta\, d\theta}{\int_0^{2\pi} \phi^2 \, d\theta}=b(\kappa),
\end{eqnarray}
while the normalized Planck constant and the normalized energy are given by equations (\ref{cos15}) and (\ref{cos16}).

We recall that the energy $E$ appearing in the time independent Schr\"odinger equation (\ref{cos1})  is generally {\it different} from the total energy per particle $E_{tot}/N$ (see Section \ref{sec_te}). The total energy is given by equation (\ref{te11}). For the HMF model, the potential energy can be expressed in terms of the magnetization by equation (\ref{fa7}). Therefore, the total energy is
\begin{equation}
\label{cos21}
E_{tot}=NE+\frac{\pi B^2}{k}.
\end{equation}
Introducing the normalized energy
\begin{equation}
\label{cos22}
\epsilon_{tot}=\frac{8\pi E_{tot}}{kM^2},
\end{equation}
we get
\begin{equation}
\label{cos23}
\epsilon_{tot}=4{\cal E}+2b^2.
\end{equation}
In the homogeneous phase where $b={\cal E}=0$, we have $\epsilon_{tot}=0$ for any value of $\chi$. This differs from the case of fermions where the energy of the homogeneous phase is given by equation (\ref{cph4}).

\begin{figure}[!h]
\begin{center}
\includegraphics[clip,scale=0.3]{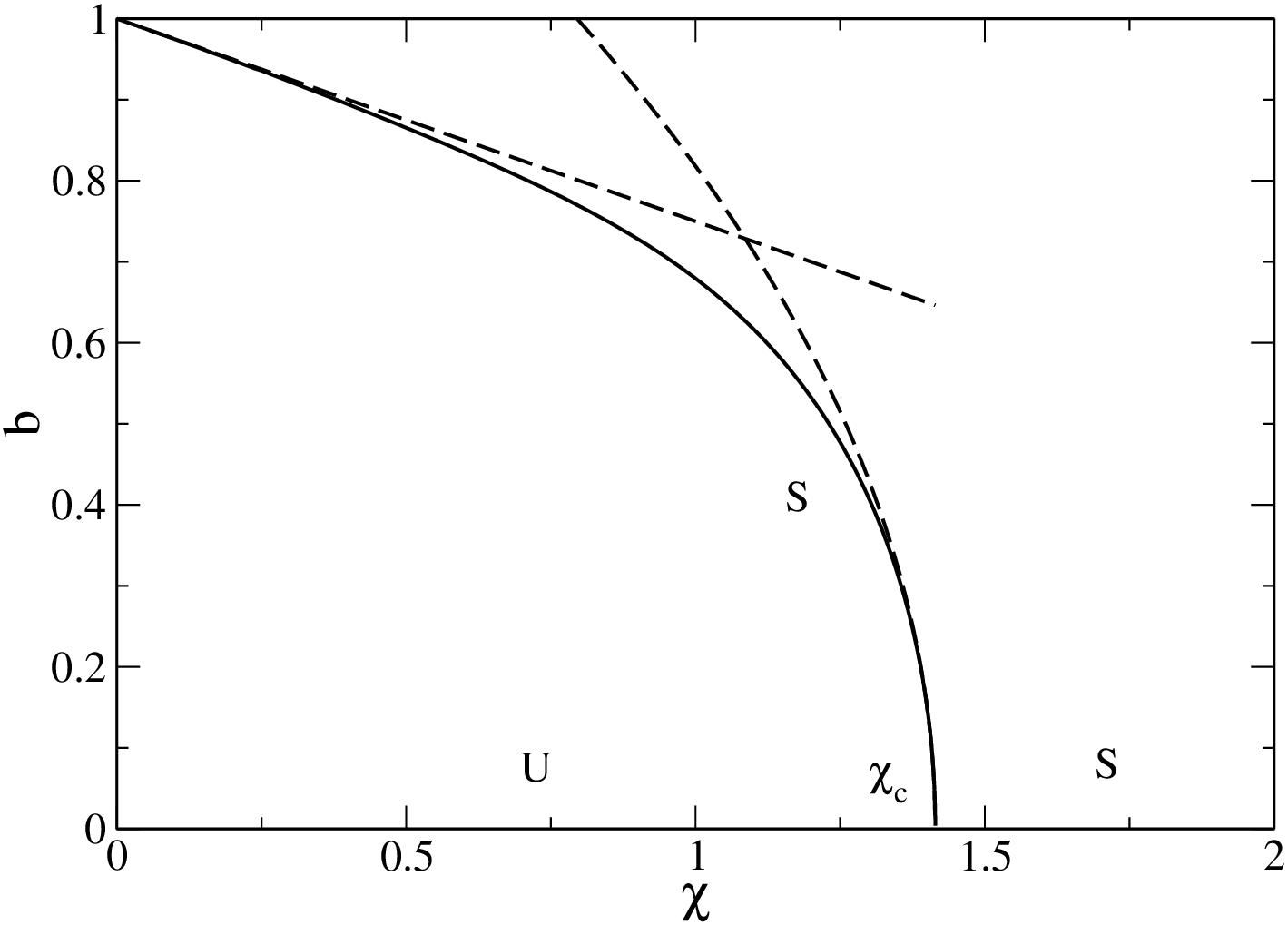}
\caption{Magnetization $b$ as a function of the normalized Planck constant $\chi$. The dashed line starting at $\chi=0$ corresponds to the harmonic approximation of Appendix \ref{sec_class}. The dashed line starting at $\chi=\chi_c=\sqrt{2}$ corresponds to the analytical expression of the magnetization close to the critical point (see Appendix \ref{sec_be}). The magnetized solution exists only for $\chi<\chi_c$ and it is stable. The homogeneous phase ($b=0$) is unstable for $\chi<\chi_c$ and stable for $\chi>\chi_c$ (see Appendix \ref{sec_sh}). This curve displays a second order phase transition between the classical regime (inhomogeneous state $b\neq 0$) and the quantum regime (homogeneous state $b=0$). It can be compared with the magnetization curve of fermions reported in Figure \ref{bVShN0.5}.}
\label{b-h}
\end{center}
\end{figure}

\begin{figure}[!h]
\begin{center}
\includegraphics[clip,scale=0.3]{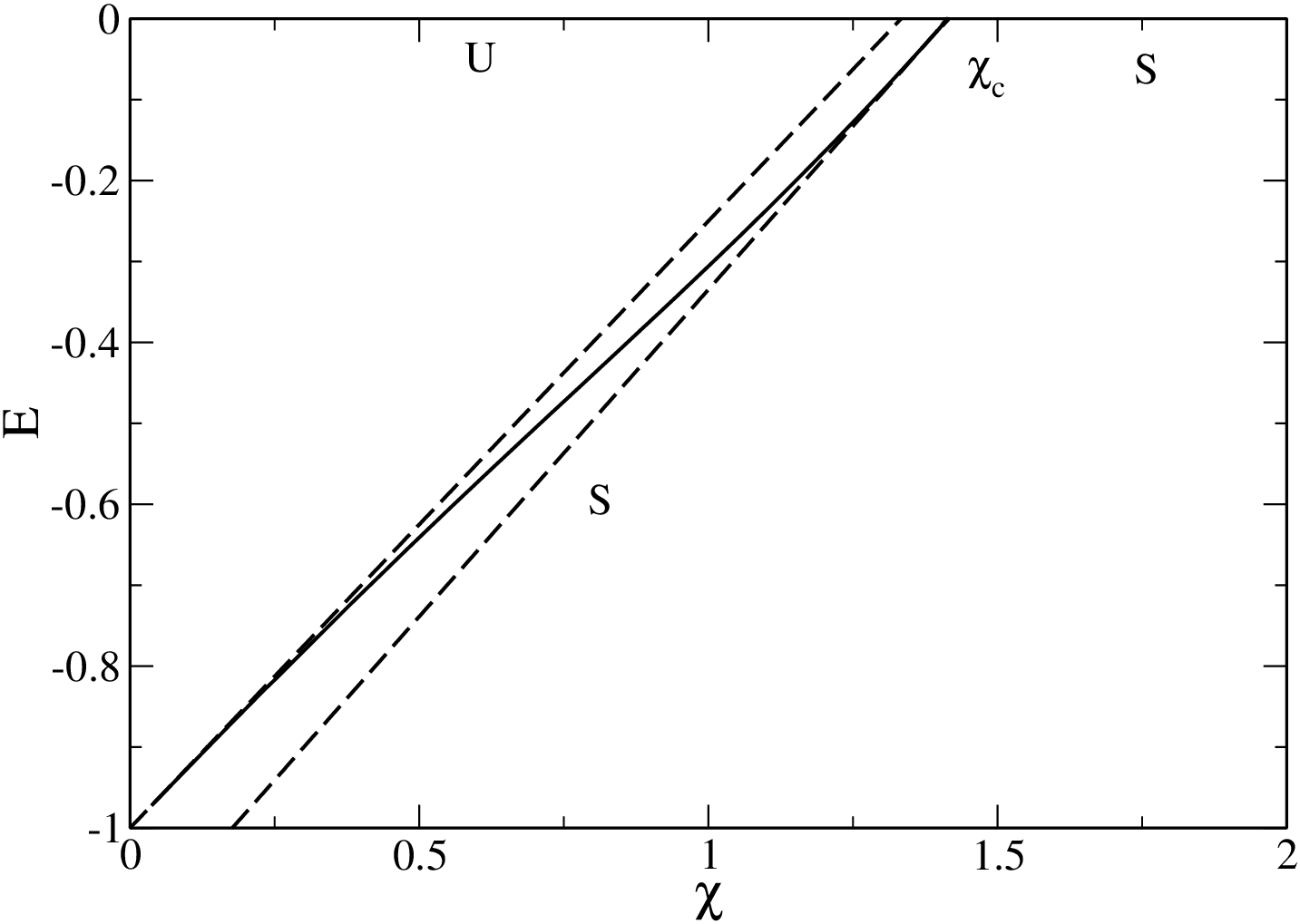}
\caption{Normalized energy ${\cal E}$ as a function of the normalized Planck constant $\chi$.}
\label{ene-h}
\end{center}
\end{figure}

\begin{figure}[!h]
\begin{center}
\includegraphics[clip,scale=0.3]{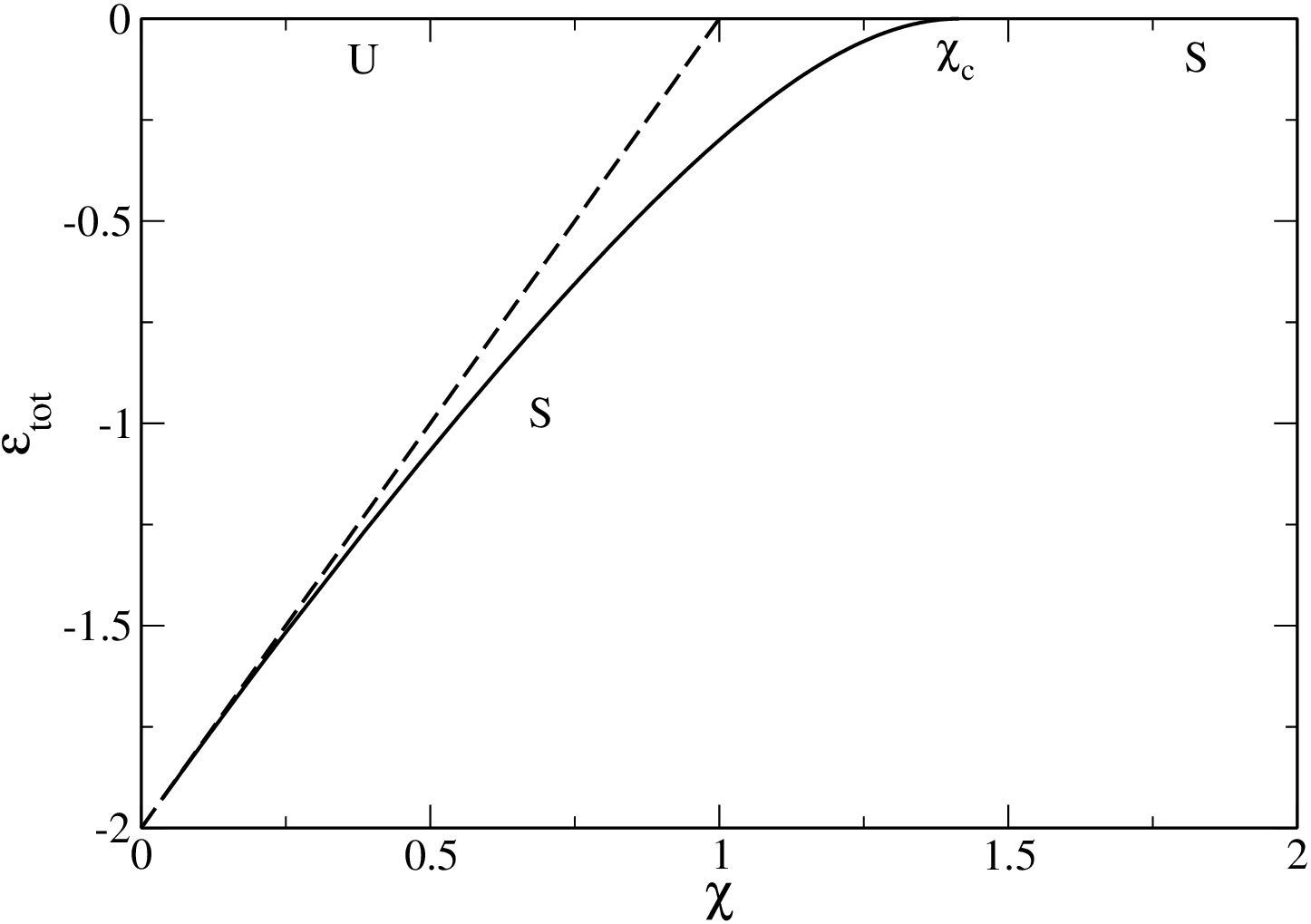}
\caption{Total normalized energy $\epsilon_{tot}$ as a function of the normalized Planck constant $\chi$. It can be compared with the energy curve of fermions reported in Figure \ref{epsilonVShN0.5GLOBALE}.}
\label{W-h}
\end{center}
\end{figure}

\begin{figure}[!h]
\begin{center}
\includegraphics[clip,scale=0.3]{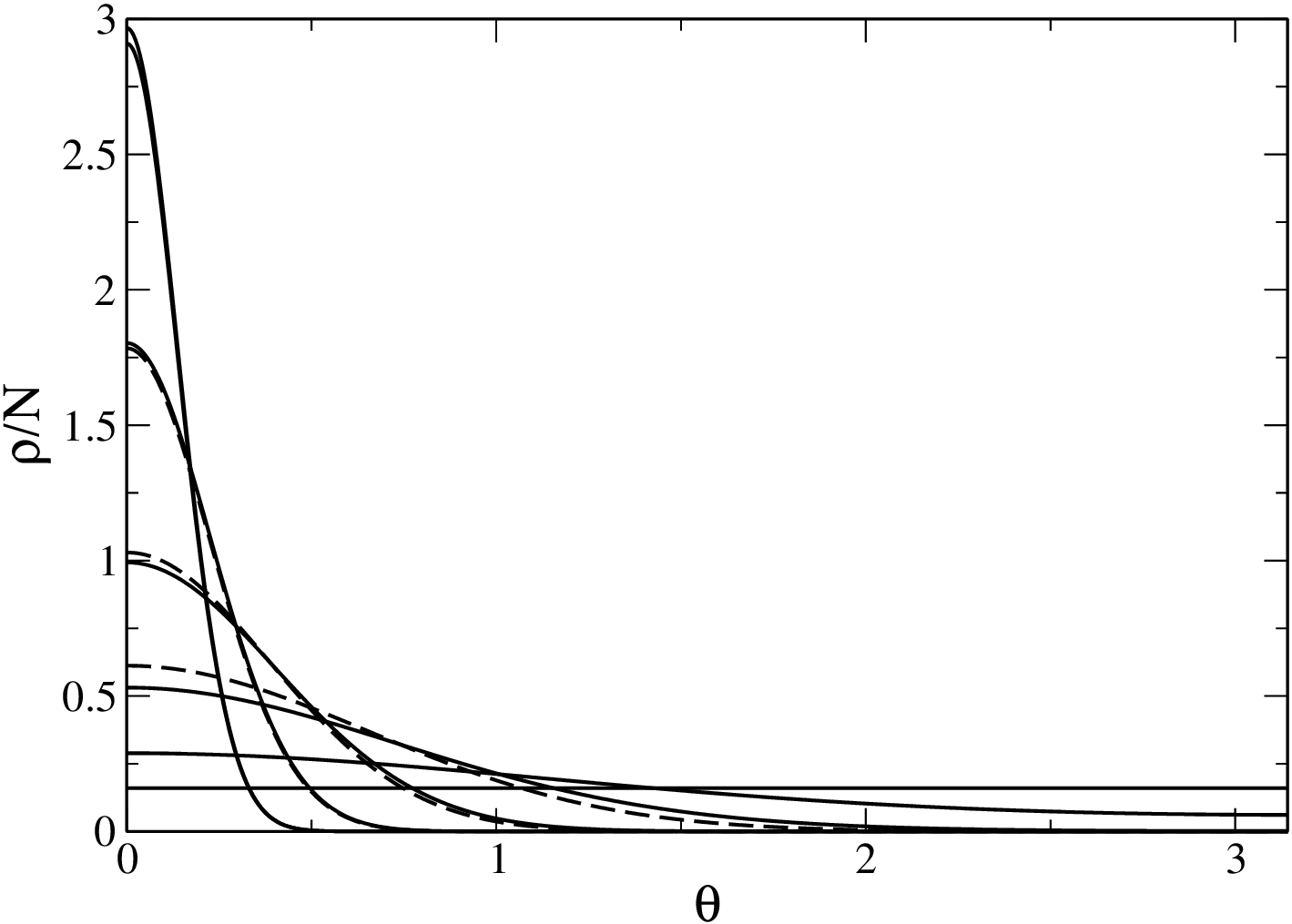}
\caption{Density profiles for $\chi=0.0376, 0.1, 0.3, 0.85, 1.33, \sqrt{2}$ (from top to bottom). The dashed lines correspond to the Gaussian approximation of Appendix \ref{sec_class}. These profiles can be compared with the density profiles of fermions reported in Figure \ref{profiles}.}
\label{ro-hNEW}
\end{center}
\end{figure}

The curves $\lambda(\kappa)$, $b(\kappa)$, $\chi(\kappa)$  and ${\cal E}(\kappa)$ are plotted in Figures \ref{lam-eps}, \ref{b-eps},
\ref{h-eps} and \ref{ene-eps}. The curves $b(\chi)$, ${\cal E}(\chi)$ and $\epsilon_{tot}(\chi)$ giving the magnetization and the normalized energies as a function of the normalized Planck constant are plotted in Figures \ref{b-h}, \ref{ene-h} and \ref{W-h}. Finally, some density profiles are shown in Figure \ref{ro-hNEW}. The homogeneous phase exists for any $\chi$. It is stable for $\chi>\chi_c=\sqrt{2}$ and unstable for $\chi<\chi_c$ (see Appendix \ref{sec_sh}). The inhomogeneous phase exists only for $\chi<\chi_c=\sqrt{2}$ where it bifurcates from the homogeneous phase. The behaviors of the magnetization and of the energy close to the bifurcation point are obtained analytically in Appendix \ref{sec_be} and compared with the numerical results in Figures \ref{b-h} and \ref{ene-h}. Analytical results are also obtained in the semi-classical limit $\chi\rightarrow 0$ by making a harmonic approximation (see Appendix \ref{sec_class}). These analytical results are compared with the numerical ones in Figures \ref{b-eps}-\ref{ro-hNEW}. According to the Poincar\'e theorem, we deduce that the inhomogeneous phase becomes stable at the point  $\chi=\chi_c=\sqrt{2}$ at which the homogeneous branch becomes unstable. Since the inhomogeneous branch is not multivalued, there is no metastable state contrary to the case of fermions. The system  displays a second order phase transition at $\chi=\chi_t$ marked by the discontinuity of the derivative of the magnetization.

Summarizing, we have determined the ground state ($T=0$) of the Bose gas with cosine interaction, forming a BEC, and investigated its thermodynamical stability. In the classical limit $\chi=0$, the ground state energy is $\epsilon_{tot}=-2$ and the magnetization is $b=1$. This corresponds to a classical inhomogeneous gas at $T=0$ whose density profile is a Dirac peak $\rho=M\delta(\theta)$. The homogeneous phase with energy $\epsilon_{tot}=0$ and magnetization $b=0$ is unstable since $T=0<T_c$. This returns the classical results \cite{cvb}. However, when quantum mechanics is taken into account, there exists a critical value of the normalized Planck constant $\chi_c=\sqrt{2}$ above which the homogeneous phase becomes stable while the inhomogeneous phase  disappears. Therefore, in the quantum regime, the homogeneous phase is stabilized against clustering by the Heisenberg uncertainty principle. This is similar to the stabilization of boson stars against gravitational collapse in astrophysics due to quantum mechanics. In the HMF model, this stabilization takes place through a {\it second order} phase transition.

\section{The mean field Gross-Pitaevskii equation}
\label{sec_mfgp}

We shall now consider a generalization of the equations of Section \ref{sec_bosons} by introducing a nonlinearity in the Schr\"odinger equation in addition to the long-range interaction. This is the so-called mean field Gross-Pitaevskii (GP) equation. As we shall see, this equation can have application for fermions and bosons. For the sake of generality, we consider an arbitrary potential of interaction in $d$ dimensions.

\subsection{The Madelung transformation}
\label{sec_g}

The mean field Gross-Pitaevskii equation can be written
\begin{equation}
\label{g1}
i\hbar \frac{\partial\psi}{\partial t}=-\frac{\hbar^2}{2m}\Delta\psi+m(\Phi+h(\rho))\psi,
\end{equation}
\begin{eqnarray}
\label{g2}
\Phi({\bf r},t)=\int \rho({\bf r}',t) u(|{\bf r}-{\bf r}'|)\, d{\bf r}',
\end{eqnarray}
\begin{eqnarray}
\label{g3}
\rho({\bf r},t)=Nm|\psi({\bf r},t)|^2,
\end{eqnarray}
\begin{eqnarray}
\label{g4}
\int |\psi({\bf r},t)|^2\, d{\bf r}=1,
\end{eqnarray}
where $h(\rho)$ is a potential depending on the density and the other quantities have been defined previously.  Adapting the Madelung transformation of Section \ref{sec_mad} to the present situation, we find that the Gross-Pitaevskii equation is equivalent to the hydrodynamic equations
\begin{equation}
\label{g5}
\frac{\partial\rho}{\partial t}+\nabla\cdot (\rho {\bf u})=0,
\end{equation}
\begin{equation}
\label{g6}
\frac{\partial {\bf u}}{\partial t}+({\bf u}\cdot \nabla){\bf u}=-\frac{1}{\rho}\nabla p-\nabla\Phi-\frac{1}{m}\nabla Q,
\end{equation}
where $Q$ is the quantum potential  (\ref{mad8}) and $p({\bf r},t)$ is a pressure that is a function of the density: $p({\bf r},t)=p\lbrack \rho({\bf r},t)\rbrack$. In this sense, the fluid is barotropic. The equation of state $p(\rho)$ is determined by the potential $h(\rho)$ through the relation
\begin{equation}
\label{g7}
h'(\rho)=\frac{p'(\rho)}{\rho},
\end{equation}
yielding $p(\rho)=\rho h(\rho)-H(\rho)$ where $H$ is a primitive of $h$. Using this relation, the  Euler equation (\ref{g6})  can be rewritten as
\begin{equation}
\label{g8}
\frac{\partial {\bf u}}{\partial t}+({\bf u}\cdot \nabla){\bf u}=-\nabla h-\nabla\Phi-\frac{1}{m}\nabla Q,
\end{equation}
which shows that the potential $h$ appearing in the GP equation can be interpreted as an enthalpy in the hydrodynamic equations. Equations (\ref{g5}), (\ref{g6}) and (\ref{g2})  will be called the quantum barotropic Euler equations. They are equivalent to the mean field GP equation.

\subsection{Interpretations of the potential}
\label{sec_int}

We can give two independent interpretations of the potential $h(\rho)$.

(i) Let us consider a BEC described by the mean field Schr\"odinger equation (\ref{mfs1})-(\ref{mfs4}) and let us assume that the potential of interaction can be written as $u=u_{LR}+u_{SR}$ where $u_{LR}$ refers to the long-range interaction and $u_{SR}$ to the short-range interaction. We assume furthermore that the short-range interaction corresponds to binary collisions that can be modeled by the effective potential $u_{SR}({\bf r}-{\bf r}')=g\delta({\bf r}-{\bf r}')$, where the coupling constant (or pseudopotential)  $g$ is related to the scattering length $a$ by $g=4\pi a\hbar^2/m^3$ (in $d=3$) \cite{revuebec}. When this form of potential is substituted in equation (\ref{mfs2}), we obtain the usual mean field  Gross-Pitaevskii equation
\begin{equation}
\label{gnez}
i\hbar \frac{\partial\psi}{\partial t}=-\frac{\hbar^2}{2m}\Delta\psi+m\Phi\psi+ N\frac{4\pi a\hbar^2}{m}|\psi|^{2}\psi.
\end{equation}
with a potential $h=g\rho=gNm|\psi|^2$. The associated equation of state is $p=\frac{1}{2}g\rho^2$ corresponding to a polytrope of index $n=1$ and polytropic constant $K=g/2$.

(ii) The mean field Gross-Pitaevskii equation (\ref{g1})-(\ref{g4}) also describes a gas of fermions\footnote{Such a description assumes that the fermions have the same probability distribution \cite{manfredi}. More fundamentally, the fermions should be described by a mixture of $N$ pure states, each with a wavefunction $\psi_i$ obeying the mean field Schr\"odinger equation without nonlinearity (except the one due to the interaction) \cite{haas}.} when one takes into account the quantum potential $Q$ which is a manifestation of the Heisenberg uncertainty principle \cite{manfredi}. In the case of fermions, we must also take into account the quantum pressure arising from the Pauli exclusion principle. It can be calculated from the Fermi-Dirac distribution function at $T=0$ (see Section \ref{sec_f}). In $d$ dimensions, it is given by $p=K\rho^{1+2/d}$ where $K=\frac{1}{d+2}(\frac{d}{2S_d})^{2/d}\frac{(2\pi \hbar)^2}{m^{2+2/d}}$ \cite{wdD}. This is the equation of state of a polytrope of index $n=d/2$ and polytropic constant $K$. This pressure term is the one that appears in the hydrodynamic equation (\ref{g6}). Using equation (\ref{g7}), it corresponds to an effective  potential of the form $h(\rho)=(d/2+1)K\rho^{2/d}$. The corresponding  Gross-Pitaevskii equation can be written
\begin{equation}
\label{gnaw}
i\hbar \frac{\partial\psi}{\partial t}=-\frac{\hbar^2}{2m}\Delta\psi+m\Phi\psi+\kappa_d N^{2/d}\frac{\hbar^2}{m}|\psi|^{4/d}\psi,
\end{equation}
where $\kappa_d=2\pi^2 (\frac{d}{2S_d})^{2/d}$ is a constant.

{\it Remark:} we note that the potential $h(\rho)\propto \rho^{2/d}$ associated to fermions becomes equivalent to the potential $h(\rho)\propto \rho$ associated to self-coupled bosons when $d=2$. In fact, the dimension $d=2$ is a critical dimension \cite{kolomeisky}. When we consider a gas of repulsive (impenetrable) bosons, the potential $h(\rho)=g\rho$ arising in the GP equation ceases to be valid for $d\le 2$ (in $d=2$ it remains marginally valid with logarithmic corrections). In particular, in $d=1$, it is replaced by $h(\rho)=\pi^2\hbar^2\rho^2/2m^4$ exactly like for spinless ($s=0$) fermions. This is a manifestation of the boson-fermion duality in one dimension \cite{girardeau}.

\subsection{The time independent Gross-Pitaevskii equation}
\label{sec_tigp}

If we consider a wavefunction of the form
\begin{equation}
\label{tigp1}
\psi({\bf r},t)=A({\bf r})e^{-i\frac{Et}{\hbar}},
\end{equation}
we obtain the  time independent GP equation
\begin{eqnarray}
\label{tigp2}
-\frac{\hbar^2}{2m}\Delta\psi({\bf r})+m(\Phi({\bf r})+h(\rho))\psi({\bf r})=E\psi({\bf r}),
\end{eqnarray}
where $\psi({\bf r})\equiv A({\bf r})$ is real and $\rho({\bf r})=Nm\psi^2({\bf r})$. Dividing equation (\ref{tigp2}) by $\psi({\bf r})$, we get
\begin{equation}
\label{tigp3}
m\Phi+mh(\rho)-\frac{\hbar^2}{2m}\frac{\Delta\sqrt{\rho}}{\sqrt{\rho}}=E.
\end{equation}
This equation can also be obtained from the quantum barotropic Euler equations (\ref{g5}), (\ref{g6}) and (\ref{g2}) since they are equivalent to the GP equation. The steady states of the Euler equation (\ref{g6}) obtained by taking $\partial_t=0$ and ${\bf u}={\bf 0}$ satisfy
\begin{equation}
\label{tigp4}
\nabla p+\rho\nabla\Phi-\frac{\hbar^2\rho}{2m^2}\nabla \left (\frac{\Delta\sqrt{\rho}}{\sqrt{\rho}}\right )={\bf 0}.
\end{equation}
This is similar to the condition of hydrostatic equilibrium with an additional quantum potential (as we have previously indicated, it can be written in the form of an anisotropic pressure). This equation is equivalent to equation (\ref{tigp3}). Indeed, integrating equation (\ref{tigp4}) and using equation (\ref{g7}), we obtain equation (\ref{tigp3}) where the energy  $E$ appears as a constant of integration.

The equations of the problem can be simplified in two important limits:

(i) {\it The non interacting limit:} the case of a BEC without short-range interactions corresponds to $h=p=0$ leading to
\begin{equation}
\label{tigp5}
\rho\nabla\Phi-\frac{\hbar^2\rho}{2m^2}\nabla \left (\frac{\Delta\sqrt{\rho}}{\sqrt{\rho}}\right )={\bf 0}.
\end{equation}
This is the situation considered in Section \ref{sec_bosons}. In that case, the equilibrium state results from a balance between the long-range potential and the quantum pressure arising from the Heisenberg principle.

(ii) {\it The Thomas-Fermi limit:} if we neglect the quantum potential in equation (\ref{tigp4}), we obtain
\begin{equation}
\label{tigp6}
\nabla p+\rho\nabla\Phi={\bf 0}.
\end{equation}
This corresponds to the standard condition of hydrostatic equilibrium describing a balance between the long-range potential and the pressure due (i) to the short-range interaction in a BEC (ii) to the quantum pressure resulting from the Pauli exclusion principle for fermions.

{\it Remark:} considering the generalized condition of hydrostatic equilibrium (\ref{tigp4}) taking into account the quantum potential, the Thomas-Fermi approximation is valid when the first term dominates over the third one. In the case of fermions, the pressure is due to the Pauli exclusion principle. Using the results of Section \ref{sec_int} and dimensional analysis, we easily see that the quantum potential (Heisenberg uncertainty principle) can be neglected in front of the Fermi pressure (Pauli exclusion principle) when $N\gg 1$. Therefore, the Thomas-Fermi approximation is {\it exact} for fermions in the thermodynamic limit $N\rightarrow +\infty$. This is precisely the situation considered in Section \ref{sec_fermions}.

\subsection{The total energy}
\label{sec_ef}

The energy functional associated with the  quantum barotropic Euler equations  (\ref{g5}), (\ref{g6}) and (\ref{g2}) is
\begin{equation}
\label{ef1}
E_{tot}=\Theta_c+\Theta_Q+U+W,
\end{equation}
where
\begin{eqnarray}
\label{ef2}
U&=&\int\rho\int^{\rho}\frac{p(\rho_1)}{\rho_1^2}\, d\rho_1\, d{\bf r}\nonumber\\
&=&\int \left\lbrack \rho h(\rho)-p(\rho)\right \rbrack\, d{\bf r}=\int H(\rho)\, d{\bf r}.
\end{eqnarray}
is the internal energy and the other functionals have already been defined in Section \ref{sec_te}. For a polytropic equation of state $p=K\rho^{\gamma}$, corresponding to a potential $h(\rho)=\frac{K\gamma}{\gamma-1}\rho^{\gamma-1}$, the internal energy takes the form
\begin{equation}
\label{ef3}
U=\frac{K}{\gamma-1}\int \rho^{\gamma}\, d{\bf r}=\frac{1}{\gamma-1}\int p\, d{\bf r}.
\end{equation}

It is easy to check that the total energy  (\ref{ef1}) and the mass $M=\int \rho\, d{\bf r}$ are conserved by the  quantum barotropic Euler equations (hence by the GP equation). This implies that a
minimum of $E_{tot}$ at fixed mass $M$ is a nonlinearly dynamically stable steady state of the mean field quantum barotropic Euler equations \cite{holm}. Writing the first variations as $\delta E_{tot}-\alpha\delta M=0$, where $\alpha$ is a Lagrange multiplier, we get ${\bf u}={\bf 0}$ and
\begin{equation}
\label{ef4}
m\Phi+mh(\rho)-\frac{\hbar^2}{2m}\frac{\Delta\sqrt{\rho}}{\sqrt{\rho}}=m\alpha.
\end{equation}
Taking the gradient of this expression and using equation (\ref{g7}) we obtain the condition of hydrostatic equilibrium (\ref{tigp4}) determining a steady state of the  quantum barotropic Euler equations. This equation is also equivalent to the time independent GP equation (\ref{tigp3}). The Lagrange multiplier $\alpha$ is related to the energy $E$ appearing in the time independent GP equation (\ref{tigp3}) by $\alpha=E/m$. This shows that the Lagrange multiplier $\alpha$  is equal to the energy $E$ by unit of mass. On the other hand, considering the second order variations of energy, we find that the distribution is dynamically stable iff
\begin{eqnarray}
\label{ef5}
\delta^2 E_{tot}\equiv \frac{1}{2}\int h'(\rho)(\delta\rho)^2\, d{\bf r}+\frac{1}{2}\int \delta\rho\delta\Phi\, d{\bf r}\nonumber\\
+\frac{\hbar^2}{8m^2}\int  \left \lbrack \nabla \left (\frac{\delta\rho}{\sqrt{\rho}}\right )\right\rbrack^2\, d{\bf r}+\frac{\hbar^2}{8m^2}\int \frac{\Delta\sqrt{\rho}}{\rho^{3/2}}(\delta\rho)^2\, d{\bf r}>0,\nonumber\\
\end{eqnarray}
for all perturbations that conserve mass: $\int \delta\rho\, d{\bf r}=0$.

At equilibrium (${\bf u}={\bf 0}, \Theta_c=0$), the total energy reduces to
\begin{equation}
\label{ef6}
E_{tot}=\Theta_Q+U+W.
\end{equation}
Let us consider the case of a polytropic equation of state for which the internal energy is given by equation (\ref{ef3}). Multiplying equation (\ref{tigp3}) by $\rho$ and integrating over the entire domain, we obtain
\begin{equation}
\label{ef7}
\gamma U+2W+\Theta_Q=NE.
\end{equation}
In the TF limit ($\Theta_Q=0$), equations (\ref{ef6}) and (\ref{ef7}) become
\begin{equation}
\label{ef8}
E_{tot}=U+W,
\end{equation}
\begin{equation}
\label{ef9}
\gamma U+2W=NE,
\end{equation}
and in the noninteracting limit, we recover equations (\ref{te9})-(\ref{te10}).

\subsection{Stability of the homogeneous phase with respect to the quantum Euler equations}
\label{sec_bel}

Let us study the linear dynamical stability of a spatially homogeneous distribution  with respect to the quantum barotropic Euler equations
\begin{eqnarray}
{\partial \rho\over\partial t}+\nabla\cdot (\rho {\bf u})=0,
\label{bel1}
\end{eqnarray}
\begin{eqnarray}
\rho\biggl \lbrack {\partial {\bf u}\over\partial t}+({\bf u}\cdot \nabla){\bf u}\biggr \rbrack=-\nabla p-\rho\nabla\Phi+\frac{\rho\hbar^2}{2m^2}\nabla\left (\frac{\Delta\sqrt{\rho}}{\sqrt{\rho}}\right ),\nonumber\\
\label{bel2}
\end{eqnarray}
\begin{eqnarray}
\Phi({\bf r},t)=\int u(|{\bf r}-{\bf r}'|)\rho({\bf r}',t)d{\bf r}'.
\label{bel3}
\end{eqnarray}
These equations are equivalent to the mean field GP equations (\ref{g1})-(\ref{g4}).  In the classical or TF limit $\hbar\rightarrow 0$, the quantum potential can be neglected and we recover the classical barotropic Euler equations considered in \cite{cvb,cd}.

Linearizing equations (\ref{bel1})-(\ref{bel3}) around a spatially homogeneous steady state ($\rho={\rm cst}$ and ${\bf u}={\bf 0}$), we obtain
\begin{eqnarray}
{\partial \delta\rho\over\partial t}+\rho\nabla\cdot \delta {\bf u}=0,
\label{bel4}
\end{eqnarray}
\begin{eqnarray}
\rho \frac{\partial \delta {\bf u}}{\partial t}=-c_s^2\nabla\delta\rho-\rho\nabla\delta\Phi+\frac{\hbar^2}{4 m^2}\Delta\delta\rho,
\label{bel5}
\end{eqnarray}
\begin{eqnarray}
\delta\Phi({\bf r},t)=\int u(|{\bf r}-{\bf r}'|)\delta \rho({\bf r}',t)d{\bf r}',
\label{bel6}
\end{eqnarray}
where we have introduced the velocity of sound
\begin{equation}
\label{sound}
c_s^2=p'(\rho)=\rho h'(\rho).
\end{equation}
Equations (\ref{bel4}) and (\ref{bel5}) can be combined to give
\begin{eqnarray}
{\partial^2\delta\rho\over\partial t^2}=c_s^2\Delta\delta\rho+\rho\Delta\delta\Phi-\frac{\hbar^2}{4 m^2}\Delta^2\delta\rho.
\label{bel7}
\end{eqnarray}
Decomposing the perturbations in normal modes $\delta\rho=\delta\hat\rho e^{i({\bf k}\cdot {\bf r}-\omega t)}$ and $\delta\Phi=\delta\hat\Phi e^{i({\bf k}\cdot {\bf r}-\omega t)}$,  we get
\begin{eqnarray}
-\omega^2\delta\hat\rho=-c_s^2 k^2\delta\hat\rho-\rho k^2\delta\hat\Phi-\frac{\hbar^2}{4 m^2}k^4\delta\hat\rho,
\label{bel8}
\end{eqnarray}
\begin{eqnarray}
\delta\hat\Phi=(2\pi)^d \hat{u}(k)\delta\hat\rho.
\label{bel9}
\end{eqnarray}
This leads  to the dispersion relation
\begin{eqnarray}
\omega^{2}=c_{s}^{2}k^{2}+(2\pi)^{d}\hat{u}({k}) \rho k^{2} +\frac{\hbar^2 k^4}{4m^2}.
\label{bel10}
\end{eqnarray}
In the classical or TF limit $\hbar\rightarrow 0$ with $c_s$ fixed, we recover the classical dispersion relation associated with the linearized Euler equation \cite{cvb,cd}. We note that the quantum pressure arising from the Heisenberg uncertainty principle (last term in equation (\ref{bel10})) becomes important at very small scales, i.e. $k\rightarrow +\infty$. According to equation (\ref{bel10}), the pulsation $\omega$ is either real ($\omega^2>0$) or purely imaginary ($\omega^2<0$). As a result, the system is linearly stable with respect to  a perturbation with wavenumber $k$ if
\begin{eqnarray}
c_{s}^{2}+(2\pi)^{d}\hat{u}({k})\rho+\frac{\hbar^2 k^2}{4m^2}>0,
\label{bel11}
\end{eqnarray}
and linearly unstable otherwise (when $\omega^2>0$, the perturbation oscillates with a pulsation $\omega$ and when $\omega^2<0$, the perturbation grows exponentially rapidly  with a growth rate $\gamma=\sqrt{-\omega^2}$).

For repulsive potentials, satisfying $\hat{u}({k})>0$, the homogeneous distribution is always stable.
For Coulombian plasmas in $d=3$ dimensions, using $(2\pi)^3 \hat{u}(k)=\frac{4\pi e^2}{m^2k^2}$, the dispersion relation can
be written
\begin{eqnarray}
\omega^{2}=\omega_p^2+c_{s}^{2}k^{2}+\frac{\hbar^2 k^4}{4m^2},
\label{bel12}
\end{eqnarray}
where $\omega_p^2=4\pi\rho e^2/m^2$ is the plasma pulsation. This corresponds to the Bogoliubov energy spectrum of the excitation of the boson ground state \cite{bogoliubov}. For large wavenumbers (small wavelengths), the quasi-particle energy tends to the kinetic energy of an individual gas particle and $\omega\sim \hbar k^2/(2m)$. For the repulsive HMF model, using $\hat{u}_n=\frac{k}{4\pi}\delta_{n,\pm 1}$,  the dispersion relation can be written
\begin{eqnarray}
\omega^{2}=c_{s}^{2}n^2+\frac{kM}{4\pi}n^2\delta_{n,\pm 1}+\frac{\hbar^2 n^4}{4}.
\label{bel13}
\end{eqnarray}
The modes $n\neq \pm 1$ oscillate with a pulsation $\omega^2=c_s^2 n^2+\hbar^2 n^4/4$. The pulsation of the mode $n=\pm 1$ is given by $\omega^{2}=\omega_p^2+c_{s}^{2}+\frac{\hbar^2}{4}$ where $\omega_p^2=\frac{kM}{4\pi}$ is the proper pulsation \cite{cd}.

For attractive potentials, satisfying $\hat{u}({k})<0$, the homogeneous distribution is linearly stable if
\begin{eqnarray}
c_{s}^{2}>(c_{s}^{2})_{crit}\equiv \max_{k}\left\lbrack (2\pi)^{d}|\hat{u}({k})|\rho-\frac{\hbar^2 k^2}{4m^2}\right\rbrack,
\label{bel14}
\end{eqnarray}
and linearly unstable otherwise.  In that
case, the unstable wavelengths are determined by the converse of inequality (\ref{bel11}).
For the gravitational
interaction in $d=3$ dimensions, using $(2\pi)^3 \hat{u}(k)=-\frac{4\pi G}{k^2}$, the
dispersion relation can be written
\begin{eqnarray}
\omega^{2}=c_{s}^{2}k^{2}-4\pi G\rho+\frac{\hbar^2 k^4}{4m^2}.
\label{bel15}
\end{eqnarray}
The system is always unstable ($(c_{s}^{2})_{crit}=\infty$) for wavenumbers $k<k_*$ where $k_*$ is the generalized Jeans wavenumber \cite{prep}:
\begin{eqnarray}
k_*^2=\frac{2m^2}{\hbar^2}\left\lbrack \sqrt{c_s^4+\frac{4\pi G\hbar^2\rho}{m^2}}-c_s^2\right \rbrack.
\label{bel15a}
\end{eqnarray}
In the classical or TF limit $\hbar\rightarrow 0$ with $c_s$ fixed, we recover the classical Jeans wavenumber
\begin{eqnarray}
k_J=\frac{\sqrt{4\pi G\rho}}{c_s}.
\label{bel15b}
\end{eqnarray}
For a non interacting BEC  ($c_s=0$), we obtain the quantum Jeans wavenumber
\begin{eqnarray}
k_Q=\left (\frac{16\pi Gm^2\rho}{\hbar^2}\right )^{1/4}.
\label{bel15c}
\end{eqnarray}
For the attractive HMF model, using $\hat{u}_n=-\frac{k}{4\pi}\delta_{n,\pm 1}$,  the dispersion relation is
\begin{eqnarray}
\omega^{2}=c_{s}^{2}n^2-\frac{kM}{4\pi}n^2\delta_{n,\pm 1}+\frac{\hbar^2 n^4}{4}.
\label{bel16}
\end{eqnarray}
The modes $n\neq \pm 1$ oscillate with a pulsation $\omega^2=c_s^2 n^2+\hbar^2 n^4/4$. The complex pulsation of the mode $n=\pm 1$ is $\omega^{2}=c_{s}^{2}-\frac{kM}{4\pi}+\frac{\hbar^2}{4}$.
The homogeneous phase  is stable for
\begin{eqnarray}
c_{s}^{2}>\frac{kM}{4\pi}-\frac{\hbar^2}{4},
\label{bel17}
\end{eqnarray}
and unstable otherwise. We note that quantum mechanics favors the stability of the homogeneous phase. If we consider a BEC without short-range interaction ($h=p=c_s=0$), the stability criterion becomes
\begin{eqnarray}
\frac{\hbar^2}{kM}>\frac{1}{\pi},
\label{bel18}
\end{eqnarray}
and we recover the result of Appendix \ref{sec_sh}. On the other hand, in the classical or TF approximation $\hbar\rightarrow 0$ with $c_s$ fixed, the stability criterion becomes
\begin{eqnarray}
c_{s}^{2}>\frac{kM}{4\pi},
\label{bel19}
\end{eqnarray}
and we recover the result of \cite{cvb}.

Let us specifically discuss the case of fermions at $T=0$ described by the Fermi distribution (\ref{f1}) leading to the polytropic equation of state (\ref{f6}). The velocity of sound is
\begin{eqnarray}
c_s=v_F=\frac{M\hbar}{4},
\label{bel20}
\end{eqnarray}
and the pulsation equation for the modes $n=\pm 1$ can be written
\begin{eqnarray}
\omega^{2}=v_{F}^{2}-\frac{kM}{4\pi}+\frac{\hbar^2}{4}.
\label{bel16z}
\end{eqnarray}
The first term on the right hand side corresponds to the Fermi pressure (Pauli exclusion principle), the second to the cosine attraction and the third term to the quantum pressure (Heisenberg uncertainty principle). In the TF approximation, the pulsation equation reduces to
\begin{eqnarray}
\omega^{2}=v_{F}^{2}-\frac{kM}{4\pi},
\label{bel16c}
\end{eqnarray}
and the stability criterion (\ref{bel19})  can be written
\begin{eqnarray}
\frac{M\hbar^2}{k}>\frac{4}{\pi}.
\label{bel21a}
\end{eqnarray}
This returns the result of Section \ref{sec_fermions}. If we now take  the quantum potential into account and consider the complete stability criterion (\ref{bel17}), we obtain
\begin{eqnarray}
\frac{M\hbar^2}{k}\left (1+\frac{4}{N}\right )>\frac{4}{\pi}.
\label{bel21}
\end{eqnarray}
Comparing equation (\ref{bel21}) with equation (\ref{bel21a}), we see that the term coming from the quantum potential has a contribution of order $1/N$. Since the mean field approximation on which our approach is based assumes that $N\rightarrow +\infty$, it is not consistent to keep terms of order $1/N$ in the calculations (other terms of order $1/N$ may appear as a deviation from the mean field limit). Therefore, the quantum potential can be neglected in the limit $N\rightarrow +\infty$. This is a general result. 
In the case of fermions, the quantum pressure arising from the Heisenberg principle is always negligible with respect to the quantum pressure arising from the Pauli principle for $N\rightarrow +\infty$ (see Section \ref{sec_tigp}). Therefore, the Thomas-Fermi approximation of Section \ref{sec_fermions} is exact in the proper thermodynamic limit defined in Appendix \ref{sec_tlfb}. Alternatively, in the case of bosons without short-range interaction, the quantum potential stabilizes the system at small scales and must be taken into account in the thermodynamic limit defined in Appendix \ref{sec_tlfb}. This is the reason why fermions and bosons have different thermodynamic limits.

\subsection{Stability of the homogeneous phase with respect to the quantum Vlasov equation}
\label{sec_w}

To complete our analysis, we study the linear dynamical stability of a spatially  homogeneous distribution function $f=f({\bf v})$  with respect to the Wigner equation. The Wigner approach is a reformulation of quantum mechanics in the classical phase space language. We recall, however, that Wigner functions can take negative values so that they are not  true probability distributions. In the case of fermions, the Wigner equation, which can be viewed as a quantum Vlasov equation, is more fundamental than the quantum Euler equation considered in Section \ref{sec_bel} which is based on approximations (see Section \ref{sec_int}). For simplicity, we restrict ourselves to the one
dimensional case. The Wigner equation reads
\begin{eqnarray}
\frac{\partial f}{\partial t}+v\frac{\partial f}{\partial x}-\frac{im^2}{2\pi\hbar}\int d\lambda dv' e^{im(v-v')\lambda}\nonumber\\
\times\left\lbrack \Phi\left (x+\frac{\lambda \hbar}{2},t\right )-\Phi\left (x-\frac{\lambda \hbar}{2},t\right )\right\rbrack f(x,v',t)=0,\nonumber\\
\label{w1}
\end{eqnarray}
\begin{eqnarray}
\Phi(x,t)=\int u(|x-x'|)\rho(x',t)dx'.
\label{w2}
\end{eqnarray}
We can check that, in the classical limit $\hbar\rightarrow 0$, equation (\ref{w1}) returns the usual Vlasov equation.  The dispersion relation corresponding to  the linearized  Wigner equation is \cite{drummond}:
\begin{eqnarray}
\epsilon(k,\omega)\equiv 1-2\pi \hat{u}(k)\int\frac{f(v)dv}{(v-\frac{\omega}{k})^2-\frac{\hbar^2 k^2}{4m^2}}=0.
\label{w3}
\end{eqnarray}
For $\hbar\rightarrow 0$, we recover the classical dispersion relation associated with the linearized Vlasov equation \cite{cvb,cd}. The point of marginal stability ($\omega=0$) is determined by the condition
\begin{eqnarray}
1-2\pi \hat{u}(k)\int\frac{f(v)dv}{v^2-\frac{\hbar^2 k^2}{4m^2}}=0.
\label{w4}
\end{eqnarray}
For the potential $2\pi\hat{u}(k)=\frac{\omega_P^2}{\rho k^2}$ of a one dimensional plasma, if neglect the Landau damping and consider the long wavelength limit $k\rightarrow 0$ of the dispersion relation (\ref{w3}), we obtain at order $k^4$:
\begin{eqnarray}
\omega^2=\omega_P^2+3\langle v^2\rangle k^2+(5\langle v^4\rangle -9\langle v^2\rangle^2)\frac{k^4}{\omega_P^2}+\frac{\hbar^2 k^4}{4m^2}+...\nonumber\\
\label{w4b}
\end{eqnarray}
This is the quantum generalization of the Langmuir wave dispersion relation. For the potential $2\pi\hat{u}(k)=-2G/k^2$ of a one dimensional gravitational system, we obtain
\begin{eqnarray}
\omega^2=-2G\rho+3\langle v^2\rangle k^2-(5\langle v^4\rangle -9\langle v^2\rangle^2)\frac{k^4}{2G\rho}+\frac{\hbar^2 k^4}{4m^2}+...\nonumber\\
\label{w4c}
\end{eqnarray} 
This gives the growth rate of the Jeans instability for small $k$ when quantum effects are taken into account. Note that for a Fermi distribution at $T=0$ (see below), the term in parenthesis in equations (\ref{w4b}) and (\ref{w4c}) vanishes.

Let us write the complex pulsation as $\omega=\omega_r+i\omega_i$. For $\omega_i=0$, the real and imaginary parts of the dielectric function $\epsilon(k,\omega_r)=\epsilon_r(k,\omega_r)+i\epsilon_i(k,\omega_r)$ are \cite{drummond}:
\begin{eqnarray}
\epsilon_r(k,\omega_r)= 1-2\pi \hat{u}(k)P\int\frac{f(v)dv}{(v-\frac{\omega_r}{k})^2-\frac{\hbar^2 k^2}{4m^2}},
\label{sun1}
\end{eqnarray}
\begin{eqnarray}
\epsilon_r(k,\omega_r)=\frac{2\pi^2 m\hat{u}(k)}{\hbar k}\left\lbrack f\left (\frac{\omega_r}{k}-\frac{\hbar k}{2m}\right )-f\left (\frac{\omega_r}{k}+\frac{\hbar k}{2m}\right )\right\rbrack,\nonumber\\
\label{sun2}
\end{eqnarray}
where $P$ stands for the principal part. The imaginary part of the dielectric function is zero when $\omega_r=k v_0$ where $v_0$ is solution of 
\begin{eqnarray}
f(v_0+{\hbar k}/{2m})=f(v_0-{\hbar k}/{2m}).
\label{sun3}
\end{eqnarray}
If $f(v)$ has a single maximum, then equation (\ref{sun3}) has a unique solution $v_0$. According to the Nyquist theorem, the system is stable if $\epsilon_r(k,kv_0)>0$ for all $k$ and it is unstable if there exists some values of $k$ for which   $\epsilon_r(k,kv_0)<0$. Therefore, the mode $k$ is stable if
\begin{eqnarray}
1-2\pi \hat{u}(k)P\int\frac{f(v)dv}{(v-v_0)^2-\frac{\hbar^2 k^2}{4m^2}}>0,
\label{sun4}
\end{eqnarray}
and unstable otherwise. This generalizes the classical criterion \cite{cd}. Equation (\ref{sun4}) can be rewritten \cite{haasnyquist}:
\begin{eqnarray}
1+2\pi \hat{u}(k)\int\frac{f(v_0+\hbar k/2m)-f(v)}{(v-v_0)^2-\frac{\hbar^2 k^2}{4m^2}}\, dv>0.
\label{sun5}
\end{eqnarray}
For the electrostatic interaction, and more generally for any repulsive potential $\hat{u}(k)>0$, a single humped distribution is always stable \cite{haasnyquist}.  For a potential $\hat{u}_n=-\frac{k}{4\pi}\delta_{n,\pm 1}$ corresponding to the (attractive) HMF model, a single humped distribution is stable iff
\begin{eqnarray}
1+\frac{k}{2} P\int\frac{f(v)dv}{(v-v_0)^2-\frac{\hbar^2}{4}}>0.
\label{sun6}
\end{eqnarray}
This generalizes the criterion obtained in \cite{cd}.

As an illustration, let us consider the linear dynamical stability of the Fermi distribution at $T=0$ with respect to the Wigner equation. The DF is given by $f=\rho/(2v_F)$ if $|v|\le v_F$ and $f=0$ if $|v|\le v_F$.  Substituting this distribution function in equation (\ref{w3}) and solving for $\omega$ we obtain
\begin{eqnarray}
\omega^2=\frac{\hbar^2k^4}{4m^2}+v_F^2k^2+\frac{\hbar k^3}{m}v_F \coth \left (\frac{\hbar v_F k}{2\pi \hat{u}(k)\rho m}\right ).
\label{w5}
\end{eqnarray}
For  $f(v)=\rho \delta (v)$, corresponding to $v_F=0$, this expression reduces to the form
\begin{eqnarray}
\omega^2=\frac{\hbar^2k^4}{4m^2}+2\pi \hat{u}(k)\rho k^2.
\label{w6}
\end{eqnarray}
It coincides with the fluid dispersion relation (\ref{bel10}) for a cold gas with $c_s=0$. In the TF approximation $\hbar\rightarrow 0$ with $v_F$ fixed, equation (\ref{w5}) becomes
\begin{eqnarray}
\omega^2=v_F^2k^2+2\pi \hat{u}(k)\rho k^2.
\label{w7}
\end{eqnarray}
It coincides with the fluid dispersion relation (\ref{bel10}) with $c_s=v_F$ and $\hbar\rightarrow 0$.
Therefore, in the TF approximation, the dispersion relations of the Fermi distribution derived from the Euler and the Vlasov equations are the same\footnote{Equation (\ref{w7}) also represents the dispersion relation of the waterbag distribution in the classical limit. Therefore, in the classical limit, the dispersion relations of the waterbag distribution derived from the Euler and the Vlasov equations are the same \cite{cd}.}. This is no more true when the quantum potential is taken into account (compare equations (\ref{w5}) and (\ref{bel10})).

For a potential $2\pi\hat{u}(k)=\frac{\omega_P^2}{\rho k^2}$ corresponding to a one dimensional plasma, we recover the results of \cite{manfredi} (the case of a gravitational plasma is obtained by reversing the sign of the interaction). On the other hand, for a potential $\hat{u}_n=-\frac{k}{4\pi}\delta_{n,\pm 1}$ corresponding to the attractive HMF model (the repulsive HMF model is obtained by the substitution $k\rightarrow -k$), only the modes $n=\pm 1$ are allowed and the quantum dispersion relation (\ref{w3}) becomes
\begin{eqnarray}
1+\frac{k}{2}\int\frac{f(v)dv}{(v-\omega)^2-\frac{\hbar^2}{4}}=0.
\label{w8}
\end{eqnarray}
For the Fermi (or waterbag) distribution, we obtain
\begin{eqnarray}
\omega^2=\frac{\hbar^2}{4}+v_F^2-\frac{kM}{4\pi} \Phi \left (\frac{4\pi \hbar v_F}{kM}\right ),
\label{w9}
\end{eqnarray}
where $\Phi(x)=x\coth(x)$. For $v_F=0$, this expression reduces to the form
\begin{eqnarray}
\omega^2=\frac{\hbar^2}{4}-\frac{kM}{4\pi}.
\label{w10}
\end{eqnarray}
In the TF approximation $\hbar\rightarrow 0$ with $v_F$ fixed, we obtain
\begin{eqnarray}
\omega^2=v_F^2-\frac{kM}{4\pi}.
\label{w11}
\end{eqnarray}
This is the classical dispersion relation of the waterbag distribution \cite{cvb}. It coincides with equation (\ref{bel16c}) as explained previously. This is no more true if we take into account the quantum pressure (compare equations (\ref{w9}) and (\ref{bel16c})).  For $\hbar\rightarrow 0$, using $\Phi(x)=1+x^2/3-x^4/45+...$ for $x\rightarrow 0$, equation (\ref{w9}) can be expanded in powers of $\hbar$ as
\begin{eqnarray}
\omega^2=\frac{\hbar^2}{4}+v_F^2-\frac{kM}{4\pi}\left (1+\frac{\pi^2\hbar^4}{3k^2}-\frac{\pi^4\hbar^8}{45k^4}+...\right ).
\label{w12}
\end{eqnarray}
This expression differs from equation (\ref{bel16c}) by terms of order $\hbar^4$ or smaller. In the limit $\hbar\rightarrow +\infty$, we get
\begin{eqnarray}
\omega^2=\frac{\hbar^2}{4}+v_F^2-\hbar v_F.
\label{w13}
\end{eqnarray}
To appreciate the effect of the number of particles $N$, we introduce the dimensionless Planck constant (\ref{cph1}) and rewrite the pulsation equation (\ref{w9}) in the form
\begin{eqnarray}
\frac{4\pi\omega^2}{kM}=\frac{2\chi^2}{N^2}+\frac{1}{2}\chi^2-\Phi \left (\frac{2\chi^2}{N}\right ).
\label{w14}
\end{eqnarray}
Similarly, the pulsation relation (\ref{bel16c}) derived from the Euler equation can be written
\begin{eqnarray}
\frac{4\pi\omega^2}{kM}=\frac{2\chi^2}{N^2}+\frac{1}{2}\chi^2-1.
\label{w15}
\end{eqnarray}
The thermodynamic limit corresponds to $N\rightarrow +\infty$ with fixed $\chi$.  In that case, the two relations take the form
\begin{eqnarray}
\frac{4\pi\omega^2}{kM}=\frac{1}{2}\chi^2-1.
\label{w16}
\end{eqnarray}
They return the critical Planck constant (\ref{cpg5}). Therefore, the TF approximation is exact in the thermodynamic limit $N\rightarrow +\infty$.

{\it Remark:} when coupled to a long-range potential of interaction, the classical Vlasov equation is known to develop filaments at smaller and smaller scales due to phase mixing and/or violent relaxation \cite{lb}. Considering (real or effective) quantum effects could be a way to regularize the Vlasov equation at small scales and serve as an alternative to coarse-graining. In that case, the (effective) Planck constant $\hbar$ could determine the scale at which the filaments are smoothed-out.

\section{Conclusion}
\label{sec_conclusion}

In this paper, we have generalized the HMF model in order to take into account quantum effects. We have considered the case of fermions and bosons with cosine interaction at $T=0$. In the classical limit $\hbar\rightarrow 0$, all the particles are located at $\theta=0$ with velocity $v=0$. This leads to a distribution function corresponding to a Dirac peak  in position and velocity space: $f(\theta,v)=M\delta(v)\delta(\theta)$. When quantum effects are taken into account, this Dirac distribution is regularized. Fermions tend to occupy the lowest energy states but cannot condensate into the state $(\theta=0,v=0)$ because they must satisfy the Pauli exclusion principle. Bosons form a condensate in velocity space (all the bosons are in the same quantum state with $v=0$) but, because of the Heisenberg principle, they are delocalized in position space so that $\rho(\theta)\neq M\delta(\theta)$.

In the classical limit, the homogeneous phase is unstable at $T=0$. When quantum mechanics is taken into account, we find that the homogeneous phase can be stabilized when the normalized Planck constant $\chi$ is sufficiently high. However, crucial differences exist between fermions and bosons. In the case of fermions, the stabilization is due to the Pauli exclusion principle while in the case of bosons, it is due to the Heisenberg principle. Therefore, the study of fermions is based on the Thomas-Fermi approximation while the study of bosons is based on the Hartree approximation. This is why the thermodynamic limit $N\rightarrow +\infty$ is different for fermions and bosons (see Appendix \ref{sec_tlfb}). For fermions, the thermodynamic limit corresponds to $N\rightarrow +\infty$ in such a way that the coupling constant scale like $k\sim N$ (yielding $E\sim N^3$). This is a new prescription. For bosons,  the thermodynamic limit corresponds to $N\rightarrow +\infty$ with $k\sim 1/N$ (yielding $E\sim N$). This corresponds to the Kac prescription like in the classical regime. In the case of fermions, the stabilization of the homogeneous phase in the quantum regime occurs through a first order phase transition (as a function of the normalized Planck constant) while the phase transition is second order in the case of bosons. On the other hand, in the semi-classical limit $\hbar\rightarrow 0$, the density profile of fermions is parabolic (see equation (\ref{class13})) while the density profile of bosons is Gaussian (see equation (\ref{class12})).

One interest of the quantum HMF model is its relative simplicity that should allow for accurate numerical simulations of quantum particles (fermions and bosons) with long-range interactions. In the present paper, we have only considered the thermodynamical equilibrium states of these systems, but the relaxation towards equilibrium is also of considerable interest. In particular, how does the system relax towards a steady state? What is the damping mechanism? These are important questions that we plan to address in the future.

\vskip0.5cm
{\it Acknowledgment:} I am grateful to L. Delfini for his assistance in some aspects of the numerical work.

\appendix

\section{Thermodynamic limits for fermions and bosons}
\label{sec_tlfb}

If we reintroduce the dimensional parameters, the Hamiltonian of the HMF model reads
\begin{eqnarray}
\label{tlfb1}
H=\frac{1}{2}\sum_{i=1}^N m v_i^2-gm^2\sum_{i<j}\cos\left ( \frac{x_i-x_j}{R}\right ),
\end{eqnarray}
where $g$ is the coupling constant (the counterpart of the gravitational constant $G$ in astrophysics), $m$ is the mass of the particles and $R$ is the system size (the particles are confined in the domain $[-\pi R,\pi R]$). By comparing the different terms of the Hamiltonian, we get the scaling
\begin{eqnarray}
\label{tlfb2}
E\sim N m v^2\sim N^2 g m^2.
\end{eqnarray}
Therefore, the dimensionless energy is
\begin{eqnarray}
\label{tlfb3}
\epsilon\sim \frac{E}{g N^2 m^2},
\end{eqnarray}
in agreement with equation (\ref{hp5}). We can also define a dynamical time $t_D\sim R/v$. Using equation (\ref{tlfb2}), we get
\begin{eqnarray}
\label{tlfb4}
t_D\sim \frac{R}{\sqrt{N g m}}.
\end{eqnarray}
Finally, in the classical regime, the temperature can be estimated by  $k_B T\sim m v^2$. Using equation (\ref{tlfb2}) to estimate $v$, we obtain the dimensionless inverse temperature
\begin{eqnarray}
\label{tlfb5}
\eta\sim \beta N g m^2,
\end{eqnarray}
in agreement with the expression $\eta=\beta kM/(4\pi)$ of \cite{cvb}.

In the case of bosons at $T=0$, the long-range interaction is balanced by the pressure arising
from  the Heisenberg uncertainty principle $\Delta x \Delta p\sim \hbar$. Writing $\Delta x\sim R$ and $\Delta p\sim m v$ and using equation (\ref{tlfb2}) to estimate $v$, we obtain the dimensionless Planck constant
\begin{eqnarray}
\label{tlfb6}
\chi_B\sim \frac{\hbar}{N^{1/2}g^{1/2}Rm^{3/2}},
\end{eqnarray}
in agreement with equation (\ref{planckb}).

In the case of fermions at $T=0$, the long-range interaction is balanced by the pressure arising
from  the Pauli exclusion principle $f\sim m^2/\hbar$. Writing $f\sim M/(R v_F)$ with $v_F\sim v$ and using equation (\ref{tlfb2}) to estimate $v$, we obtain the dimensionless Planck constant
\begin{eqnarray}
\label{tlfb7}
\chi_F\sim \frac{\hbar N^{1/2}}{Rg^{1/2}m^{3/2}},
\end{eqnarray}
in agreement with equation (\ref{cph1}).

These scalings allow us to correctly define the thermodynamic limit. In the classical regime, the only dimensionless parameters are $\epsilon$ and $\eta$. The classical thermodynamic limit corresponds to $N\rightarrow +\infty$ in such a way that $\epsilon$ and $\eta$ remain of order unity. If we take $m\sim R\sim 1$, we see that the choice  $g\sim 1/N$ (Kac prescription) leads to an extensive scaling $E\sim N$ of the energy and to an intensive scaling $\beta\sim 1$ of the inverse temperature. On the other hand, the dynamical time scales like $t_D\sim 1$.
We could also take $g\sim 1$ and $m\sim 1/\sqrt{N}$. This leads to the same scaling  $E/N\sim \beta\sim 1$ of the energy and inverse temperature but in that case $t_D\sim R/N^{1/4}$ (we may then take $R\sim N^{1/4}$ in order to have $t_D\sim 1$).

In the quantum regime, the dimensionless parameters are $\epsilon$, $\eta$ and $\chi$. The quantum thermodynamic limit corresponds to $N\rightarrow +\infty$ in such a way that $\epsilon$, $\eta$ and $\chi$ remain of order unity. Let us take $\hbar\sim m\sim R\sim 1$. In the case of bosons, we find from equations (\ref{tlfb6}) that the coupling constant scales like $g\sim 1/N$  which coincides with the Kac prescription. In that case, we have $E/N\sim \beta\sim t_D\sim 1$. Therefore, the bosonic quantum thermodynamic limit coincides with the classical thermodynamic limit. In the case of fermions, we find from equation (\ref{tlfb7}) that the coupling constant scales like $g\sim N$ which is different from the Kac prescription. In that case, the energy scales like $E\sim N^3$ which is different from the extensive scaling\footnote{The $N^3$ scaling of the ground state energy can be  directly seen on equation (\ref{hp4b}).} and the inverse temperature scales like $\beta\sim N^{-2}$ which is different from the intensive scaling. On the other hand, the dynamical time scales like $t_D\sim 1/N$. Therefore, the fermionic quantum thermodynamic limit differs from the classical and bosonic thermodynamic limits. The fact that the bosonic and fermionic thermodynamic limits differ is not surprising since the quantum pressure which balances the long-range interaction corresponds to the Heisenberg uncertainty principle for bosons and to the Pauli exclusion principle for fermions which are of a completely different nature.

We note that other scalings are possible. For example, it makes sense to take $\hbar\sim m\sim g\sim 1$ since these quantities are fixed physical constants that should not depend on $N$. In the classical limit, this implies $E\sim N^2$, $\beta\sim N^{-1}$ and $t_D\sim R/\sqrt{N}$ (we may then take $R\sim \sqrt{N}$ in order to have $t_D\sim 1$). In the case of fermions, using equations (\ref{tlfb7}) and (\ref{tlfb3})-(\ref{tlfb5}), we obtain  $R\sim N^{1/2}$, $E\sim N^2$, $\beta\sim N^{-1}$ and $t_D\sim 1$.
In the case of bosons, using equations (\ref{tlfb6}) and (\ref{tlfb3})-(\ref{tlfb5}), we get $R\sim N^{-1/2}$, $E\sim N^2$, $\beta\sim N^{-1}$ and $t_D\sim N^{-1}$. We could also impose the Kac prescription $R\sim m\sim g N\sim 1$ to fermions by taking $\hbar\sim N^{-1}$. In that case $E/N\sim \beta \sim t_D\sim 1$. Although this limit is mathematically conceivable, there is no physical reason why $\hbar$ should depend on $N$. Therefore, we could impose the Kac prescription under the form $\hbar\sim m\sim gN\sim 1$ by taking $R\sim N$. In that case $E/N\sim \beta \sim 1$ but $t_D\sim \sqrt{N}$.

These strange scalings should not cause surprise. They arise due to the long-range nature of the interactions making the energy non-additive. Similar (unusual) scalings are found for classical and quantum self-gravitating systems \cite{ijmpb,ht,messer}.

\section{Stability of the homogeneous phase for bosons}
\label{sec_sh}

In order to study the stability of the homogeneous phase, we shall use the Madelung representation of the Schr\"odinger equation in terms of hydrodynamical equations. This will allow us to draw a close parallel with the stability analysis of the classical HMF model performed in \cite{cvb}.  For the bosonic HMF model, the quantum Euler equations (\ref{mad6}), (\ref{mad7}) and (\ref{mfs2}) take the form
\begin{equation}
\label{sh1}
\frac{\partial\rho}{\partial t}+\frac{\partial}{\partial\theta}(\rho  u)=0,
\end{equation}
\begin{equation}
\label{sh2}
\frac{\partial {u}}{\partial t}+u\frac{\partial u}{\partial\theta}=-\frac{\partial\Phi}{\partial\theta}+\frac{\hbar^2}{2}\frac{\partial}{\partial\theta}\left (\frac{(\sqrt{\rho})''}{\sqrt{\rho}}\right ),
\end{equation}
\begin{eqnarray}
\label{sh3}
\Phi(\theta)=-\frac{k}{2\pi}\int_0^{2\pi} \rho(\theta') \cos(\theta-\theta')\, d\theta'.
\end{eqnarray}
We stress that these equations are equivalent to the mean field Schr\"odinger equation (\ref{mfs1})-(\ref{mfs4}). In the classical limit $\hbar\rightarrow 0$, the quantum potential can be neglected and  we recover the Euler equations at $T=0$ considered in \cite{cvb}. 

Linearizing these equations around the homogeneous solution $\rho=\frac{M}{2\pi}$, we obtain
\begin{equation}
\label{sh4}
\frac{\partial\delta\rho}{\partial t}+\rho\frac{\partial\delta u}{\partial\theta}=0,
\end{equation}
\begin{equation}
\label{sh5}
\frac{\partial {\delta u}}{\partial t}=-\frac{\partial\delta\Phi}{\partial\theta}+\frac{\hbar^2}{4\rho}
\frac{\partial^3\delta\rho}{\partial\theta^3},
\end{equation}
\begin{eqnarray}
\label{sh6}
\delta\Phi(\theta,t)=-\frac{k}{2\pi}\int_0^{2\pi} \delta\rho(\theta',t) \cos(\theta-\theta')\, d\theta'.
\end{eqnarray}
Eliminating $\delta u$ between the first two equations, we get
\begin{equation}
\label{sh7}
\frac{\partial^2\delta\rho}{\partial t^2}=\rho\frac{\partial^2\delta\Phi}{\partial\theta^2}-\frac{\hbar^2}{4}
\frac{\partial^4\delta\rho}{\partial\theta^4}.
\end{equation}
Let us consider a perturbation of the density profile  of the form
\begin{equation}
\label{sh8}
\delta\rho(\theta,t)={\rm Re} \left\lbrack Ae^{i(n\theta-\omega t)}\right\rbrack.
\end{equation}
Using equation (\ref{sh6}), the corresponding perturbation of the potential is
\begin{equation}
\label{sh9}
\delta\Phi(\theta,t)={\rm Re} \left\lbrack -\frac{kA}{2}(\delta_{n,1}+\delta_{n,-1})e^{i(n\theta-\omega t)}\right\rbrack.
\end{equation}
Substituting equations (\ref{sh8}) and (\ref{sh9}) in equation (\ref{sh7}), we obtain the dispersion relation
\begin{equation}
\label{sh10}
\omega^2=-\frac{kM}{4\pi}n^2(\delta_{n,1}+\delta_{n,-1})+\frac{\hbar^2}{4}n^4.
\end{equation}
For $n\neq \pm 1$, we get
\begin{equation}
\label{sh11}
\omega^2=\frac{\hbar^2}{4}n^4,
\end{equation}
so that these modes are stable: the perturbation oscillates with a pulsation $\omega=\hbar n^2/2$. For $n=\pm 1$, we obtain
\begin{equation}
\label{sh12}
\omega^2=-\frac{kM}{4\pi}+\frac{\hbar^2}{4}.
\end{equation}
In terms of the normalized Planck constant (\ref{planckb}) for bosons, we find that the homogeneous phase is stable iff
\begin{equation}
\label{sh14}
\chi\equiv \hbar \left (\frac{2\pi}{kM}\right )^{1/2}>\chi_c=\sqrt{2}.
\end{equation}
In particular, the homogeneous solution is unstable in the classical regime $\chi\rightarrow 0$ \cite{cvb} but it becomes stable in the quantum regime for $\chi>\chi_c$. In that case, it is stabilized by the Heisenberg principle or, equivalently, by the Bohm quantum potential. Note that the critical Planck constant $\chi_c=\sqrt{2}$ precisely corresponds to the bifurcation point at which the inhomogeneous branch appears (see Appendix \ref{sec_be}). For $\chi<\chi_c$, the inhomogeneous solutions with $B\neq 0$ are stable while the homogeneous solutions are unstable.

\section{Bifurcation analysis for bosons}
\label{sec_be}

In this Appendix, we determine the behavior of the magnetization and of the energy close to the critical point at which the bifurcation from the homogeneous (quantum) to the inhomogeneous (classical) phase takes place.

The time independent mean field Schr\"odinger equation with a cosine potential can be written in dimensionless form (see Section \ref{sec_cos}):
\begin{eqnarray}
\label{be1}
-\frac{1}{2}\chi^2\psi''-b\cos\theta\psi={\cal E}\psi,
\end{eqnarray}
\begin{eqnarray}
\label{be2}
b=\int_{0}^{2\pi}\psi^2\cos\theta\, d\theta,
\end{eqnarray}
\begin{eqnarray}
\label{be3}
\int_{0}^{2\pi}\psi^2\, d\theta=1,
\end{eqnarray}
\begin{eqnarray}
\label{be4}
\rho(\theta)=N\psi^2(\theta).
\end{eqnarray}
We assume that the wavefunction is symmetric with respect to $\theta=0$ so that $\psi(-\theta)=\psi(\theta)$. Therefore, the boundary conditions are
\begin{eqnarray}
\label{be5}
\psi'(0)=\psi'(\pi)=0.
\end{eqnarray}

The homogeneous solution corresponds to $\psi_{0}=1/\sqrt{2\pi}$ and $b_0={\cal E}_0=0$. Close to the bifurcation, we make the expansion
\begin{eqnarray}
\label{be6}
\psi=\frac{1}{\sqrt{2\pi}}+\epsilon\psi_1+\epsilon^2\psi_2+\epsilon^3\psi_3+...
\end{eqnarray}
\begin{eqnarray}
\label{be7}
b=\epsilon b_1+\epsilon^2 b_2+\epsilon^3 b_3+...
\end{eqnarray}
\begin{eqnarray}
\label{be8}
{\cal E}=\epsilon {\cal E}_1+\epsilon^2 {\cal E}_2+\epsilon^3 {\cal E}_3+...
\end{eqnarray}
\begin{eqnarray}
\label{be9}
\chi=\chi_0+\epsilon \chi_1+\epsilon^2 \chi_2+\epsilon^3 \chi_3+...
\end{eqnarray}
where $\epsilon\ll 1$ is a small parameter.

At the order $\epsilon$, we obtain
\begin{eqnarray}
\label{be10}
-\frac{1}{2}\chi_0^2\psi_1''-\frac{1}{\sqrt{2\pi}}b_1\cos\theta=\frac{1}{\sqrt{2\pi}}{\cal E}_1,
\end{eqnarray}
\begin{eqnarray}
\label{be11}
b_1=\frac{2}{\sqrt{2\pi}}\int_{0}^{2\pi}\psi_1\cos\theta\, d\theta,
\end{eqnarray}
\begin{eqnarray}
\label{be12}
\int_{0}^{2\pi}\psi_1\, d\theta=0.
\end{eqnarray}
Solving these equations with the boundary conditions (\ref{be5}), we find that ${\cal E}_1=0$, $\chi_{0}=\sqrt{2}$ and
\begin{eqnarray}
\label{be13}
\psi_1=\frac{1}{\sqrt{2\pi}}b_1\cos\theta.
\end{eqnarray}

At the order $\epsilon^2$, we obtain
\begin{eqnarray}
\label{be14}
-\psi_2''-\sqrt{2}\chi_1\psi''_1-\left (b_1\psi_1+\frac{b_2}{\sqrt{2\pi}}\right )\cos\theta=\frac{{\cal E}_2}{\sqrt{2\pi}},\nonumber\\
\end{eqnarray}
\begin{eqnarray}
\label{be15}
b_2=\int_{0}^{2\pi}\left (\frac{2}{\sqrt{2\pi}}\psi_2+\psi_1^2\right ) \cos\theta\, d\theta,
\end{eqnarray}
\begin{eqnarray}
\label{be16}
\int_{0}^{2\pi}\psi_2\, d\theta=-\frac{\sqrt{2\pi}}{2}\int_{0}^{2\pi}\psi_1^2\, d\theta.
\end{eqnarray}
Solving these equations with the boundary conditions (\ref{be5}), we find that ${\cal E}_2=-b_1^2/2$, $\chi_{1}=0$ and
\begin{eqnarray}
\label{be17}
\psi_2=\frac{1}{\sqrt{2\pi}}b_2\cos\theta+\frac{1}{8\sqrt{2\pi}}b_1^2\cos(2\theta)-\frac{b_1^2}{4\sqrt{2\pi}}.
\end{eqnarray}

At the order $\epsilon^3$, we obtain
\begin{eqnarray}
\label{be18}
-\psi_3''-\sqrt{2}\chi_2\psi''_1-\left (b_1\psi_2+b_2\psi_1+\frac{b_3}{\sqrt{2\pi}}\right )\cos\theta\nonumber\\
={\cal E}_2\psi_1+\frac{{\cal E}_3}{\sqrt{2\pi}},
\end{eqnarray}
\begin{eqnarray}
\label{be19}
b_3=\int_{0}^{2\pi}\left (\frac{2}{\sqrt{2\pi}}\psi_3+2\psi_1\psi_2\right )\cos\theta\, d\theta.
\end{eqnarray}
Solving these equations with the boundary conditions (\ref{be5}), we find that ${\cal E}_3=-b_1b_2$, $\chi_2=-7 b_1^2/(8\sqrt{2})$ and
\begin{eqnarray}
\label{be20}
\sqrt{2\pi}\psi_3=\left (\frac{3b_1^3}{16}+b_3\right )\cos\theta+\frac{b_1 b_2}{4}\cos(2\theta)\nonumber\\
+\frac{b_1^3}{144}\cos(3\theta)+C,
\end{eqnarray}
where the constant $C$ could be obtained by writing the normalization condition at order $\epsilon^3$ (but we shall not need it here).

Combining the previous results, we find that the bifurcation takes place at $\chi=\chi_c=\sqrt{2}$, i.e. at the point where the homogeneous phase becomes unstable. Close to the critical point, the magnetization, the energy and the wavefunction behave like
\begin{eqnarray}
\label{be21}
b^2\sim\frac{8\sqrt{2}}{7}(\chi_c-\chi),
\end{eqnarray}
\begin{eqnarray}
\label{be22}
{\cal E}\sim -\frac{1}{2}b^2\sim -\frac{4\sqrt{2}}{7}(\chi_c-\chi),
\end{eqnarray}
\begin{eqnarray}
\psi(\theta)\simeq \frac{1}{\sqrt{2\pi}}\left\lbrack 1+b(\chi)\cos\theta\right \rbrack.
\end{eqnarray}
We also find from our analysis that the total energy $\epsilon_{tot}$ vanishes at the order $\epsilon^3$ so that it scales like $\epsilon_{tot}\propto (\chi_c-\chi)^2$. The prefactor could be obtained by extending the asymptotic analysis to next order.

\section{The semi-classical limit for bosons and fermions: harmonic approximation}
\label{sec_class}

In the classical limit $\hbar=0$, the system is equivalent to a classical gas at $T=0$. In that case, all the particles have collapsed at $\theta=0$ and the density profile is a Dirac distribution $\rho(\theta)=M\delta(\theta)$ with magnetization $b=1$. According to equations (\ref{te12}) and (\ref{fa7}), the total energy is $\epsilon_{tot}=-2$ and  the energy appearing in the Schr\"odinger equation is ${\cal E}=-1$.

In the semi-classical limit $\hbar\rightarrow 0$, the distribution is strongly peaked around $\theta=0$. We can therefore replace $\cos\theta$ by $1-\theta^2/2$ since $\theta\ll 1$ and extend the integrals over the angles to infinity. This is valid in the limit $\kappa\rightarrow 0$. In that limit, equations (\ref{cos12}) and (\ref{cos13}) become
\begin{eqnarray}
\label{class1}
-\frac{\kappa^2}{2}\psi''+\frac{1}{2}\theta^2\psi=(\lambda+1)\psi,
\end{eqnarray}
\begin{eqnarray}
\label{class2}
\int_{-\infty}^{+\infty} \psi^2\, d\theta=1.
\end{eqnarray}
For given $\kappa$, equation (\ref{class1})  is just the ordinary Schr\"odinger equation of a harmonic oscillator. The fundamental eigenvalue is
\begin{eqnarray}
\label{class3}
\lambda=-1+\frac{1}{2}\kappa,
\end{eqnarray}
and the corresponding wavefunction is
\begin{eqnarray}
\label{class4}
\psi(\theta)=\left (\frac{1}{\pi\kappa}\right )^{1/4}e^{-\frac{\theta^2}{2\kappa}}.
\end{eqnarray}
The magnetization can be approximated by
\begin{eqnarray}
\label{class5}
b=\int_{-\infty}^{+\infty} \psi^2 \left (1-\frac{\theta^2}{2}\right )\, d\theta.
\end{eqnarray}
Substituting equation (\ref{class4}) in equation (\ref{class5}), we get
\begin{eqnarray}
\label{class6}
b=1-\frac{1}{4}\kappa.
\end{eqnarray}
Finally, substituting the expansions (\ref{class3}) and (\ref{class6}) in equations (\ref{cos15}) and (\ref{cos16}), we obtain
\begin{eqnarray}
\label{class7}
\chi\simeq \kappa,\qquad {\cal E}\simeq  -1+\frac{3}{4}\kappa.
\end{eqnarray}
We can now express the results in terms of the normalized Planck constant $\chi$, defined by equation (\ref{planckb}), in the limit $\chi\rightarrow 0$. Eliminating $\kappa$ between equations (\ref{class6}) and (\ref{class7}), we find that
\begin{eqnarray}
\label{class8}
b=1-\frac{1}{4}\chi,
\end{eqnarray}
and
\begin{eqnarray}
\label{class9}
{\cal E}\simeq  -1+\frac{3}{4}\chi.
\end{eqnarray}
Using equation (\ref{cos23}), the total energy is
\begin{eqnarray}
\label{class10}
\epsilon_{tot}\simeq  -2+2\chi.
\end{eqnarray}
Finally, the wave function and the density profile are given by
\begin{eqnarray}
\label{class11}
\psi(\theta)=\left (\frac{1}{\pi\chi}\right )^{1/4}e^{-\frac{\theta^2}{2\chi}},
\end{eqnarray}
and
\begin{eqnarray}
\label{class12}
\rho(\theta)=N\left (\frac{1}{\pi\chi}\right )^{1/2}e^{-\frac{\theta^2}{\chi}}.
\end{eqnarray}
These asymptotic expansions are compared with the numerical results in Section \ref{sec_cos}.

We can also consider the semi-classical limit for fermions. This corresponds to the limit  $x\rightarrow x_c=3/2$ in the equations of Section \ref{sec_ipB}. For $x=x_c$, the density profile (\ref{ip16}) is a Dirac peak $\rho(\theta)=M\delta(\theta)$. For $x\rightarrow x_c$, we can make the harmonic approximation  $\cos\theta\simeq 1-\theta^2/2$ and we obtain
\begin{eqnarray}
\label{class13}
\rho(\theta)=\frac{\sqrt{3}M}{\pi(x-x_c)^{1/2}}\left \lbrack 1-\frac{3\theta^2}{4(x-x_c)}\right \rbrack^{1/2},
\end{eqnarray}
for $\theta\le\theta_c=\sqrt{4(x-x_c)/3}$ and $\rho(\theta)=0$ for $\theta_c\le \theta\le \pi$. In this limit, the integrals (\ref{ip19}) can be approximated by 
\begin{eqnarray}
\label{class14}
{\cal I}_{3,m}(x)=\frac{1}{3\sqrt{2}}(x-x_c)-\frac{1}{18\sqrt{2}}m^2 (x-x_c)^2.
\end{eqnarray}
Therefore, when $x\rightarrow x_c$, the magnetization, the inverse polytropic temperature and the total energy are given by
\begin{eqnarray}
\label{class15}
b\simeq 1-\frac{1}{6}(x-x_c),
\end{eqnarray}
\begin{eqnarray}
\label{class16}
\eta\sim \frac{27}{2(x-x_c)^2},
\end{eqnarray}
\begin{eqnarray}
\label{class17}
\epsilon \simeq -2+\frac{4}{3}(x-x_c).
\end{eqnarray}
Finally, using equations (\ref{mr2}) and (\ref{cph3}), the parameter $x$ can be related to the maximal value of the distribution $\mu$ or to the Planck constant $\chi$ by the relation
\begin{eqnarray}
\label{class18}
\mu= \frac{1}{\chi}\sim \frac{3}{2(x-x_c)}.
\end{eqnarray}

\end{document}